%% file: main.tex
\pdfoutput=1
\documentclass[10pt,conference]{IEEEtran} 
\IEEEoverridecommandlockouts

\usepackage{xcolor}
\usepackage{xspace}
\usepackage{booktabs}
\usepackage{longtable}
\usepackage{array}
\usepackage{ragged2e}
\usepackage{tabularx}
\usepackage{enumitem}
\usepackage[switch]{lineno}
\usepackage[numbers,compress]{natbib}
\usepackage{url}

 \usepackage{tikz}
  
\usepackage{pgfkeys}

\def\BibTeX{{\rm B\kern-.05em{\sc i\kern-.025em b}\kern-.08em
    T\kern-.1667em\lower.7ex\hbox{E}\kern-.125emX}}

\usepackage[most]{tcolorbox}

\newtcolorbox{findingsbox}[1][]{
  enhanced,
  colback=gray!5,
  colframe=black!70,
  colbacktitle=black!70,
  coltitle=white,
  boxrule=0.5pt,
  arc=2pt,
  left=4pt, right=4pt, top=4pt, bottom=4pt,
  fonttitle=\bfseries\small,
  title={Key findings: #1},
}

\newcommand{\val}[1]{\pgfkeysvalueof{/ed/#1}}
\newcommand{\hmval}[1]{\pgfkeysvalueof{/hp/#1}}
\newcommand{\cmpval}[1]{\pgfkeysvalueof{/cmp/skill/#1}}
\newcommand{\screenval}[1]{\pgfkeysvalueof{/screening/#1}}

\newcommand{\valpct}[2][0]{\pgfmathprintnumber[fixed,precision=#1]{\pgfkeysvalueof{/ed/#2}}}
\newcommand{\hmvalpct}[2][0]{\pgfmathprintnumber[fixed,precision=#1]{\pgfkeysvalueof{/hp/#2}}}
\newcommand{\cmpvalpct}[2][0]{\pgfmathprintnumber[fixed,precision=#1]{\pgfkeysvalueof{/cmp/skill/#2}}}

\newcommand{\checkNum}[1]{\textcolor{orange}{\textit{#1}}}

\renewcommand{\checkNum}[1]{#1}

\input{tables/educators-question}
\input{tables/hiring-questions}
\input{tables/skills}


\input{values/educator-demographics-values}
\input{values/hiring-demographics-values}
\input{values/educator-q3-values}
\input{values/educator-q3-4-values}
\input{values/educator-q4-values}
\input{values/educator-q3-7-values}
\input{values/educator-q4-3-values}
\input{values/hiring-q3-values}
\input{values/hiring-q4-values}
\input{values/educator-q5-values}
\input{values/hiring-q5-values}
\input{values/skill-comparison-values}
\input{values/screening-values}
\input{values/hiring-q3-6-values}
\input{values/hiring-q4-12-values}

\input{figures/q33-stacked}
\input{figures/q47-stacked}
\input{figures/skills-importance}
\input{figures/skills-readiness}

\input{figures/skills-importance-likert}
\input{figures/skills-readiness-likert}
\input{findings/findingsrq1}
\input{findings/findingsrq2}
\input{findings/findingsrq3}

\begin{document}
\title{Four Years of GenAI: How Educators and Industry Adapted Their Assessment Strategies}

\author{\IEEEauthorblockN{May Mahmoud}
\IEEEauthorblockA{\textit{New York University Abu Dhabi}\\
Abu Dhabi, United Arab Emirates \\
m.mahmoud@nyu.edu}
\and
\IEEEauthorblockN{Nisa Shahid}
\IEEEauthorblockA{\textit{New York University Abu Dhabi}\\
Abu Dhabi, United Arab Emirates \\
ns5376@nyu.edu}
\and
\IEEEauthorblockN{Izah Sohail}
\IEEEauthorblockA{\textit{New York University Abu Dhabi}\\
Abu Dhabi, United Arab Emirates \\
is2587@nyu.edu}
\and
\IEEEauthorblockN{Gulshan Sharma}
\IEEEauthorblockA{\textit{New York University Abu Dhabi}\\
Abu Dhabi, United Arab Emirates \\
gs4615@nyu.edu}
\and
\IEEEauthorblockN{Hanan Salam}
\IEEEauthorblockA{\textit{New York University Abu Dhabi}\\
Abu Dhabi, United Arab Emirates \\
hanan.salam@nyu.edu}
\and
\IEEEauthorblockN{Sarah Nadi}
\IEEEauthorblockA{\textit{New York University Abu Dhabi}\\
Abu Dhabi, United Arab Emirates \\
sarah.nadi@nyu.edu}
}


\maketitle

\begin{abstract}
GenAI's ability to solve a wide range of software engineering tasks is reshaping the software industry, raising the question of what an entry-level software engineer looks like today. In this work, we investigate whether the core skills for entry-level engineers have changed with GenAI and how the assessment of those skills has changed. Through two surveys, we consider both the educational and hiring perspectives, drawing on responses from 56 educators and 24 hiring professionals across diverse geographic regions to identify gaps and misaligned expectations. We find that educators frame their GenAI policies primarily around academic integrity, whereas GenAI use in industry is near-universal, with organizational policy governing how it should be used rather than whether to use it. Policies on GenAI use during interviews are still emerging, and respondents are split on whether they prefer candidates who demonstrate GenAI skills. Both populations are moving toward GenAI-resistant assessments built on observable real-time interactions and higher-order tasks, and both agree on which skills matter most, rating critical evaluation of AI-generated output, responsible and effective use of GenAI tools, and the ability to learn and adapt independently highly. However, they disagree on graduate readiness in these skills: hiring professionals report larger gaps, with the disagreement concentrated on foundational, non-GenAI-specific skills.
\end{abstract}


\section{Introduction}
Large Language Models (LLMs), and especially LLM-based agents, are reshaping how software systems are currently built. These tools have shown strong capability in software engineering tasks such as code development and debugging \cite{hassan2024towards, hou2024large}.
With this new reality, there are concerns about the changing role and expectations of entry-level software engineers, or fresh computer science graduates in general \cite{Per25,feng2026junior}.
Educators are already raising strong concerns about students' overreliance on Generative AI (GenAI) tools and the possibility that graduates may struggle to perform what has long been considered core software engineering skills independently \cite{Pra23,Che24,Mah26}. The current situation raises two key questions: (1) \textit{have core skills required of entry-level software engineers changed}? and (2) \textit{has assessment of these skills evolved in response to the advent of GenAI?}
Addressing these questions requires consideration of both educational and industry perspectives. Hiring professionals define the skills expected of candidates and educators prepare students to meet them, with both designing assessments to evaluate those skills from their respective sides. Since graduating seniors are effectively entry-level candidates, alignment between both sides determines whether computer science graduates are adequately prepared for entry-level software engineering roles.

However, research on GenAI in computing has largely examined educational and industry perspectives in isolation.
A growing body of computing education literature examined how educators are adapting to GenAI, including changes in teaching practices and the suitability of different assessment approaches  \cite{Pra23, Lau23, Pet24, Kan24, Ohm25, Fer25}. In parallel, studies have examined the impact of GenAI on industry, including how developers use GenAI, evolving expectations for entry-level software engineers, and emerging changes in hiring practices \cite{Ste22, Lou25, Che25c}. Most of these studies do not consider both perspectives or examine their alignment, and many are further limited to specific institutional or regional contexts. Moreover, pre-GenAI research identified an academia–industry gap in computing, particularly regarding workplace readiness \cite{Oca25, Rad14,Ext18}.
The rapid pace of GenAI development further complicates the issue, with even relatively recent studies capturing only a snapshot of an increasingly dynamic landscape. 

This gap in the literature motivates our work. We examine how computer science and software engineering educators, as well as industry professionals, have adapted their assessment and evaluation practices in response to GenAI, and whether there is misalignment between the competencies emphasized in higher education and those prioritized in entry-level software engineering hiring. We address this through two surveys. The first targets educators in undergraduate computer science and related programs and examines changes in teaching and assessment practices, perceptions of GenAI's impact, and views on essential software engineering skills and on graduating students' proficiency in those skills. The second targets industry professionals involved in hiring and examines changes in hiring and candidate evaluation practices, observed skill gaps, and assessments of the readiness of newly graduating software engineers. Both surveys include the same set of twelve software engineering skills, enabling direct comparison of responses across the two populations and identification of areas of alignment and divergence.
Our study is organized around three research themes, each comprising multiple research questions:

\noindent \textbf{Theme 1: Higher Education Perspective:}\\
\noindent
\textbf{RQ1.1} What policies govern the use of GenAI in higher-education coursework?\\
\textbf{RQ1.2} How have educators adapted their assessment practices in response to GenAI?

\noindent \textbf{Theme 2: Industry Perspective:}\\ 
\textbf{RQ2.1} What are organizational policies and industry expectations regarding the use of GenAI in software development?
\textbf{RQ2.2} How have hiring professionals adapted candidate-evaluation practices and expectations of entry-level software engineers in response to GenAI?

\noindent \textbf{Theme 3: The Higher Education-Industry Alignment}\\
\textbf{RQ3.1} How do educators and hiring professionals perceive graduate preparedness and the importance of specific skills in a GenAI-assisted environment?\\
\textbf{RQ3.2} To what extent do educators' and hiring professionals' competency expectations align, and do gaps in graduate preparedness exist?

Our results draw on \screenval{ed/final/n} faculty across \val{d5/n-countries} countries and \screenval{hp/final/n} industry professionals across \hmval{d4/n-countries} countries. Both groups redesign assessments to be GenAI-proof, using live assessments to monitor or exclude GenAI or shifting to higher-order, open-ended tasks that GenAI finds hard. Educators frame policy around academic integrity, while industry focuses on governance, security, and privacy; policies on candidate GenAI use in interviews are still emerging. The two populations agree closely on which skills matter most, with higher-order and GenAI-specific skills such as critical evaluation of AI-generated output topping both rankings. They diverge on graduate readiness: hiring professionals perceive larger gaps than educators, with a concentration on foundational skills. We include the surveys, scripts, and results in our replication package \footnote{Replication Package: \url{https://figshare.com/s/43531ffb4ab924ea8d8c}}.



\section{Related Work}
 We review research related to  GenAI and CS education, and GenAI and industry practice, including skill gaps.
\paragraph{GenAI and CS Education}
A growing body of work examines benefits and challenges of GenAI for computing education. We focus on educators' perspectives rather than students' experiences~\cite{Mah26,Han23,Cho25,Amo24}. Studies from 2023--2024 show that computing educators are still negotiating how GenAI should be governed and assessed in their courses. Early work found that institutional policies were often unclear, producing disagreement between instructors and students over acceptable use~\cite{Pra23}. Ethical use was a recurring concern~\cite{Pra23,Lau23,She24}; in one survey, most participants reported that nearly all of their students used GenAI tools unethically~\cite{Pra23}. Consistent with this uncertainty, \citet{Ali24} found substantial variation in GenAI guidance across syllabi from U.S. R1 institutions: policies were typically embedded in academic-integrity statements, more often restrictive than permissive, and half prohibited use outright. Although some educators saw benefits for debugging, code comprehension, and language support, many worried that over-reliance would erode foundational skills~\cite{Lau23,She24}. Educators' positions ranged from banning GenAI to embracing it as preparation for industry, with broad agreement that assessment practices must change~\cite{Lau23,She24}. We draw on survey instruments from this prior work to inform our survey design. 

A second line of work integrates GenAI into computing courses through interventions and experience reports~\cite{Vad24,Pet24,Ran24,Kha26,Sal25}. These studies commonly emphasize that students still need foundational understanding to use GenAI effectively and to avoid overreliance. A related body of work redesigns assessments in response to GenAI, including oral exams~\cite{Ohm25}, code interviews~\cite{Kan24}, conversational exams~\cite{Bar26}, and assignment-driven written quizzes under an open GenAI policy~\cite{Chu26}. 
These studies suggest that conventional artifact-based assessment is increasingly insufficient for verifying student understanding, motivating a shift toward assessment settings in which students must demonstrate competence without fully offloading work to GenAI.
\citet{Chir26} surveyed approximately 95,000 students from all disciplines across 20 U.S. public research institutions. They found that computer science students reported the highest GenAI use (62\%). They argue that disciplines should reform assessment around learning outcomes rather than simply permitting or banning GenAI, either constraining its use through assessment design or incorporating it to teach responsible practice.

\paragraph{GenAI and Industry Practices}

Industry-focused research has examined how software developers use GenAI in day-to-day practice~\cite{Kho24,Ser24,Str25,Tab25,Per25,Wei24,But24}, documenting common use cases and areas of high adoption, including testing, documentation, and other development tasks~\cite{Kho24,Ser24,Tab25,Per25}. Other studies examined productivity effects and shifts in developers' attitudes toward GenAI-assisted work~\cite{Wei24,But24,Par24b}. However, this literature focuses primarily on adoption and use among practicing developers, rather than on hiring expectations or the skills required of entry-level software engineers in the GenAI era. A notable exception is the 2024 interview study by \citet{Che25c}, who interviewed 32 recruiters in Hawaii to examine how GenAI is affecting candidate screening and evaluation. They found that, although most organizations had not formally changed their hiring criteria, recruiters valued candidates who could use GenAI effectively while still demonstrating fundamental software engineering skills. We adapt some of their survey questions, while broadening the focus beyond one regional labor market. We further examine skill importance and readiness gaps for entry-level software engineers.

\educatorsqstab
Prior work documented substantial academia-industry gaps in computing before the rise of GenAI, particularly around workplace readiness~\cite{Oca25}. Through interviews with hiring professionals, primarily in the United States, \citet{Rad14} found that common deficiencies among entry-level candidates included limited project experience, difficulty using configuration management tools, weak communication, and insufficient problem-solving skills. They also reported gaps in software testing and code commenting. Similarly, \citet{Ext18} found that industry most valued problem solving, independent learning, and communication, while undergraduate programs varied in their coverage of these skills.  \citet{Lou25} studied what hiring professionals expect from entry-level software engineers and how these expectations appear in job postings, without focusing on GenAI.

The academia-industry gap, therefore, predates GenAI, but its current shape remains unclear. Existing work typically examines academic assessment practices, industry adoption of GenAI, or pre-existing skill gaps in isolation, rarely connecting academic and industry perspectives. We address this gap by studying how assessment and evaluation practices are adapting to GenAI, whether educators and industry professionals prioritize different skills for entry-level engineers, and how both groups perceive graduate readiness. By surveying participants across multiple regions, we provide an updated view of the academia-industry gap in the GenAI era.

\section{Methods}

We develop two surveys: one for educators and one for hiring professionals. Eligible educators are those who teach in a computing-relevant program and had taught an undergraduate course within the past two years. We focus on undergraduate education, since we are interested in preparation for entry-level software engineering roles. Eligible hiring professionals are those involved in hiring or evaluating software engineering candidates. Our survey design is informed by prior GenAI-related surveys of educators, and hiring professionals~\cite{Pra23,Lau23,She24,Che25c}, which helped identify common topics and support partial comparison with prior findings. We implemented both surveys in Qualtrics.
We piloted each survey to estimate completion time and identify ambiguous wording; both participants finished in under 15 minutes and recommended making several open-ended questions optional.

\subsection{Survey Distribution}
We recruit participants using a combination of convenience and snowball sampling~\cite{Goo61}. We contacted educators and industry professionals within the authors' professional networks via email and LinkedIn. We encouraged recipients to share the survey with eligible colleagues. To broaden participation, we also advertised the survey through professional social-media channels and relevant online communities. Because our recruitment relied on network-based dissemination and the survey was anonymous, the total number of individuals exposed to the survey is unknown. Consequently, we cannot calculate an exact response rate. We obtained IRB approval from our institution prior to conducting the surveys.

\subsection{Educator Survey (ED)}
Table~\ref{tab:educator-questions} summarizes the educator survey and its relationship to the research questions.
For easier referencing, we assign each question a label (e.g., ED-Q3.1).
The survey begins with demographic and screening questions, which we omit from the table for brevity.
The survey then comprises three question sets.
Question set 1 examines educators' use of GenAI tools, their influence on teaching-material preparation, and perceived changes to instructional practice. Question set 2 focuses on course policies governing GenAI use, assessment adaptations introduced in response to GenAI, and educators' views regarding the role of GenAI in assessment.
\hiringsqstab
Question set 3 covers educators' perceptions of student GenAI use, curriculum preparedness, and agreement with nine benefit- and concern-framed statements about student adoption. We deliberately mix statement framing to reduce agreement bias~\cite{Cro46}.

A central component of the educator survey w.r.t curriculum preparedness is evaluating the importance of twelve software-engineering competencies and the proficiency of graduating students in each. These competencies are
(1) problem decomposition, (2) code comprehension, (3) code development, (4) code review, (5) debugging, (6) documentation, (7) system-level design, (8) algorithmic thinking, (9) critical evaluation of AI-generated output, (10) the ability to learn and adapt independently, (11) collaboration and communication, and (12) responsible and effective use of GenAI tools.

\subsection{Hiring Professional Survey (HP)} 

Table~\ref{tab:hiring-questions} summarizes the hiring-professional survey. Following demographic and screening questions, the survey again consists of three question sets. 
Question set 1 covers participants' professional and organizational use of GenAI. Question set 2 focuses on hiring and evaluation practices, including whether organizations modified candidate-assessment processes, interview expectations, or evaluation criteria in response to GenAI.
We also ask participants whether they permit GenAI usage during technical interviews and their confidence in assessing candidates' underlying skills in a GenAI-enabled environment.
Finally, Question set 3 explores perceptions of graduate readiness for contemporary software-engineering practice, including views on curriculum preparation, the importance of GenAI-related education, and perceived skill gaps among recent graduates.
As in the educator survey, participants evaluate the same set of twelve software-engineering competencies. Using a shared competency framework across both surveys enables direct comparison between educator and hiring-professional perspectives.

\subsection{Data analysis}
We analyze closed-ended questions quantitatively using descriptive statistics. For optional questions, we compute percentages over substantive responses unless stated otherwise; responses are considered substantive if they were nonblank and, for open-ended items, interpretable. We analyze open-ended responses using inductive open coding~\cite{Cor90}, allowing codes to emerge from the data. 
After consolidating initial codes, we use Claude Opus 4.7
to support codebook refinement by suggesting more granular codes or grouping related ones, while reviewing and finalizing the codebooks ourselves. For each of the 12 skills, we test differences between educators and hiring professionals using Mann-Whitney U tests~\cite{mann1947test} with Benjamini-Hochberg false-discovery-rate corrections~\cite{benjamini1995controlling} applied.
We run the tests separately for importance (ED-Q5.4 vs. HP-Q4.7) and graduate readiness (ED-Q5.5 vs. HP-Q5.3). We report rank-biserial correlations as effect sizes~\cite{kerby2014simple}. 


\section{Results}
\subsection{Demographics}
We summarize the main demographic characteristics here and report complete distributions in the online artifact.
\paragraph{Educators} After excluding incomplete responses and participants who did not meet our inclusion criteria, we retained \screenval{ed/final/n} valid educator responses from \val{d5/n-countries} countries. The most common countries were Canada (\valpct{d5/canada/pct}\%), the United Arab Emirates (\valpct{d5/uae/pct}\%), and the United States (\valpct{d5/us/pct}\%). Most participants held faculty positions (\valpct{d3/tenure-track/pct}\%), with a median of \val{d4/median} years of teaching experience (range: \val{d4/min}--\val{d4/max}). They taught a median of two topics, most commonly software engineering (\valpct{d6/se/pct}\%), introductory programming (\valpct{d6/intro-programming/pct}\%), object-oriented programming (\valpct{d6/oop/pct}\%), and web or mobile development (\valpct{d6/web-mobile/pct}\%). Participants taught across undergraduate levels, most often junior and senior students (\valpct{d7/junior/pct}\% and \valpct{d7/senior/pct}\%, respectively).

\paragraph{Hiring professionals} After excluding incomplete responses and participants who did not meet our inclusion criteria, we retained \screenval{hp/final/n} valid hiring-professional responses from \hmval{d4/n-countries} countries. The largest groups were from the United States (\hmvalpct{d4/us/pct}\%), the United Arab Emirates (\hmvalpct{d4/uae/pct}\%), Germany (\hmvalpct{d4/germany/pct}\%), and Canada (\hmvalpct{d4/canada/pct}\%). Respondents are experienced practitioners, with a median of \hmval{d3/median} years of industry experience (range: \hmval{d3/min}--\hmval{d3/max}). Most hold senior or leadership roles and are directly involved in hiring, including \hmvalpct{d5/conduct-interviews/pct}\% who conduct technical interviews and \hmvalpct{d5/final-decision/pct}\% who make final hiring decisions. Participants mainly work in the technology sector (\hmvalpct{d8/tech/pct}\%), and organization sizes ranged from startups to companies with more than 1,000 employees.

\subsection{RQ1. Changes in Teaching and Assessment Practices}
For RQ1, we analyze responses to the first two question sets of the educators' survey.
\subsubsection{RQ1.1 GenAI in Teaching and Policy Changes}
\paragraph{GenAI in Teaching}
Most educators (\valpct{q3/q31/users/pct}\%) use at least one GenAI tool for teaching preparation, with ChatGPT dominating both actual use (\valpct{q3/q31/chatgpt/pct-of-users}\% of users) and stated top choice (\valpct{q3/q32/chatgpt/pct}\%).  Their usage is selective, concentrating mainly on code- and assessment-artifact tasks (e.g., generating examples or sample solutions), and drops sharply for student-facing tasks such as giving feedback or grading (ED-Q3.3). The same pattern appears in perceived value, where educators most often point to assessment-oriented tasks as where GenAI helps most (ED-Q3.4). Educators describe a shift in their role as instructors (ED-Q3.9), with the most common framing being teaching AI literacy (\checkNum{25\%}), and becoming a facilitator or mentor (\checkNum{9.1\%}). As one educator put it, \textit{``My role is at least 50\% teaching them when, why, and how to use AI responsibly.'' (PE45)} A notable group reports no change in their role (\checkNum{18.2\%}). Some emphasize a pedagogical shift toward harder-to-automate skills such as system design and code review: \textit{``the goal is no longer to teach students how to code, it is teaching them how to read/trace code'' (PE40).}

\edqfourseventfigure

\paragraph{Policy and guidance (ED-Q4.1--ED-Q4.3)}
Every educator reports providing at least some guidance on student GenAI use: \valpct{q4/q41/yes-formal/pct}\% provide a clear, formally documented policy, \valpct{q4/q41/yes-informal/pct}\% communicate informally, and \valpct{q4/q41/some/pct}\% provide some non-explicit guidance. Among those who provide guidance, policies most often address allowed vs.\ prohibited uses (\valpct{q4/q42/allowed-prohibited/pct}\%), academic integrity (\valpct{q4/q42/integrity/pct}\%), and consequences of misuse (\valpct{q4/q42/consequences/pct}\%).
Citation or disclosure requirements (\valpct{q4/q42/citation-disclosure/pct}\%), assessment-specific rules (\valpct{q4/q42/assessment-rules/pct}\%), and tool-specific guidance (\valpct{q4/q42/tool-guidance/pct}\%) are less common. A subset of \val{q43/n} educators shared an excerpt or summary of their policy (ED-Q4.3), from which seven themes emerge. Most specify allowed vs. prohibited usage (\valpct{q43/code/permission-scope/pct}\%), often varying by task or assignment type: \textit{``You may use generative AI technologies such as Gemini or Claude for assisting in code development. We Very Strongly Discourage using AI to generate writing responses, but we encourage its use for editing your writing'' (PE15).} Disclosure requirements appear in over half (\valpct{q43/code/disclosure/pct}\%) of shared policies, though only \valpct{q43/code/citation-format/pct}\% specify a concrete citation format. Many place responsibility on the student for the correctness and understanding of GenAI-generated content (\valpct{q43/code/accountability/pct}\%), and around \valpct{q43/code/integrity-framing/pct}\% invoke academic integrity or specify penalties. Fewer frame rules with pedagogical rationale (\valpct{q43/code/educative/pct}\%), e.g., \textit{``over-reliance on AI can hinder independent thinking, creativity, and your overall learning'' (PE54)}, or provide worked examples of acceptable vs.\ unacceptable use (\valpct{q43/code/worked-examples/pct}\%). 

\subsubsection{RQ1.2 GenAI and Assessment}
\paragraph{Assessment Modification}
The majority of participants have modified their assessments (ED-Q4.4), with \valpct{q4/q44/yes-substantially/pct}\% modifying them substantially and \valpct{q4/q44/yes-slightly/pct}\% modifying them slightly. Only \valpct{q4/q44/no-considering/pct}\% of participants did not change their assessment yet but are considering it, while \valpct{q4/q44/no-wont/pct}\% did not modify it and do not intend to. Around \valpct{q4/q49/yes/pct}\% report redistributing grade weights across assessments, \valpct{q4/q49/no-considering/pct}\% report that they did not but are considering redistribution, and \valpct{q4/q49/no-wont/pct}\% did not and do not intend to. The two main reasons for participants to modify their assessments (ED-Q4.5) are that students are relying heavily on GenAI to complete their work without understanding it, and that GenAI made some assessments invalid, as students can easily solve them without assessing the real skill. A small number of educators modified their assessments to adapt to GenAI to require \textit{``more ambitious projects''} because students \textit{``will be using AI'' (PE44).} For those who did not modify their assessment, we note that they already report using assessments that mitigate GenAI-related issues, such as in-person, closed-book exams. Some still mention in ED-Q4.6 that they are considering changes, as one realized that \textit{``some students are not learning anything at all, even though they are submitting pull requests with the outcomes of the assignments'' (PE46)}.

Figure~\ref{fig:ed-q47-stacked} summarizes how educators changed specific assessment formats (ED-Q4.7).  
Oral exams and interviews are the most changed assessments, where \valpct{q4/q47/oral/substantive/increased-pct}\% report increasing or beginning to use this format, and no respondent reported decreasing it. The next top changed format is in-class or supervised assessment, where \valpct{q4/q47/in-class/substantive/increased-pct}\% mentioned they increased or began using this format, as opposed to \valpct{q4/q47/in-class/substantive/decreased-pct}\% who mentioned decreasing such format.
For reflective or justification-based components, \valpct{q4/q47/reflective/substantive/increased-pct}\% of respondents increased or began using them, while \valpct{q4/q47/reflective/substantive/decreased-pct}\% decreased or removed them. The remaining two formats show smaller movement, with \valpct{q4/q47/process-based/substantive/increased-pct}\% increasing process-based assessment (and \valpct{q4/q47/process-based/substantive/no-change-pct}\% reporting no change) and \valpct{q4/q47/group-work/substantive/increased-pct}\% increasing group work (and \valpct{q4/q47/group-work/substantive/no-change-pct}\% reporting no change). We can see from these results that educators are shifting toward assessment formats where it is more difficult for students to delegate or completely do with GenAI, such as oral exams or supervised assessments.
Some educators mention further assessment changes in ED-Q4.8 where they redesigned tasks such that GenAI cannot fully help with, or changed the tasks' nature (e.g., from writing code to debugging code or using case studies). Others mention tracking document edit history, while one participant actually tests the assessment by running it through a GenAI tool to see if it can solve it; if not, it is a suitable assessment.

\paragraph{Educator Acceptance and Practices Regarding GenAI-written code Submissions}
When it comes to educators' acceptance and practices regarding students' submissions of GenAI-written code, \valpct{q4/q410/positive/pct}\% agreed to some degree that they will accept it if students fully understand it, as opposed to \valpct{q4/q410/neutral/pct}\% who were neutral and \valpct{q4/q410/negative/pct}\% who disagreed to some extent. \valpct{q4/q411/positive/pct}\% reported that they actively check submitted assignments for GenAI-written code often or always, with another \valpct{q4/q411/neutral/pct}\% checking sometimes and \valpct{q4/q411/negative/pct}\% rarely or never.  Only \valpct{q4/q412/positive/pct}\% of educators felt confident they can reliably detect GenAI-written code often or always, with \valpct{q4/q412/neutral/pct}\% reporting they can do so sometimes and \valpct{q4/q412/negative/pct}\% rarely or never.

\vspace{-0.1cm}
\findingrqone 
\vspace{-0.4cm}

\subsection{RQ2. Changes in Hiring Practices}
For RQ2, we analyze responses to the first two question sets of the hiring-professional survey.
\subsubsection{RQ2.1 Professional and Organizational Use of GenAI}

\paragraph{GenAI Tool Adoption} 
On average, respondents are familiar with \hmval{q3/q31/mean-tools-per-respondent} GenAI tools per person. 
Among the participants, \hmvalpct{q3/q32/top-box/pct}\% report that they use GenAI for work tasks \textit{Often} or \textit{Always} (\hmvalpct{q3/q32/always/pct}\% \textit{Always}), \hmvalpct{q3/q32/middle/pct}\% report \textit{Sometimes}, with no respondent selecting \textit{Never} or \textit{Rarely}.
We find that most common choice as the preferred tool.
GenAI adoption at the organization level is also extensive: \hmvalpct{q3/q34/high-use/pct}\% of respondents report that their organization uses GenAI either \textit{Widely} or as \textit{Standard practice}, \hmvalpct{q3/q34/limited/pct}\% report \textit{Limited} and \hmvalpct{q3/q34/some-teams/pct}\% \textit{Used by some teams};  only \hmvalpct{q3/q34/not-used/pct}\% report that GenAI is \textit{Not used} in their organization. 

\paragraph{Policy and Guidelines for GenAI use} Most organizations have a formal policy on use of GenAI in software development (HP-Q3.5); \hmvalpct{q3/q35/clear-formal/pct}\% report a clear and formal policy, \hmvalpct{q3/q35/informal/pct}\% report informal guidance, and only \hmvalpct{q3/q35/none/pct}\% report no guidance. 
Our open coding of the shared policy details (HP-Q3.6) produced \checkNum{9} codes organized across two dimensions: (1) \emph{Mandate} which represents how the organization positions GenAI use, whether required, encouraged, or prohibited, and (2) \emph{Governance}, which represents how the organization regulates GenAI use, for example, by listing approved tools or establishing accountability. We add an \emph{Other} dimension for responses that do not fit into these dimensions. 
\emph{Governance} themes are more prevalent than \emph{Mandate} themes (\hmvalpct{q36/group/governance/pct}\% versus \hmvalpct{q36/group/mandate/pct}\%). This suggests that when organizations articulate a GenAI policy, they more often spell out how use is regulated than whether use is required. On the Mandate side, \hmvalpct{q36/code/encouraged-use/pct}\% of policies actively encourage GenAI use and \hmvalpct{q36/code/required-use/pct}\% require it, while only \hmvalpct{q36/code/mandatory-training/pct}\% link adoption with mandatory training or onboarding. On the Governance side, \hmvalpct{q36/code/accountability/pct}\% of policies place explicit responsibility on developers for the correctness of AI-generated output, \hmvalpct{q36/code/privacy-security/pct}\% specify privacy, security, or intellectual-property controls, and \hmvalpct{q36/code/approved-tools/pct}\% restrict use to an approved set of tools. The \hmvalpct{q36/code/tool-inventory/pct}\% remaining responses under \emph{Other} merely list which tools are in use. 

\subsubsection{RQ2.2 Current Interview Practices}

\paragraph{Policies and Changing Criteria}
We find that only \hmvalpct{q4/q41/yes/pct}\% of participants report that their organization has developed official guidelines or policies for evaluating candidates in light of GenAI tools (HP-Q4.1), while \hmvalpct{q4/q43/yes/pct}\% report having changed the criteria they use to evaluate a candidate's software development skills (HP-Q4.3) and \hmvalpct{q4/q45/yes/pct}\% report permitting candidates to use GenAI tools during a technical interview (HP-Q4.5). This shows a split in opinion, suggesting that organizational and professional policies regarding candidate evaluation in the presence of GenAI are still evolving. 

In HP-Q4.4, 12 of the \checkNum{14} respondents who report changing their criteria in HP-Q4.3 provide a summary of the changes, revealing four patterns.
The first and most common pattern is to observe candidates using GenAI in real time. Rather than testing unaided coding, interviewers watch how candidates work with AI tools on realistic tasks. One interviewer asks candidates to \textit{``share their screen and do some vibe coding to see how they prompt'' (PHP3)}, another \textit{``moved from whiteboard interviews to pair programming exercises where we can watch how the candidate uses their AI tools'' (PHP14)}, and others assess candidates on \textit{``brownfield questions (fix a bug/implement a feature) or code review questions'' (PHP7)} carried out with AI assistance. The second pattern shifts emphasis toward higher-order skills that GenAI does not readily replace. Interviewers place \textit{``less emphasis on routine coding tasks'' (PHP12)} and instead look for \textit{``critical thinking skills and less on the actual technical code writing skills'' (PHP8)}, or for creativity in designing broader architectures and flows that can be effectively executed with, and by, GenAI tools (PHP23).
The third pattern assesses what candidates know about AI itself, rather than watching them use it. The focus is knowledge, ranging from conceptual understanding to familiarity with concrete AI-assisted workflows: whether a candidate can \textit{``understand the full lifecycle of using AI in coding'' (PHP13)}; others probe ML and LLM knowledge or concrete AI-assisted workflows such as refactoring and PR review (PHP17, PHP20, PHP22). 
A fourth, less common pattern shifts evaluation towards real work, with one participant relying on \textit{``on[-]the[-]job testing, [with] no attention to pre-employment assessments or portfolio ''} (PHP21).

The most common reason for allowing GenAI use during an interview (HP-Q4.6) is that candidates are expected to use these tools on the job, so they should not be barred from using them during the interview.  As one participant puts it, \textit{``it would be weird to forbid them during the interview process''} when the tools will be mandated on the job (PHP7), and another permits them because \textit{``it's more realistic to allow them to use tools they'd be using in their day-to-day work''} (PHP12). Another reason mentioned is to allow the interviewer to assess the candidate's GenAI proficiency itself. For example, one participant mentions that they allow GenAI tools \textit{``just to see how they [prompt], and what their methodology''} is (PHP3), or to check for \textit{``responsible use of the tools, specifically that the candidate is reviewing the outputs and understands them''} (PHP14).  We also see that AI-resistant question design already makes allowing GenAI in interviews low-stakes. For example, PHP18 says that they ask \textit{``open-ended ''} problems such that using \textit{``AI is not a clear advantage''} (PHP18). 

Those who do not permit the use of GenAI mention two main reasons. One is that they want to see the candidate's raw skills unassisted with AI. As one participant mentions, they want to assess \textit{``baseline skills''} which may be \textit{``difficult to do [...] with GenAI tools in the loop''} (PHP1), reasoning that \textit{``people without these skills cannot develop good software just using GenAI tools''} (PHP6). Others mentioned that they do not allow them yet, but indicated that this may change in the future (PHP10) or that \textit{``technical interview questions are still not updated to match this''} (PHP13).

\paragraph{Candidates Assessments and GenAI} 
\findingrqtwo
Regarding asking candidates about their personal experience with GenAI tools (HP-Q4.8), \hmvalpct{q4/q48/positive/pct}\% report doing so \textit{Often} or \textit{Always}, 
\hmvalpct{q4/q48/neutral/pct}\% \textit{Sometimes}, and \hmvalpct{q4/q48/negative/pct}\% \textit{Rarely} or \textit{Never}. 
When it comes to their confidence in detecting AI-assisted work versus the genuine candidate skills during the interview process (HP-Q4.9), only \hmvalpct{q4/q49/positive/pct}\% report being \textit{Very} or \textit{Extremely confident}, \hmvalpct{q4/q49/neutral/pct}\% report \textit{Moderately confident}, and \hmvalpct{q4/q49/negative/pct}\% report \textit{Slightly} or \textit{Not at all confident}. This distribution closely mirrors that reported by educators regarding detecting AI-generated work in submitted assignments. In terms of preferring a candidate who can demonstrate GenAI skills (HP-Q4.10), we find that participants are split in opinion: \hmvalpct{q4/q410/positive/pct}\% report a \textit{Strong Preference} or treat such skills as \textit{Always a Priority}, \hmvalpct{q4/q410/neutral/pct}\% report \textit{Moderate Preference}, and \hmvalpct{q4/q410/negative/pct}\% report \textit{No Preference}. 
At the same time, \hmvalpct{q4/q411/yes/pct}\%  of respondents report facing challenges when assessing candidates in the presence of GenAI (HP-Q4.11).
Respondents who elaborate in HP-Q4.12 either describe challenges they face (\checkNum{60}\% of substantive respondents) or indicate lack of challenges due to mitigation strategies they adopted (\checkNum{44}\% of substantive responses); some responses included both challenges and mitigation strategies. The most commonly mentioned difficulty is that GenAI makes it hard to verify a candidate's true ability. Responses describe candidates who present well but cannot back it up later; one notes that a candidate can use GenAI \textit{``to be very prepared for the interview''} yet, once hired, their \textit{``real work skills [do] not match the expected level''} (PHP5). Another response discusses how the participant can sense AI-generated material but cannot act on it, since \textit{``we have a hunch they're generated, but of course it's hard to prove''} (PHP6). A subtler form of the same problem is that GenAI can obscure strength as well as weakness, as one respondent puts it: \textit{``it's easy to judge negatives, but hard to assess true positive from false positive''} (PHP23). A different kind of challenge emerges where participants describe worry about candidates' over-reliance on GenAI. Some of these participants mention that candidates are \textit{``unable to explain their approaches''} and \textit{``aren't able to dive deeper into a problem''} without GenAI (PHP1). The remaining challenges appear less often, namely difficulty enforcing no-tool rules during interviews (PHP11) and inexperience with GenAI-era assessment, including the observation that \textit{``diversity of multiple GenAI tools makes evaluation harder''} (PHP17).
Among those who assess candidates effectively, two strategies emerge. The first keeps the assessment live and interactive, so GenAI use becomes visible, since asking a candidate to explain their work is \textit{``hard to fake, even with AI assistance''} (PHP4, PHP6). The second shifts the interview toward higher-order skills GenAI does not readily supply, so the \textit{``narrative of the interview changes from technical questions to system design and critical thinking questions''} (PHP8). Less often, respondents track candidates' AI prompts (PHP7) or add questions assessing GenAI proficiency as a skill in its own right (PHP13). These adaptations mirror those of educators: oral exams and in-class work that keep performance observable, or a shift toward higher-order skills that are harder to delegate to GenAI.

\subsection{RQ3. Graduate Preparedness in a GenAI-Assisted Context: Educators and Hiring Professionals' Perspectives}
For RQ3, we use the third question set of each survey and HP-Q4.11. We first discuss educators' views on graduate readiness and the importance of skills, then move on to hiring professionals' views, and finally compare the two.
\subsubsection{RQ3.1 Educators' and Hiring Professionals' views}
\paragraph{Educator Views}
We find that educators are skeptical when it comes to current curricula preparing students to use GenAI in professional settings (ED-Q5.3).  Only \valpct{q5/q53/positive/pct}\% of educators rate the curriculum as preparing students \textit{Well} or \textit{Very Well}, while \valpct{q5/q53/negative/pct}\% rated it as preparing students \textit{Not at All}  or only \textit{Slightly}, and the remaining \valpct{q5/q53/neutral/pct}\% considered preparation \textit{Moderate}. Educators acknowledge the wide use of GenAI by students:  \valpct{q5/q51/positive/pct}\% of educators believe \textit{Many} or \textit{Almost Everyone} of their students use GenAI on assignments (ED-Q5.1) and \valpct{q5/q52/positive/pct}\% report \textit{Many} or \textit{Almost Everyone} students use GenAI \textit{before} attempting work independently (ED-Q5.2).

When asked about skill importance (ED-Q5.4), educators' top three skills, rated as \textit{Very} or \textit{Extremely important} are: 
\val{q5/q54/top1/label} (\valpct{q5/q54/top1/pct}\%), \val{q5/q54/top2/label} (\valpct{q5/q54/top2/pct}\%), and \val{q5/q54/top3/label} (\valpct{q5/q54/top3/pct}\%). 
Overall, while none of the twelve skills have a majority voting of \emph{Slightly} or \emph{Not at all important}, there are still skills that educators see as less important.
For example, only \valpct{q5/q54/documentation/positive/pct}\% of educators consider Documentation as highly important, while \valpct{q5/q54/documentation/neutral/pct}\% are neutral about it. Similarly, for Code development, only \valpct{q5/q54/code-development/positive/pct}\% of educators consider it as highly important, while \valpct{q5/q54/code-development/neutral/pct}\% are neutral. When it comes to their views on graduating students having \textit{developed} the skills (ED-Q5.5), educators see proficiency as moderate at best. The top skills, rated as \textit{Proficient} or \textit{Highly proficient}, are \valpct{q5/q55/code-development/positive/pct}\% for \val{q5/q55/code-development/label}, \valpct{q5/q55/code-comprehension/positive/pct}\% for \val{q5/q55/code-comprehension/label}, and \valpct{q5/q55/algorithmic-thinking/positive/pct}\% for \val{q5/q55/algorithmic-thinking/label}. The skills educators rate as least proficient are Responsible and effective use of GenAI tools (only around \valpct{q5/q55/responsible-genai/positive/pct}\% of educators see graduates as proficient and \valpct{q5/q55/responsible-genai/neutral/pct}\% see them as moderately proficient, while \valpct{q5/q55/responsible-genai/negative/pct}\% see them as low proficiency), Critical evaluation of AI-generated output (\valpct{q5/q55/critical-eval-ai/positive/pct}\%,\valpct{q5/q55/critical-eval-ai/neutral/pct}\%, \valpct{q5/q55/critical-eval-ai/negative/pct}\%), and Documentation (\valpct{q5/q55/documentation/positive/pct}\%, \valpct{q5/q55/documentation/neutral/pct}\%, \valpct{q5/q55/documentation/negative/pct}\%). Interestingly, Critical evaluation of AI-generated output is rated as one of the most important for students to have and simultaneously rated as one of the weakest in graduating students. 

We find that \valpct{q5/q56/s3/positive/pct}\% agree (\textit{Agree} or \textit{Strongly agree}) that students submit GenAI-generated work without understanding it, with \valpct{q5/q56/s3/positive/extreme/pct}\% selecting \textit{Strongly agree}. Similarly, \valpct{q5/q56/s1/positive/pct}\% agree that students will become overly reliant on GenAI to solve their assignments, and \valpct{q5/q56/s7/positive/pct}\% agree that students cannot distinguish correct from incorrect GenAI-generated answers. A second cluster of concern-framed statements focuses on students not reporting their use of GenAI properly: \valpct{q5/q56/s9/positive/pct}\% agree that students use GenAI more than they report, and \valpct{q5/q56/s6/positive/pct}\% agree that GenAI has increased the frequency of academic integrity violations. 
In terms of benefits, \valpct{q5/q56/s2/positive/pct}\% agree that GenAI can help students explore topics more deeply and independently, \valpct{q5/q56/s8/positive/pct}\% agree that GenAI can enhance learning by providing personalized feedback, and only \valpct{q5/q56/s4/positive/pct}\% agree that GenAI can help students develop critical thinking skills when used appropriately. 

\skillimpfigure

\paragraph{Hiring Professionals' Views}
Based on HP-Q5.1, we find that only \hmvalpct{q5/q51/positive/pct}\% of professionals see that current curricula prepare graduates for a Gen-AI assisted industry \textit{to a large extent} or \textit{to a very large extent}, while \hmvalpct{q5/q51/negative/pct}\% rated it as occurring \textit{not at all} or only \textit{to a small extent}, and \hmvalpct{q5/q51/neutral/pct}\% considered preparation \textit{moderate}. 
This suggests that hiring professionals see a clear gap between what current curricula deliver and what GenAI-assisted industry expects. 
Consistent with this observation, \hmvalpct{q5/q52/positive/pct}\% of the participants see integrating GenAI into universities' curricula (HP-Q5.2) as important (\hmvalpct{q5/q52/positive/extreme/pct}\% \textit{Extremely}, \hmvalpct{q5/q52/positive/base/pct}\% \textit{Very}), \hmvalpct{q5/q52/neutral/pct}\% rated it as \textit{Moderately important}, and only \hmvalpct{q5/q52/negative/pct}\% rated it as \textit{Slightly important} or below. 
Regarding the importance of skills from professionals' perspective, three top skills draw near-universal agreement as \textit{Very} or \textit{Extremely important}: \hmval{q4/q47/top1/label} (\hmvalpct{q4/q47/top1/pct}\%), \hmval{q4/q47/top2/label} (\hmvalpct{q4/q47/top2/pct}\%), and \hmval{q4/q47/top3/label} (\hmvalpct{q4/q47/top3/pct}\%). \hmval{q4/q47/collaboration/label} (\hmvalpct{q4/q47/collaboration/positive/pct}\%) and \hmval{q4/q47/problem-decomposition/label} (\hmvalpct{q4/q47/problem-decomposition/positive/pct}\%) follow closely.
The skill that has the least agreement on its importance is Documentation, with around only \hmvalpct{q4/q47/documentation/positive/pct}\% having an important rating, while \hmvalpct{q4/q47/documentation/neutral/pct}\%\ have a neutral rating. 
Interestingly, Documentation is also the skill hiring professionals observe with the least gap (around \hmvalpct{q5/q53/documentation/negative/pct}\% see low gaps, and around \hmvalpct{q5/q53/documentation/neutral/pct}\% see moderate gaps). Following that, participants also see Code Development having a low gap (\hmvalpct{q5/q53/code-development/negative/pct}\% low gaps, \hmvalpct{q5/q53/code-development/neutral/pct}\% moderate gaps).
The skills they see as having the largest gaps are Code Review (\hmvalpct{q5/q53/code-review/negative/pct}\% see low gap, \hmvalpct{q5/q53/code-review/neutral/pct}\%  see moderate gap, and  \hmvalpct{q5/q53/code-review/positive/pct}\% see high gap), followed by Debugging (\hmvalpct{q5/q53/debugging/negative/pct}\%,   \hmvalpct{q5/q53/debugging/neutral/pct}\%, \hmvalpct{q5/q53/debugging/positive/pct}\%) followed by Critical evaluation of AI-generated output (\hmvalpct{q5/q53/critical-eval-ai/negative/pct}\%, \hmvalpct{q5/q53/critical-eval-ai/neutral/pct}\%, \hmvalpct{q5/q53/critical-eval-ai/positive/pct}\%), followed by .
\subsubsection{RQ3.2 Cross-Survey Comparison}

\skillreadyfigure
\paragraph{Skill Importance Alignment}  
Figure \ref{fig:skills-imp} compares the two populations' importance ratings per skill, ordered by the absolute gap. 
A smaller gap indicates alignment between the two groups. We sort the skills by this gap in descending order. Accordingly, skills at the top of the graph show the skills on which the two populations disagree in terms of importance, while those at the bottom are skills on which the populations agree. 
From the top of the graph, we can see that the two populations disagree on skills that are primarily core to software engineering and not GenAI-related. The industry rates System-level design, Code comprehension, and Code development as more important than educators do. The skills they both agree on at the bottom of the graph are Ability to learn and adapt independently, Critical evaluation of AI-generated output, and Documentation. Interestingly, Critical evaluation of AI-generated output and Ability to learn and adapt independently are among the most important skills rated by both populations. Both populations also agree to some extent (96\% importance for industry vs. 87\% for educators, $\Delta$~=~8.6) on the importance of Responsible and effective use of GenAI tools, which is also among the most important skills. The mean absolute difference between the two populations is \cmpvalpct[1]{importance/mean-abs-gap}~percentage points.
Additionally, a Mann-Whitney U test with Benjamini-Hochberg correction shows no statistically significant difference in importance ratings for any skill. Overall, this shows that the two populations generally agree on the importance of the skills.

\paragraph{Graduate-Readiness Alignment (ED-Q5.5 vs.\ HP-Q5.3)} When it comes to readiness, we asked educators to rate the proficiency of graduates regarding the skills, while we asked hiring professionals to identify gaps in the skills of entry-level hires. To make the numbers from the two surveys comparable, we convert educator-rated proficiency\emph{educator-reported weaknesses} by calculating the share of educators who rate recent graduates as \textit{Very weak} or \textit{Weak} for each skill. This allows us to compare it to the \emph{industry-reported gap}, calculated as the share of hiring managers reporting a \textit{Significant} or \textit{Very significant} gap. 

Figure~\ref{fig:skills-ready} shows the dumbbell chart comparison. Interpreting these results, we see a different picture of graduates' readiness across the two perspectives, with an observable misalignment. The mean absolute difference between the two populations across all skills is \cmpvalpct{readiness/mean-abs-gap}~percentage points, more than twice the importance difference (around \checkNum{12\%}). On \cmpval{readiness/n-hp-higher} of 12 skills, hiring professionals report a larger problem than educators perceive in their graduates, whereas educators never reported a larger gap than industry, except for Responsible use of GenAI tools. The two skills both populations agree on the extent they are lacking are Responsible and effective use of GenAI tools and Documentation.
Interestingly, both populations viewed Responsible and effective use of GenAI tools as important but both viewed documentation as less important.

The largest misalignment is on \cmpval{readiness/max-div/label}, where educators report \cmpvalpct{readiness/code-review/ed-pct}\% weakness while industry reports \cmpvalpct{readiness/code-review/hp-pct}\% ($\Delta$~=~\cmpvalpct{readiness/max-div/abs-gap}~pp). We also see that Debugging, Problem decomposition, and Code comprehension all show educator-vs-industry gaps exceeding 40\% points. All four are core software engineering skills, essential even before GenAI. The misalignment also persists for GenAI-specific skills such as Critical evaluation of AI-generated output (\cmpvalpct{readiness/critical-eval-ai/ed-pct}\% educators vs. \cmpvalpct{readiness/critical-eval-ai/hp-pct}\% industry gap). 
Statistically, the Mann-Whitney U test with Benjamini-Hochberg correction finds significant differences ($p<0.05$) between graduate readiness ratings of the two populations on six of the 12 skills, all in technical fundamentals: 
Problem decomposition, Code comprehension, Debugging, System-level design, Code review, and Code development.
All effect sizes were negative, medium to large in magnitude, indicating that hiring professionals report a larger graduate gap on these skills than educators report weakness.

\findingrqthree
\section{Implications of Results}

Across both sides of the hiring pipeline, the two populations adopt similar GenAI-resistant strategies but differ in speed. Educators have already substantially revised their assessments, while industry interview practices are still emerging. Their policies on GenAI use also diverge, with educators framing use in terms of academic integrity and industry in terms of governance, though both hold the individual accountable for AI-assisted output. The two groups agree closely on which skills matter most, yet disagree on graduate readiness: hiring professionals perceive substantially larger gaps, spanning both foundational skills, such as code review and debugging, and the GenAI-specific skills, both rate as most important. Notably, the majority of educators now accept GenAI-written code when students understand it, reversing an earlier finding in which 60\% of instructors viewed it as unethical~\cite{Lau23}.

\subsection{Implications for Entry-Level Software Engineers}
\paragraph{Foundational software-engineering skills remain essential}
Even as GenAI takes over routine coding, hiring professionals continue to expect competence in foundational skills and report the largest graduate gaps in core activities such as code review and debugging. They also still expect competencies in code development. Graduates should keep building these skills rather than delegating them entirely to GenAI, since they remain the basis on which industry evaluates entry-level engineers.
\paragraph{Critical evaluation and responsible use of GenAI are now core competencies}
Both classroom and workplace policies make the individual responsible for the correctness of AI-generated output, so graduates should understand the code and content they are held accountable for.

\paragraph{Graduates are accountable for the GenAI-assisted work they produce}
Both classroom and workplace policies make the individual responsible for the correctness of AI-generated output, so graduates should understand the code and content they will be held accountable for.

\subsection{Implications for Educators}
\paragraph{Teach and assess critical evaluation and responsible use explicitly}
The skills both populations rate most important, critical evaluation of AI output and responsible use, are also those that graduates are weakest in. Educators should teach and assess these explicitly.

\paragraph{Strengthen delivery of the skills both sides already value}
Because both populations agree on which skills matter, the gap is about graduate readiness, not goals. This agreement gives educators an industry-endorsed target: strengthen the foundational skills where industry reports the largest gaps, such as code review and debugging, reframed toward reading, reviewing, and debugging code, including AI-generated code.
\paragraph{Align classroom policies with industry governance}
Industry frames responsible GenAI use in terms of governance, including approved tools, privacy and security, and accountability, while classroom policies focus on permission and academic integrity. Aligning classroom guidance with the governance practices students will meet on the job can make academic policies feel less restrictive and prepare students for responsible professional use.

\subsection{Implications for Hiring Professionals}

\paragraph{Distinguish informal GenAI use from formal training when assessing readiness}
Hiring professionals should expect graduates to arrive already familiar with GenAI tools, since their use in coursework is widespread, but that familiarity is largely self-directed rather than formally taught. Many programs restrict or prohibit GenAI in assessments, and few teach its responsible and effective use explicitly, so graduates may use these tools regularly without having developed the judgment to evaluate their output critically or apply them responsibly. Onboarding and evaluation should target this specific gap.

\paragraph{Formalize GenAI policy for candidate evaluation}
Policy regarding candidate evaluation practices around GenAI remains largely improvised, with only a minority of organizations having official guidelines, even as interviewers individually adjust criteria and permit the use of GenAI. Hiring professionals should formalize how GenAI is treated in interviews, specifying whether and how candidates may use it and how that use is evaluated.

\section{Threats to Validity}
\paragraph{Construct Validity} Our study focuses on graduating students who will seek a software development or software engineering-related job. However, some students may pursue other options, such as an academic career, for example. However, research and statistics show that the most common job for CS graduates is, in one way or another, related to software development \cite{Ste22}. We also select 12 skills based on a review of previous work and personal experience. Our list is not exhaustive, and some participants may interpret the skills differently. However, we include open-ended questions for respondents to elaborate on their choices, when applicable, to help us better understand their views. 

\paragraph{Internal Validity}
We consider anonymous populations from different regions. Accordingly, the educators are not necessarily teaching or training the same candidates that the industry professionals are assessing. We are, however, looking for observable patterns rather than directly linking target students/candidates across the two populations.  

\paragraph{External Validity}
Our study is an important first step toward understanding the changing skill set required of entry-level software engineers, how the assessment of those skills is evolving, and the alignment between education and industry on skill importance and graduate readiness. Although modest in size, our sample is diverse, spanning educators and hiring professionals across multiple regions; a larger sample would strengthen generalizability, but our findings already offer actionable insights for entry-level engineers, educators, and hiring professionals.

\section{Conclusion}
Through two surveys covering the same twelve software-engineering skills, we examine how educators and hiring professionals have adapted their assessment practices to GenAI and whether the competencies emphasized in higher education align with those prized in entry-level hiring. Both populations are reshaping assessment, and both agree on which skills matter, yet they diverge on graduate readiness: hiring professionals report larger gaps than educators, focused on foundational skills. The academia--industry gap is therefore a problem of graduate readiness against a shared set of goals, not a disagreement over those goals. We close with take-home messages for educators, hiring professionals, and entry-level software engineers.
\bibliographystyle{IEEEtranN}
\bibliography{paper}

\end{document}

%% file: tables/educators-question.tex
\newcommand{\educatorsqstab}{%
\begin{table*}[htbp]
  \centering
  \caption{Educator survey questions organized by section and mapped to research questions.}
  \label{tab:educator-questions}
  \scriptsize
  \setlength{\tabcolsep}{4pt}
  \renewcommand{\arraystretch}{0.95}
  \begin{tabularx}{\textwidth}{@{}l >{\RaggedRight\arraybackslash}X >{\RaggedRight\arraybackslash}p{0.24\textwidth}@{}}
    \toprule
    \textbf{Label} & \textbf{Question} & \textbf{Type} \\
    \midrule
    \multicolumn{3}{@{}l}{\textbf{RQ1.1 -- Experience with GenAI in Teaching}} \\
    \midrule
    \textbf{ED-Q3.1*} & Which GenAI tools have you used in teaching preparation? & Multiple choice \\
    \textbf{ED-Q3.2*} & What is your top-choice GenAI tool for teaching prep, if any? & Single choice \\
    \textbf{ED-Q3.3*} & How often do you use GenAI for each teaching task? (12 tasks) & Likert matrix (Never--Always) \\
    \textbf{ED-Q3.4} & For which tasks does GenAI add the most value, and why? & Open-ended \\
    \textbf{ED-Q3.5*} & How experienced are you with AI code-generation tools? & 4-point Likert (None--Extensive) \\
    \textbf{ED-Q3.6*} & Have GenAI tools changed how you prepare teaching materials? & Yes / No \\
    \textbf{ED-Q3.7} & How have GenAI tools affected, or not affected, that preparation? & Open-ended \\
    \textbf{ED-Q3.8*} & Since GenAI, what have you changed in your teaching? & Multiple choice \\
    \textbf{ED-Q3.9*} & How has your role as an instructor changed, if at all, due to GenAI? & Open-ended \\
    \midrule
    \multicolumn{3}{@{}l}{\textbf{RQ1.1--1.2 -- GenAI and Assessments}} \\
    \midrule
    \textbf{ED-Q4.1*}  & Do you give students explicit guidance on GenAI use in assignments? & Single choice \\
    \textbf{ED-Q4.2*}  & If so, what does that guidance typically address? & Multiple choice \\
    \textbf{ED-Q4.3}  & If willing, paste a sample GenAI policy from an assignment. & Open-ended \\
    \textbf{ED-Q4.4*}  & Have you modified your assessments in response to GenAI? & Single choice \\
    \textbf{ED-Q4.5--4.6*}  & Why did you modify (or not modify) them, and any challenges? \textit{(based on ED-Q4.4)} & Open-ended \\
    \textbf{ED-Q4.7*}  & Which assessment changes have you made or are considering? (5 formats) & Likert matrix \\
    \textbf{ED-Q4.8}  & What other assessment changes have you made or are considering? & Open-ended \\
    \textbf{ED-Q4.9*}  & Have you redistributed grade weights in response to GenAI? & Single choice \\
    \textbf{ED-Q4.10*} & I would accept GenAI-written code if students fully understand it. & Likert (Strongly disagree--Strongly agree) \\
    \textbf{ED-Q4.11*} & I actively check for GenAI-written code in submissions. & Likert (Never--Always) \\
    \textbf{ED-Q4.12*} & I can reliably detect GenAI-written code in submissions. & Likert (Never--Always) \\
    \midrule
    \multicolumn{3}{@{}l}{\textbf{RQ3 -- Skills, Graduate Readiness, and Student GenAI Use}} \\
    \midrule
    \textbf{ED-Q5.1*} & How often do you think students use GenAI on assignments? & Likert (None--Almost everyone) \\
    \textbf{ED-Q5.2*} & How many students use GenAI before trying a problem independently? & Likert (None--Almost everyone) \\
    \textbf{ED-Q5.3*} & How well does the curriculum prepare students for GenAI in practice? & Likert (Not at all--Very well prepared) \\
    \textbf{ED-Q5.4*} & How important are the following skills for undergraduate CS students? (12 skills) & Likert matrix (Not at all--Extremely important) \\
    \textbf{ED-Q5.5*} & How well have graduates of the last three years developed these skills? (12 skills) & Likert matrix (Very weak--Highly proficient) \\
    \textbf{ED-Q5.6*} & Rate your agreement with the following statements on student GenAI use. (9 statements) & Likert matrix (Strongly disagree--Strongly agree) \\
    \textbf{ED-Q5.7} & What impact, if any, does GenAI have on students' skill development? & Open-ended \\
    \bottomrule
  \end{tabularx}

  \vspace{2pt}
  \footnotesize\RaggedRight
  \textit{Note.} Some questions are reworded for space. All matrix items use a 5-point scale (ED-Q3.5 uses 4 points). ``Single choice'' allows an \textit{Other} entry; Likert items offer a non-substantive option (\textit{Not sure}/\textit{Not applicable}) where appropriate. Only substantive endpoints are listed. * denotes required questions.
\end{table*}
}

%% file: tables/hiring-questions.tex
\newcommand{\hiringsqstab}{%
\begin{table*}[htbp]
  \centering
  \caption{Hiring-professional survey questions organized by section and mapped to research questions.}
  \label{tab:hiring-questions}
  \footnotesize
  \setlength{\tabcolsep}{4pt}
  \renewcommand{\arraystretch}{0.95}
  \begin{tabularx}{\textwidth}{@{}l >{\RaggedRight\arraybackslash}X >{\RaggedRight\arraybackslash}p{0.22\textwidth}@{}}
    \toprule
    \textbf{Label} & \textbf{Question} & \textbf{Type} \\
    \midrule
    \multicolumn{3}{@{}l}{\textbf{RQ2.1 -- Professional and Organizational Use of GenAI}} \\
    \midrule
    \textbf{HP-Q3.1*} & Which GenAI tools are you familiar with? & Multiple choice \\
    \textbf{HP-Q3.2*} & How often do you use each of seven GenAI tools for work? & Likert matrix (Never--Always) \\
    \textbf{HP-Q3.3*} & What is your top-choice GenAI tool for work, if any? & Single choice \\
    \textbf{HP-Q3.4*} & How prevalent is GenAI in your organization's development workflows? & Single choice (Not used--Standard) \\
    \textbf{HP-Q3.5*} & Does your organization provide formal guidance on GenAI use in development? & Single choice (None--Formal) \\
    \textbf{HP-Q3.6} & Can you summarize your organization's GenAI policy? & Open-ended \\
    \midrule
    \multicolumn{3}{@{}l}{\textbf{RQ2.2 -- Current Interview Practices}} \\
    \midrule
    \textbf{HP-Q4.1*}  & Has your organization set official policies on GenAI when evaluating candidates? & Yes / No \\
    \textbf{HP-Q4.2*}  & Can you summarize those updated policies? \textit{(if HP-Q4.1 = Yes)} & Open-ended \\
    \textbf{HP-Q4.3*}  & Have you changed your candidate-evaluation criteria because of GenAI? & Yes / No \\
    \textbf{HP-Q4.4*}  & Can you summarize your updated criteria? \textit{(if HP-Q4.3 = Yes)} & Open-ended \\
    \textbf{HP-Q4.5*}  & Do you permit candidates to use GenAI in a technical interview? & Yes / No \\
    \textbf{HP-Q4.6*}  & Why do you permit or forbid GenAI during a technical interview? & Open-ended \\
    \textbf{HP-Q4.7*}  & How important are the following skills for entry-level engineers? (12 skills) & Likert matrix (Not at all--Extremely important) \\
    \textbf{HP-Q4.8*}  & How often do you ask candidates about their GenAI experience? & Likert (Never--Always) \\
    \textbf{HP-Q4.9*}  & How confident are you that interviews separate genuine skill from AI-assisted work? & Likert (Not at all--Extremely confident) \\
    \textbf{HP-Q4.10*} & How much do you prefer candidates who demonstrate GenAI skills? & Likert (No preference--Always a priority) \\
    \textbf{HP-Q4.11*} & Does GenAI make it harder to assess candidates' true proficiency and experience? & Yes / No \\
    \textbf{HP-Q4.12*} & What challenges do you face, or avoid, when assessing candidates amid GenAI? & Open-ended \\
    \midrule
    \multicolumn{3}{@{}l}{\textbf{RQ3 -- GenAI Readiness of Fresh Graduates}} \\
    \midrule
    \textbf{HP-Q5.1*} & How well do current CS/SE curricula prepare graduates for GenAI-assisted industry? & Likert (Not at all--Very large extent) \\
    \textbf{HP-Q5.2*} & How important is it for universities to teach GenAI tool use in their curricula? & Likert (Not at all--Extremely important) \\
    \textbf{HP-Q5.3*} & Over the last three years, have you seen graduate-readiness gaps in the following skills? (12 skills) & Likert matrix (No gap--Very significant gap) \\
    \textbf{HP-Q5.4} & For your highest-gap skills, how does the gap affect day-to-day work and GenAI use? & Open-ended \\
    \textbf{HP-Q5.5} & How might GenAI tools be integrated into courses to better prepare students for industry? & Open-ended \\
    \bottomrule
  \end{tabularx}

  \vspace{2pt}
  \footnotesize\RaggedRight
  \textit{Note.} Some questions are reworded for space. All matrix items use a 5-point scale. ``Single choice'' allows an \textit{Other (please specify)} entry where applicable; Likert items offer a non-substantive option (\textit{Not sure}) where appropriate. The Type column lists only substantive scale endpoints. * denotes required questions.
\end{table*}
}

%% file: values/educator-demographics-values.tex
\pgfkeyssetvalue{/ed/n}{56}
\pgfkeyssetvalue{/ed/d3/full-prof/n}{17}
\pgfkeyssetvalue{/ed/d3/full-prof/pct}{30.4}
\pgfkeyssetvalue{/ed/d3/assoc-prof/n}{13}
\pgfkeyssetvalue{/ed/d3/assoc-prof/pct}{23.2}
\pgfkeyssetvalue{/ed/d3/asst-prof/n}{13}
\pgfkeyssetvalue{/ed/d3/asst-prof/pct}{23.2}
\pgfkeyssetvalue{/ed/d3/lecturer/n}{7}
\pgfkeyssetvalue{/ed/d3/lecturer/pct}{12.5}
\pgfkeyssetvalue{/ed/d3/ta/n}{3}
\pgfkeyssetvalue{/ed/d3/ta/pct}{5.4}
\pgfkeyssetvalue{/ed/d3/adjunct/n}{1}
\pgfkeyssetvalue{/ed/d3/adjunct/pct}{1.8}
\pgfkeyssetvalue{/ed/d3/other/n}{2}
\pgfkeyssetvalue{/ed/d3/other/pct}{3.6}
\pgfkeyssetvalue{/ed/d3/tenure-track/n}{43}
\pgfkeyssetvalue{/ed/d3/tenure-track/pct}{76.8}
\pgfkeyssetvalue{/ed/d4/mean}{14.5}
\pgfkeyssetvalue{/ed/d4/median}{14.5}
\pgfkeyssetvalue{/ed/d4/sd}{9.2}
\pgfkeyssetvalue{/ed/d4/min}{2}
\pgfkeyssetvalue{/ed/d4/max}{40}
\pgfkeyssetvalue{/ed/d4/q1}{7.8}
\pgfkeyssetvalue{/ed/d4/q3}{20.0}
\pgfkeyssetvalue{/ed/d4/iqr}{12.2}
\pgfkeyssetvalue{/ed/d4/band/1-5/n}{12}
\pgfkeyssetvalue{/ed/d4/band/1-5/pct}{21.4}
\pgfkeyssetvalue{/ed/d4/band/6-10/n}{12}
\pgfkeyssetvalue{/ed/d4/band/6-10/pct}{21.4}
\pgfkeyssetvalue{/ed/d4/band/11-20/n}{20}
\pgfkeyssetvalue{/ed/d4/band/11-20/pct}{35.7}
\pgfkeyssetvalue{/ed/d4/band/21plus/n}{12}
\pgfkeyssetvalue{/ed/d4/band/21plus/pct}{21.4}
\pgfkeyssetvalue{/ed/d5/n-countries}{17}
\pgfkeyssetvalue{/ed/d5/canada/n}{20}
\pgfkeyssetvalue{/ed/d5/canada/pct}{35.7}
\pgfkeyssetvalue{/ed/d5/us/n}{5}
\pgfkeyssetvalue{/ed/d5/us/pct}{8.9}
\pgfkeyssetvalue{/ed/d5/uae/n}{12}
\pgfkeyssetvalue{/ed/d5/uae/pct}{21.4}
\pgfkeyssetvalue{/ed/d5/uk/n}{2}
\pgfkeyssetvalue{/ed/d5/uk/pct}{3.6}
\pgfkeyssetvalue{/ed/d5/egypt/n}{1}
\pgfkeyssetvalue{/ed/d5/egypt/pct}{1.8}
\pgfkeyssetvalue{/ed/d5/pakistan/n}{1}
\pgfkeyssetvalue{/ed/d5/pakistan/pct}{1.8}
\pgfkeyssetvalue{/ed/d5/india/n}{1}
\pgfkeyssetvalue{/ed/d5/india/pct}{1.8}
\pgfkeyssetvalue{/ed/d5/singapore/n}{2}
\pgfkeyssetvalue{/ed/d5/singapore/pct}{3.6}
\pgfkeyssetvalue{/ed/d5/romania/n}{1}
\pgfkeyssetvalue{/ed/d5/romania/pct}{1.8}
\pgfkeyssetvalue{/ed/d5/sweden/n}{1}
\pgfkeyssetvalue{/ed/d5/sweden/pct}{1.8}
\pgfkeyssetvalue{/ed/d5/switzerland/n}{1}
\pgfkeyssetvalue{/ed/d5/switzerland/pct}{1.8}
\pgfkeyssetvalue{/ed/d5/brazil/n}{4}
\pgfkeyssetvalue{/ed/d5/brazil/pct}{7.1}
\pgfkeyssetvalue{/ed/d5/myanmar/n}{1}
\pgfkeyssetvalue{/ed/d5/myanmar/pct}{1.8}
\pgfkeyssetvalue{/ed/d5/region/north-america/n}{25}
\pgfkeyssetvalue{/ed/d5/region/north-america/pct}{44.6}
\pgfkeyssetvalue{/ed/d5/region/europe/n}{5}
\pgfkeyssetvalue{/ed/d5/region/europe/pct}{8.9}
\pgfkeyssetvalue{/ed/d5/region/mena/n}{13}
\pgfkeyssetvalue{/ed/d5/region/mena/pct}{23.2}
\pgfkeyssetvalue{/ed/d5/region/sse-asia/n}{5}
\pgfkeyssetvalue{/ed/d5/region/sse-asia/pct}{8.9}
\pgfkeyssetvalue{/ed/d5/region/latin-america/n}{4}
\pgfkeyssetvalue{/ed/d5/region/latin-america/pct}{7.1}
\pgfkeyssetvalue{/ed/d6/mean-per-respondent}{2.5}
\pgfkeyssetvalue{/ed/d6/n-distinct}{18}
\pgfkeyssetvalue{/ed/d6/intro-programming/n}{17}
\pgfkeyssetvalue{/ed/d6/intro-programming/pct}{30.4}
\pgfkeyssetvalue{/ed/d6/dsa/n}{12}
\pgfkeyssetvalue{/ed/d6/dsa/pct}{21.4}
\pgfkeyssetvalue{/ed/d6/oop/n}{16}
\pgfkeyssetvalue{/ed/d6/oop/pct}{28.6}
\pgfkeyssetvalue{/ed/d6/se/n}{29}
\pgfkeyssetvalue{/ed/d6/se/pct}{51.8}
\pgfkeyssetvalue{/ed/d6/databases/n}{6}
\pgfkeyssetvalue{/ed/d6/databases/pct}{10.7}
\pgfkeyssetvalue{/ed/d6/os/n}{1}
\pgfkeyssetvalue{/ed/d6/os/pct}{1.8}
\pgfkeyssetvalue{/ed/d6/networks/n}{6}
\pgfkeyssetvalue{/ed/d6/networks/pct}{10.7}
\pgfkeyssetvalue{/ed/d6/web-mobile/n}{12}
\pgfkeyssetvalue{/ed/d6/web-mobile/pct}{21.4}
\pgfkeyssetvalue{/ed/d6/compilers/n}{5}
\pgfkeyssetvalue{/ed/d6/compilers/pct}{8.9}
\pgfkeyssetvalue{/ed/d6/theory/n}{3}
\pgfkeyssetvalue{/ed/d6/theory/pct}{5.4}
\pgfkeyssetvalue{/ed/d6/ai-ml/n}{8}
\pgfkeyssetvalue{/ed/d6/ai-ml/pct}{14.3}
\pgfkeyssetvalue{/ed/d6/plp/n}{4}
\pgfkeyssetvalue{/ed/d6/plp/pct}{7.1}
\pgfkeyssetvalue{/ed/d6/architecture/n}{2}
\pgfkeyssetvalue{/ed/d6/architecture/pct}{3.6}
\pgfkeyssetvalue{/ed/d6/cybersec/n}{6}
\pgfkeyssetvalue{/ed/d6/cybersec/pct}{10.7}
\pgfkeyssetvalue{/ed/d6/data-science/n}{3}
\pgfkeyssetvalue{/ed/d6/data-science/pct}{5.4}
\pgfkeyssetvalue{/ed/d6/hci/n}{3}
\pgfkeyssetvalue{/ed/d6/hci/pct}{5.4}
\pgfkeyssetvalue{/ed/d6/game-dev/n}{1}
\pgfkeyssetvalue{/ed/d6/game-dev/pct}{1.8}
\pgfkeyssetvalue{/ed/d6/other/n}{7}
\pgfkeyssetvalue{/ed/d6/other/pct}{12.5}
\pgfkeyssetvalue{/ed/d7/freshman/n}{23}
\pgfkeyssetvalue{/ed/d7/freshman/pct}{41.1}
\pgfkeyssetvalue{/ed/d7/sophomore/n}{28}
\pgfkeyssetvalue{/ed/d7/sophomore/pct}{50.0}
\pgfkeyssetvalue{/ed/d7/junior/n}{37}
\pgfkeyssetvalue{/ed/d7/junior/pct}{66.1}
\pgfkeyssetvalue{/ed/d7/senior/n}{35}
\pgfkeyssetvalue{/ed/d7/senior/pct}{62.5}
\pgfkeyssetvalue{/ed/d7/mean-per-respondent}{2.2}
\pgfkeyssetvalue{/ed/d8/lt50/n}{6}
\pgfkeyssetvalue{/ed/d8/lt50/pct}{10.7}
\pgfkeyssetvalue{/ed/d8/51-100/n}{5}
\pgfkeyssetvalue{/ed/d8/51-100/pct}{8.9}
\pgfkeyssetvalue{/ed/d8/101-500/n}{18}
\pgfkeyssetvalue{/ed/d8/101-500/pct}{32.1}
\pgfkeyssetvalue{/ed/d8/501-1000/n}{8}
\pgfkeyssetvalue{/ed/d8/501-1000/pct}{14.3}
\pgfkeyssetvalue{/ed/d8/gt1000/n}{14}
\pgfkeyssetvalue{/ed/d8/gt1000/pct}{25.0}
\pgfkeyssetvalue{/ed/d8/not-sure/n}{5}
\pgfkeyssetvalue{/ed/d8/not-sure/pct}{8.9}
\pgfkeyssetvalue{/ed/d8/lt500/n}{29}
\pgfkeyssetvalue{/ed/d8/lt500/pct}{51.8}
\pgfkeyssetvalue{/ed/d8/gte500/n}{22}
\pgfkeyssetvalue{/ed/d8/gte500/pct}{39.3}

%% file: values/hiring-demographics-values.tex
\pgfkeyssetvalue{/hp/n}{24}
\pgfkeyssetvalue{/hp/d2/senior-dev/n}{5}
\pgfkeyssetvalue{/hp/d2/senior-dev/pct}{20.8}
\pgfkeyssetvalue{/hp/d2/architect/n}{2}
\pgfkeyssetvalue{/hp/d2/architect/pct}{8.3}
\pgfkeyssetvalue{/hp/d2/tech-lead/n}{3}
\pgfkeyssetvalue{/hp/d2/tech-lead/pct}{12.5}
\pgfkeyssetvalue{/hp/d2/pm/n}{1}
\pgfkeyssetvalue{/hp/d2/pm/pct}{4.2}
\pgfkeyssetvalue{/hp/d2/it-mgr/n}{1}
\pgfkeyssetvalue{/hp/d2/it-mgr/pct}{4.2}
\pgfkeyssetvalue{/hp/d2/director/n}{4}
\pgfkeyssetvalue{/hp/d2/director/pct}{16.7}
\pgfkeyssetvalue{/hp/d2/ba/n}{1}
\pgfkeyssetvalue{/hp/d2/ba/pct}{4.2}
\pgfkeyssetvalue{/hp/d2/c-level/n}{5}
\pgfkeyssetvalue{/hp/d2/c-level/pct}{20.8}
\pgfkeyssetvalue{/hp/d2/other/n}{2}
\pgfkeyssetvalue{/hp/d2/other/pct}{8.3}
\pgfkeyssetvalue{/hp/d2/senior-leadership/n}{16}
\pgfkeyssetvalue{/hp/d2/senior-leadership/pct}{66.7}
\pgfkeyssetvalue{/hp/d3/mean}{15.1}
\pgfkeyssetvalue{/hp/d3/median}{15.5}
\pgfkeyssetvalue{/hp/d3/sd}{8.1}
\pgfkeyssetvalue{/hp/d3/min}{3}
\pgfkeyssetvalue{/hp/d3/max}{35}
\pgfkeyssetvalue{/hp/d3/q1}{8.0}
\pgfkeyssetvalue{/hp/d3/q3}{20.0}
\pgfkeyssetvalue{/hp/d3/iqr}{12.0}
\pgfkeyssetvalue{/hp/d3/band/1-5/n}{3}
\pgfkeyssetvalue{/hp/d3/band/1-5/pct}{12.5}
\pgfkeyssetvalue{/hp/d3/band/6-10/n}{5}
\pgfkeyssetvalue{/hp/d3/band/6-10/pct}{20.8}
\pgfkeyssetvalue{/hp/d3/band/11-20/n}{12}
\pgfkeyssetvalue{/hp/d3/band/11-20/pct}{50.0}
\pgfkeyssetvalue{/hp/d3/band/21plus/n}{4}
\pgfkeyssetvalue{/hp/d3/band/21plus/pct}{16.7}
\pgfkeyssetvalue{/hp/d4/n-countries}{10}
\pgfkeyssetvalue{/hp/d4/us/n}{7}
\pgfkeyssetvalue{/hp/d4/us/pct}{29.2}
\pgfkeyssetvalue{/hp/d4/canada/n}{2}
\pgfkeyssetvalue{/hp/d4/canada/pct}{8.3}
\pgfkeyssetvalue{/hp/d4/uae/n}{6}
\pgfkeyssetvalue{/hp/d4/uae/pct}{25.0}
\pgfkeyssetvalue{/hp/d4/germany/n}{3}
\pgfkeyssetvalue{/hp/d4/germany/pct}{12.5}
\pgfkeyssetvalue{/hp/d4/uk/n}{1}
\pgfkeyssetvalue{/hp/d4/uk/pct}{4.2}
\pgfkeyssetvalue{/hp/d4/australia/n}{1}
\pgfkeyssetvalue{/hp/d4/australia/pct}{4.2}
\pgfkeyssetvalue{/hp/d4/saudi/n}{1}
\pgfkeyssetvalue{/hp/d4/saudi/pct}{4.2}
\pgfkeyssetvalue{/hp/d4/region/north-america/n}{9}
\pgfkeyssetvalue{/hp/d4/region/north-america/pct}{37.5}
\pgfkeyssetvalue{/hp/d4/region/europe/n}{4}
\pgfkeyssetvalue{/hp/d4/region/europe/pct}{16.7}
\pgfkeyssetvalue{/hp/d4/region/mena/n}{7}
\pgfkeyssetvalue{/hp/d4/region/mena/pct}{29.2}
\pgfkeyssetvalue{/hp/d4/region/oceania/n}{1}
\pgfkeyssetvalue{/hp/d4/region/oceania/pct}{4.2}
\pgfkeyssetvalue{/hp/d5/mean-per-respondent}{2.9}
\pgfkeyssetvalue{/hp/d5/review-resumes/n}{12}
\pgfkeyssetvalue{/hp/d5/review-resumes/pct}{50.0}
\pgfkeyssetvalue{/hp/d5/design-questions/n}{13}
\pgfkeyssetvalue{/hp/d5/design-questions/pct}{54.2}
\pgfkeyssetvalue{/hp/d5/conduct-interviews/n}{16}
\pgfkeyssetvalue{/hp/d5/conduct-interviews/pct}{66.7}
\pgfkeyssetvalue{/hp/d5/final-decision/n}{13}
\pgfkeyssetvalue{/hp/d5/final-decision/pct}{54.2}
\pgfkeyssetvalue{/hp/d5/onboard/n}{13}
\pgfkeyssetvalue{/hp/d5/onboard/pct}{54.2}
\pgfkeyssetvalue{/hp/d5/other/n}{2}
\pgfkeyssetvalue{/hp/d5/other/pct}{8.3}
\pgfkeyssetvalue{/hp/d6/yes/n}{19}
\pgfkeyssetvalue{/hp/d6/yes/pct}{79.2}
\pgfkeyssetvalue{/hp/d6/no/n}{5}
\pgfkeyssetvalue{/hp/d6/no/pct}{20.8}
\pgfkeyssetvalue{/hp/d7/n-managers}{19}
\pgfkeyssetvalue{/hp/d7/1-5/n}{6}
\pgfkeyssetvalue{/hp/d7/1-5/pct}{31.6}
\pgfkeyssetvalue{/hp/d7/6-10/n}{5}
\pgfkeyssetvalue{/hp/d7/6-10/pct}{26.3}
\pgfkeyssetvalue{/hp/d7/11-20/n}{4}
\pgfkeyssetvalue{/hp/d7/11-20/pct}{21.1}
\pgfkeyssetvalue{/hp/d7/21-50/n}{4}
\pgfkeyssetvalue{/hp/d7/21-50/pct}{21.1}
\pgfkeyssetvalue{/hp/d7/gt50/n}{0}
\pgfkeyssetvalue{/hp/d7/gt50/pct}{0.0}
\pgfkeyssetvalue{/hp/d8/finance/n}{3}
\pgfkeyssetvalue{/hp/d8/finance/pct}{12.5}
\pgfkeyssetvalue{/hp/d8/tech/n}{19}
\pgfkeyssetvalue{/hp/d8/tech/pct}{79.2}
\pgfkeyssetvalue{/hp/d8/healthcare/n}{1}
\pgfkeyssetvalue{/hp/d8/healthcare/pct}{4.2}
\pgfkeyssetvalue{/hp/d8/hr/n}{1}
\pgfkeyssetvalue{/hp/d8/hr/pct}{4.2}
\pgfkeyssetvalue{/hp/d9/lt50/n}{6}
\pgfkeyssetvalue{/hp/d9/lt50/pct}{25.0}
\pgfkeyssetvalue{/hp/d9/50-100/n}{3}
\pgfkeyssetvalue{/hp/d9/50-100/pct}{12.5}
\pgfkeyssetvalue{/hp/d9/101-500/n}{3}
\pgfkeyssetvalue{/hp/d9/101-500/pct}{12.5}
\pgfkeyssetvalue{/hp/d9/501-1000/n}{2}
\pgfkeyssetvalue{/hp/d9/501-1000/pct}{8.3}
\pgfkeyssetvalue{/hp/d9/gt1000/n}{10}
\pgfkeyssetvalue{/hp/d9/gt1000/pct}{41.7}
\pgfkeyssetvalue{/hp/d9/not-sure/n}{0}
\pgfkeyssetvalue{/hp/d9/not-sure/pct}{0.0}
\pgfkeyssetvalue{/hp/d9/lt100/n}{9}
\pgfkeyssetvalue{/hp/d9/lt100/pct}{37.5}
\pgfkeyssetvalue{/hp/d9/gt1000-only/n}{10}
\pgfkeyssetvalue{/hp/d9/gt1000-only/pct}{41.7}

%% file: values/educator-q3-values.tex
\pgfkeyssetvalue{/ed/q3/n}{56}
\pgfkeyssetvalue{/ed/q3/q31/users/n}{44}
\pgfkeyssetvalue{/ed/q3/q31/users/pct}{78.6}
\pgfkeyssetvalue{/ed/q3/q31/nonusers/n}{12}
\pgfkeyssetvalue{/ed/q3/q31/nonusers/pct}{21.4}
\pgfkeyssetvalue{/ed/q3/q31/mean-tools-per-user}{2.4}
\pgfkeyssetvalue{/ed/q3/q31/chatgpt/n}{39}
\pgfkeyssetvalue{/ed/q3/q31/chatgpt/pct}{69.6}
\pgfkeyssetvalue{/ed/q3/q31/chatgpt/pct-of-users}{88.6}
\pgfkeyssetvalue{/ed/q3/q31/claude/n}{22}
\pgfkeyssetvalue{/ed/q3/q31/claude/pct}{39.3}
\pgfkeyssetvalue{/ed/q3/q31/claude/pct-of-users}{50.0}
\pgfkeyssetvalue{/ed/q3/q31/gemini/n}{23}
\pgfkeyssetvalue{/ed/q3/q31/gemini/pct}{41.1}
\pgfkeyssetvalue{/ed/q3/q31/gemini/pct-of-users}{52.3}
\pgfkeyssetvalue{/ed/q3/q31/copilot/n}{14}
\pgfkeyssetvalue{/ed/q3/q31/copilot/pct}{25.0}
\pgfkeyssetvalue{/ed/q3/q31/copilot/pct-of-users}{31.8}
\pgfkeyssetvalue{/ed/q3/q31/perplexity/n}{4}
\pgfkeyssetvalue{/ed/q3/q31/perplexity/pct}{7.1}
\pgfkeyssetvalue{/ed/q3/q31/perplexity/pct-of-users}{9.1}
\pgfkeyssetvalue{/ed/q3/q31/cursor/n}{2}
\pgfkeyssetvalue{/ed/q3/q31/cursor/pct}{3.6}
\pgfkeyssetvalue{/ed/q3/q31/cursor/pct-of-users}{4.5}
\pgfkeyssetvalue{/ed/q3/q31/other/n}{1}
\pgfkeyssetvalue{/ed/q3/q31/other/pct}{1.8}
\pgfkeyssetvalue{/ed/q3/q31/other/pct-of-users}{2.3}
\pgfkeyssetvalue{/ed/q3/q32/chatgpt/n}{24}
\pgfkeyssetvalue{/ed/q3/q32/chatgpt/pct}{42.9}
\pgfkeyssetvalue{/ed/q3/q32/claude/n}{9}
\pgfkeyssetvalue{/ed/q3/q32/claude/pct}{16.1}
\pgfkeyssetvalue{/ed/q3/q32/gemini/n}{8}
\pgfkeyssetvalue{/ed/q3/q32/gemini/pct}{14.3}
\pgfkeyssetvalue{/ed/q3/q32/copilot/n}{4}
\pgfkeyssetvalue{/ed/q3/q32/copilot/pct}{7.1}
\pgfkeyssetvalue{/ed/q3/q32/other/n}{2}
\pgfkeyssetvalue{/ed/q3/q32/other/pct}{3.6}
\pgfkeyssetvalue{/ed/q3/q32/no-response/n}{9}
\pgfkeyssetvalue{/ed/q3/q32/no-response/pct}{16.1}
\pgfkeyssetvalue{/ed/q3/q32/named/n}{47}
\pgfkeyssetvalue{/ed/q3/q32/named/pct}{83.9}
\pgfkeyssetvalue{/ed/q3/q33/task1/label}{Preparing lecture slides}
\pgfkeyssetvalue{/ed/q3/q33/task1/n-na}{0}
\pgfkeyssetvalue{/ed/q3/q33/task1/never-pct}{41.1}
\pgfkeyssetvalue{/ed/q3/q33/task1/rarely-pct}{21.4}
\pgfkeyssetvalue{/ed/q3/q33/task1/sometimes-pct}{19.6}
\pgfkeyssetvalue{/ed/q3/q33/task1/often-pct}{16.1}
\pgfkeyssetvalue{/ed/q3/q33/task1/always-pct}{1.8}
\pgfkeyssetvalue{/ed/q3/q33/task1/high-usage-pct}{17.9}
\pgfkeyssetvalue{/ed/q3/q33/task1/occasional-usage-pct}{19.6}
\pgfkeyssetvalue{/ed/q3/q33/task1/low-usage-pct}{62.5}
\pgfkeyssetvalue{/ed/q3/q33/task2/label}{Preparing teaching notes or explanations}
\pgfkeyssetvalue{/ed/q3/q33/task2/n-na}{3}
\pgfkeyssetvalue{/ed/q3/q33/task2/never-pct}{41.5}
\pgfkeyssetvalue{/ed/q3/q33/task2/rarely-pct}{17.0}
\pgfkeyssetvalue{/ed/q3/q33/task2/sometimes-pct}{20.8}
\pgfkeyssetvalue{/ed/q3/q33/task2/often-pct}{17.0}
\pgfkeyssetvalue{/ed/q3/q33/task2/always-pct}{3.8}
\pgfkeyssetvalue{/ed/q3/q33/task2/high-usage-pct}{20.8}
\pgfkeyssetvalue{/ed/q3/q33/task2/occasional-usage-pct}{20.8}
\pgfkeyssetvalue{/ed/q3/q33/task2/low-usage-pct}{58.5}
\pgfkeyssetvalue{/ed/q3/q33/task3/label}{Designing assignments or projects}
\pgfkeyssetvalue{/ed/q3/q33/task3/n-na}{0}
\pgfkeyssetvalue{/ed/q3/q33/task3/never-pct}{23.2}
\pgfkeyssetvalue{/ed/q3/q33/task3/rarely-pct}{21.4}
\pgfkeyssetvalue{/ed/q3/q33/task3/sometimes-pct}{32.1}
\pgfkeyssetvalue{/ed/q3/q33/task3/often-pct}{21.4}
\pgfkeyssetvalue{/ed/q3/q33/task3/always-pct}{1.8}
\pgfkeyssetvalue{/ed/q3/q33/task3/high-usage-pct}{23.2}
\pgfkeyssetvalue{/ed/q3/q33/task3/occasional-usage-pct}{32.1}
\pgfkeyssetvalue{/ed/q3/q33/task3/low-usage-pct}{44.6}
\pgfkeyssetvalue{/ed/q3/q33/task4/label}{Creating quiz or exam questions}
\pgfkeyssetvalue{/ed/q3/q33/task4/n-na}{1}
\pgfkeyssetvalue{/ed/q3/q33/task4/never-pct}{27.3}
\pgfkeyssetvalue{/ed/q3/q33/task4/rarely-pct}{21.8}
\pgfkeyssetvalue{/ed/q3/q33/task4/sometimes-pct}{25.5}
\pgfkeyssetvalue{/ed/q3/q33/task4/often-pct}{23.6}
\pgfkeyssetvalue{/ed/q3/q33/task4/always-pct}{1.8}
\pgfkeyssetvalue{/ed/q3/q33/task4/high-usage-pct}{25.5}
\pgfkeyssetvalue{/ed/q3/q33/task4/occasional-usage-pct}{25.5}
\pgfkeyssetvalue{/ed/q3/q33/task4/low-usage-pct}{49.1}
\pgfkeyssetvalue{/ed/q3/q33/task5/label}{Writing sample solutions}
\pgfkeyssetvalue{/ed/q3/q33/task5/n-na}{2}
\pgfkeyssetvalue{/ed/q3/q33/task5/never-pct}{35.2}
\pgfkeyssetvalue{/ed/q3/q33/task5/rarely-pct}{18.5}
\pgfkeyssetvalue{/ed/q3/q33/task5/sometimes-pct}{20.4}
\pgfkeyssetvalue{/ed/q3/q33/task5/often-pct}{24.1}
\pgfkeyssetvalue{/ed/q3/q33/task5/always-pct}{1.9}
\pgfkeyssetvalue{/ed/q3/q33/task5/high-usage-pct}{25.9}
\pgfkeyssetvalue{/ed/q3/q33/task5/occasional-usage-pct}{20.4}
\pgfkeyssetvalue{/ed/q3/q33/task5/low-usage-pct}{53.7}
\pgfkeyssetvalue{/ed/q3/q33/task6/label}{Drafting feedback to students}
\pgfkeyssetvalue{/ed/q3/q33/task6/n-na}{2}
\pgfkeyssetvalue{/ed/q3/q33/task6/never-pct}{61.1}
\pgfkeyssetvalue{/ed/q3/q33/task6/rarely-pct}{20.4}
\pgfkeyssetvalue{/ed/q3/q33/task6/sometimes-pct}{11.1}
\pgfkeyssetvalue{/ed/q3/q33/task6/often-pct}{3.7}
\pgfkeyssetvalue{/ed/q3/q33/task6/always-pct}{3.7}
\pgfkeyssetvalue{/ed/q3/q33/task6/high-usage-pct}{7.4}
\pgfkeyssetvalue{/ed/q3/q33/task6/occasional-usage-pct}{11.1}
\pgfkeyssetvalue{/ed/q3/q33/task6/low-usage-pct}{81.5}
\pgfkeyssetvalue{/ed/q3/q33/task7/label}{Grading student assignments}
\pgfkeyssetvalue{/ed/q3/q33/task7/n-na}{3}
\pgfkeyssetvalue{/ed/q3/q33/task7/never-pct}{77.4}
\pgfkeyssetvalue{/ed/q3/q33/task7/rarely-pct}{17.0}
\pgfkeyssetvalue{/ed/q3/q33/task7/sometimes-pct}{1.9}
\pgfkeyssetvalue{/ed/q3/q33/task7/often-pct}{3.8}
\pgfkeyssetvalue{/ed/q3/q33/task7/always-pct}{0.0}
\pgfkeyssetvalue{/ed/q3/q33/task7/high-usage-pct}{3.8}
\pgfkeyssetvalue{/ed/q3/q33/task7/occasional-usage-pct}{1.9}
\pgfkeyssetvalue{/ed/q3/q33/task7/low-usage-pct}{94.3}
\pgfkeyssetvalue{/ed/q3/q33/task8/label}{Generating examples or code snippets}
\pgfkeyssetvalue{/ed/q3/q33/task8/n-na}{1}
\pgfkeyssetvalue{/ed/q3/q33/task8/never-pct}{25.5}
\pgfkeyssetvalue{/ed/q3/q33/task8/rarely-pct}{12.7}
\pgfkeyssetvalue{/ed/q3/q33/task8/sometimes-pct}{32.7}
\pgfkeyssetvalue{/ed/q3/q33/task8/often-pct}{25.5}
\pgfkeyssetvalue{/ed/q3/q33/task8/always-pct}{3.6}
\pgfkeyssetvalue{/ed/q3/q33/task8/high-usage-pct}{29.1}
\pgfkeyssetvalue{/ed/q3/q33/task8/occasional-usage-pct}{32.7}
\pgfkeyssetvalue{/ed/q3/q33/task8/low-usage-pct}{38.2}
\pgfkeyssetvalue{/ed/q3/q33/task9/label}{Debugging or refining instructional code}
\pgfkeyssetvalue{/ed/q3/q33/task9/n-na}{4}
\pgfkeyssetvalue{/ed/q3/q33/task9/never-pct}{34.6}
\pgfkeyssetvalue{/ed/q3/q33/task9/rarely-pct}{19.2}
\pgfkeyssetvalue{/ed/q3/q33/task9/sometimes-pct}{23.1}
\pgfkeyssetvalue{/ed/q3/q33/task9/often-pct}{23.1}
\pgfkeyssetvalue{/ed/q3/q33/task9/always-pct}{0.0}
\pgfkeyssetvalue{/ed/q3/q33/task9/high-usage-pct}{23.1}
\pgfkeyssetvalue{/ed/q3/q33/task9/occasional-usage-pct}{23.1}
\pgfkeyssetvalue{/ed/q3/q33/task9/low-usage-pct}{53.8}
\pgfkeyssetvalue{/ed/q3/q33/task10/label}{Summarizing technical content}
\pgfkeyssetvalue{/ed/q3/q33/task10/n-na}{2}
\pgfkeyssetvalue{/ed/q3/q33/task10/never-pct}{42.6}
\pgfkeyssetvalue{/ed/q3/q33/task10/rarely-pct}{11.1}
\pgfkeyssetvalue{/ed/q3/q33/task10/sometimes-pct}{27.8}
\pgfkeyssetvalue{/ed/q3/q33/task10/often-pct}{14.8}
\pgfkeyssetvalue{/ed/q3/q33/task10/always-pct}{3.7}
\pgfkeyssetvalue{/ed/q3/q33/task10/high-usage-pct}{18.5}
\pgfkeyssetvalue{/ed/q3/q33/task10/occasional-usage-pct}{27.8}
\pgfkeyssetvalue{/ed/q3/q33/task10/low-usage-pct}{53.7}
\pgfkeyssetvalue{/ed/q3/q33/task11/label}{Drafting course announcements or communications}
\pgfkeyssetvalue{/ed/q3/q33/task11/n-na}{2}
\pgfkeyssetvalue{/ed/q3/q33/task11/never-pct}{42.6}
\pgfkeyssetvalue{/ed/q3/q33/task11/rarely-pct}{22.2}
\pgfkeyssetvalue{/ed/q3/q33/task11/sometimes-pct}{13.0}
\pgfkeyssetvalue{/ed/q3/q33/task11/often-pct}{18.5}
\pgfkeyssetvalue{/ed/q3/q33/task11/always-pct}{3.7}
\pgfkeyssetvalue{/ed/q3/q33/task11/high-usage-pct}{22.2}
\pgfkeyssetvalue{/ed/q3/q33/task11/occasional-usage-pct}{13.0}
\pgfkeyssetvalue{/ed/q3/q33/task11/low-usage-pct}{64.8}
\pgfkeyssetvalue{/ed/q3/q33/task12/label}{Generating starter code for exercises or assignments}
\pgfkeyssetvalue{/ed/q3/q33/task12/n-na}{3}
\pgfkeyssetvalue{/ed/q3/q33/task12/never-pct}{34.0}
\pgfkeyssetvalue{/ed/q3/q33/task12/rarely-pct}{26.4}
\pgfkeyssetvalue{/ed/q3/q33/task12/sometimes-pct}{17.0}
\pgfkeyssetvalue{/ed/q3/q33/task12/often-pct}{18.9}
\pgfkeyssetvalue{/ed/q3/q33/task12/always-pct}{3.8}
\pgfkeyssetvalue{/ed/q3/q33/task12/high-usage-pct}{22.6}
\pgfkeyssetvalue{/ed/q3/q33/task12/occasional-usage-pct}{17.0}
\pgfkeyssetvalue{/ed/q3/q33/task12/low-usage-pct}{60.4}
\pgfkeyssetvalue{/ed/q3/q33/top1/label}{Generating examples or code snippets}
\pgfkeyssetvalue{/ed/q3/q33/top1/high-usage-pct}{29.1}
\pgfkeyssetvalue{/ed/q3/q33/top1/low-usage-pct}{38.2}
\pgfkeyssetvalue{/ed/q3/q33/top1/task-num}{8}
\pgfkeyssetvalue{/ed/q3/q33/top2/label}{Writing sample solutions}
\pgfkeyssetvalue{/ed/q3/q33/top2/high-usage-pct}{25.9}
\pgfkeyssetvalue{/ed/q3/q33/top2/low-usage-pct}{53.7}
\pgfkeyssetvalue{/ed/q3/q33/top2/task-num}{5}
\pgfkeyssetvalue{/ed/q3/q33/top3/label}{Creating quiz or exam questions}
\pgfkeyssetvalue{/ed/q3/q33/top3/high-usage-pct}{25.5}
\pgfkeyssetvalue{/ed/q3/q33/top3/low-usage-pct}{49.1}
\pgfkeyssetvalue{/ed/q3/q33/top3/task-num}{4}
\pgfkeyssetvalue{/ed/q3/q33/bottom1/label}{Grading student assignments}
\pgfkeyssetvalue{/ed/q3/q33/bottom1/high-usage-pct}{3.8}
\pgfkeyssetvalue{/ed/q3/q33/bottom1/low-usage-pct}{94.3}
\pgfkeyssetvalue{/ed/q3/q33/bottom1/task-num}{7}
\pgfkeyssetvalue{/ed/q3/q33/bottom2/label}{Drafting feedback to students}
\pgfkeyssetvalue{/ed/q3/q33/bottom2/high-usage-pct}{7.4}
\pgfkeyssetvalue{/ed/q3/q33/bottom2/low-usage-pct}{81.5}
\pgfkeyssetvalue{/ed/q3/q33/bottom2/task-num}{6}
\pgfkeyssetvalue{/ed/q3/q33/bottom3/label}{Preparing lecture slides}
\pgfkeyssetvalue{/ed/q3/q33/bottom3/high-usage-pct}{17.9}
\pgfkeyssetvalue{/ed/q3/q33/bottom3/low-usage-pct}{62.5}
\pgfkeyssetvalue{/ed/q3/q33/bottom3/task-num}{1}
\pgfkeyssetvalue{/ed/q3/q35/n-substantive}{52}
\pgfkeyssetvalue{/ed/q3/q35/n-missing}{4}
\pgfkeyssetvalue{/ed/q3/q35/median-label}{Moderate}
\pgfkeyssetvalue{/ed/q3/q35/iqr}{0.5}
\pgfkeyssetvalue{/ed/q3/q35/top-box-pct}{75.0}
\pgfkeyssetvalue{/ed/q3/q35/none/n}{0}
\pgfkeyssetvalue{/ed/q3/q35/none/pct}{0.0}
\pgfkeyssetvalue{/ed/q3/q35/minimal/n}{13}
\pgfkeyssetvalue{/ed/q3/q35/minimal/pct}{25.0}
\pgfkeyssetvalue{/ed/q3/q35/moderate/n}{26}
\pgfkeyssetvalue{/ed/q3/q35/moderate/pct}{50.0}
\pgfkeyssetvalue{/ed/q3/q35/extensive/n}{13}
\pgfkeyssetvalue{/ed/q3/q35/extensive/pct}{25.0}
\pgfkeyssetvalue{/ed/q3/q36/yes/n}{39}
\pgfkeyssetvalue{/ed/q3/q36/yes/pct}{69.6}
\pgfkeyssetvalue{/ed/q3/q36/no/n}{17}
\pgfkeyssetvalue{/ed/q3/q36/no/pct}{30.4}
\pgfkeyssetvalue{/ed/q3/q38/changers/n}{49}
\pgfkeyssetvalue{/ed/q3/q38/changers/pct}{87.5}
\pgfkeyssetvalue{/ed/q3/q38/no-change/n}{7}
\pgfkeyssetvalue{/ed/q3/q38/no-change/pct}{12.5}
\pgfkeyssetvalue{/ed/q3/q38/mean-changes-per-changer}{2.3}
\pgfkeyssetvalue{/ed/q3/q38/lecture-content/n}{25}
\pgfkeyssetvalue{/ed/q3/q38/lecture-content/pct}{44.6}
\pgfkeyssetvalue{/ed/q3/q38/lecture-content/pct-of-changers}{51.0}
\pgfkeyssetvalue{/ed/q3/q38/in-class/n}{32}
\pgfkeyssetvalue{/ed/q3/q38/in-class/pct}{57.1}
\pgfkeyssetvalue{/ed/q3/q38/in-class/pct-of-changers}{65.3}
\pgfkeyssetvalue{/ed/q3/q38/examples-demos/n}{25}
\pgfkeyssetvalue{/ed/q3/q38/examples-demos/pct}{44.6}
\pgfkeyssetvalue{/ed/q3/q38/examples-demos/pct-of-changers}{51.0}
\pgfkeyssetvalue{/ed/q3/q38/student-guidance/n}{29}
\pgfkeyssetvalue{/ed/q3/q38/student-guidance/pct}{51.8}
\pgfkeyssetvalue{/ed/q3/q38/student-guidance/pct-of-changers}{59.2}

%% file: values/educator-q3-4-values.tex
\pgfkeyssetvalue{/ed/q34/n}{56}
\pgfkeyssetvalue{/ed/q34/named-task/n}{40}
\pgfkeyssetvalue{/ed/q34/named-task/pct}{71.4}
\pgfkeyssetvalue{/ed/q34/annotated/n}{18}
\pgfkeyssetvalue{/ed/q34/annotated/pct}{32.1}
\pgfkeyssetvalue{/ed/q34/none-response/n}{8}
\pgfkeyssetvalue{/ed/q34/none-response/pct}{14.3}
\pgfkeyssetvalue{/ed/q34/total-task-mentions}{59}
\pgfkeyssetvalue{/ed/q34/mean-tasks-per-namer}{1.5}
\pgfkeyssetvalue{/ed/q34/task/designing-assignments/n}{11}
\pgfkeyssetvalue{/ed/q34/task/designing-assignments/pct-of-namers}{27.5}
\pgfkeyssetvalue{/ed/q34/task/designing-assignments/pct}{19.6}
\pgfkeyssetvalue{/ed/q34/task/starter-code/n}{8}
\pgfkeyssetvalue{/ed/q34/task/starter-code/pct-of-namers}{20.0}
\pgfkeyssetvalue{/ed/q34/task/starter-code/pct}{14.3}
\pgfkeyssetvalue{/ed/q34/task/sample-solutions/n}{7}
\pgfkeyssetvalue{/ed/q34/task/sample-solutions/pct-of-namers}{17.5}
\pgfkeyssetvalue{/ed/q34/task/sample-solutions/pct}{12.5}
\pgfkeyssetvalue{/ed/q34/task/examples-snippets/n}{7}
\pgfkeyssetvalue{/ed/q34/task/examples-snippets/pct-of-namers}{17.5}
\pgfkeyssetvalue{/ed/q34/task/examples-snippets/pct}{12.5}
\pgfkeyssetvalue{/ed/q34/task/quiz-exam/n}{7}
\pgfkeyssetvalue{/ed/q34/task/quiz-exam/pct-of-namers}{17.5}
\pgfkeyssetvalue{/ed/q34/task/quiz-exam/pct}{12.5}
\pgfkeyssetvalue{/ed/q34/task/summarizing/n}{4}
\pgfkeyssetvalue{/ed/q34/task/summarizing/pct-of-namers}{10.0}
\pgfkeyssetvalue{/ed/q34/task/summarizing/pct}{7.1}
\pgfkeyssetvalue{/ed/q34/task/teaching-notes/n}{4}
\pgfkeyssetvalue{/ed/q34/task/teaching-notes/pct-of-namers}{10.0}
\pgfkeyssetvalue{/ed/q34/task/teaching-notes/pct}{7.1}
\pgfkeyssetvalue{/ed/q34/task/lecture-slides/n}{4}
\pgfkeyssetvalue{/ed/q34/task/lecture-slides/pct-of-namers}{10.0}
\pgfkeyssetvalue{/ed/q34/task/lecture-slides/pct}{7.1}
\pgfkeyssetvalue{/ed/q34/task/communications/n}{3}
\pgfkeyssetvalue{/ed/q34/task/communications/pct-of-namers}{7.5}
\pgfkeyssetvalue{/ed/q34/task/communications/pct}{5.4}
\pgfkeyssetvalue{/ed/q34/task/feedback/n}{3}
\pgfkeyssetvalue{/ed/q34/task/feedback/pct-of-namers}{7.5}
\pgfkeyssetvalue{/ed/q34/task/feedback/pct}{5.4}
\pgfkeyssetvalue{/ed/q34/task/debugging/n}{1}
\pgfkeyssetvalue{/ed/q34/task/debugging/pct-of-namers}{2.5}
\pgfkeyssetvalue{/ed/q34/task/debugging/pct}{1.8}
\pgfkeyssetvalue{/ed/q34/task/top1/label}{Designing assignments or projects}
\pgfkeyssetvalue{/ed/q34/task/top1/n}{11}
\pgfkeyssetvalue{/ed/q34/task/top1/pct-of-namers}{27.5}
\pgfkeyssetvalue{/ed/q34/task/top2/label}{Generating starter code for exercises or assignments}
\pgfkeyssetvalue{/ed/q34/task/top2/n}{8}
\pgfkeyssetvalue{/ed/q34/task/top2/pct-of-namers}{20.0}
\pgfkeyssetvalue{/ed/q34/task/top3/label}{Writing sample solutions}
\pgfkeyssetvalue{/ed/q34/task/top3/n}{7}
\pgfkeyssetvalue{/ed/q34/task/top3/pct-of-namers}{17.5}
\pgfkeyssetvalue{/ed/q34/themes/n-distinct}{6}
\pgfkeyssetvalue{/ed/q34/theme/time-saving/n}{8}
\pgfkeyssetvalue{/ed/q34/theme/time-saving/pct-of-annotated}{44.4}
\pgfkeyssetvalue{/ed/q34/theme/time-saving/pct}{14.3}
\pgfkeyssetvalue{/ed/q34/theme/assessment-variation/n}{6}
\pgfkeyssetvalue{/ed/q34/theme/assessment-variation/pct-of-annotated}{33.3}
\pgfkeyssetvalue{/ed/q34/theme/assessment-variation/pct}{10.7}
\pgfkeyssetvalue{/ed/q34/theme/content-verification/n}{3}
\pgfkeyssetvalue{/ed/q34/theme/content-verification/pct-of-annotated}{16.7}
\pgfkeyssetvalue{/ed/q34/theme/content-verification/pct}{5.4}
\pgfkeyssetvalue{/ed/q34/theme/material-refinement/n}{3}
\pgfkeyssetvalue{/ed/q34/theme/material-refinement/pct-of-annotated}{16.7}
\pgfkeyssetvalue{/ed/q34/theme/material-refinement/pct}{5.4}
\pgfkeyssetvalue{/ed/q34/theme/grading-support/n}{1}
\pgfkeyssetvalue{/ed/q34/theme/grading-support/pct-of-annotated}{5.6}
\pgfkeyssetvalue{/ed/q34/theme/grading-support/pct}{1.8}
\pgfkeyssetvalue{/ed/q34/theme/ai-capability/n}{1}
\pgfkeyssetvalue{/ed/q34/theme/ai-capability/pct-of-annotated}{5.6}
\pgfkeyssetvalue{/ed/q34/theme/ai-capability/pct}{1.8}
\pgfkeyssetvalue{/ed/q34/theme/top1/label}{Time saving / efficiency}
\pgfkeyssetvalue{/ed/q34/theme/top1/slug}{time-saving}
\pgfkeyssetvalue{/ed/q34/theme/top1/n}{8}
\pgfkeyssetvalue{/ed/q34/theme/top1/pct-of-annotated}{44.4}
\pgfkeyssetvalue{/ed/q34/theme/top2/label}{Variation \& assessment design}
\pgfkeyssetvalue{/ed/q34/theme/top2/slug}{assessment-variation}
\pgfkeyssetvalue{/ed/q34/theme/top2/n}{6}
\pgfkeyssetvalue{/ed/q34/theme/top2/pct-of-annotated}{33.3}
\pgfkeyssetvalue{/ed/q34/theme/top3/label}{Content checking / verification}
\pgfkeyssetvalue{/ed/q34/theme/top3/slug}{content-verification}
\pgfkeyssetvalue{/ed/q34/theme/top3/n}{3}
\pgfkeyssetvalue{/ed/q34/theme/top3/pct-of-annotated}{16.7}

%% file: values/educator-q4-values.tex
\pgfkeyssetvalue{/ed/q4/n}{56}
\pgfkeyssetvalue{/ed/q4/q41/yes-formal/n}{40}
\pgfkeyssetvalue{/ed/q4/q41/yes-formal/pct}{71.4}
\pgfkeyssetvalue{/ed/q4/q41/yes-informal/n}{9}
\pgfkeyssetvalue{/ed/q4/q41/yes-informal/pct}{16.1}
\pgfkeyssetvalue{/ed/q4/q41/some/n}{7}
\pgfkeyssetvalue{/ed/q4/q41/some/pct}{12.5}
\pgfkeyssetvalue{/ed/q4/q41/none/n}{0}
\pgfkeyssetvalue{/ed/q4/q41/none/pct}{0.0}
\pgfkeyssetvalue{/ed/q4/q41/any-guidance/n}{56}
\pgfkeyssetvalue{/ed/q4/q41/any-guidance/pct}{100.0}
\pgfkeyssetvalue{/ed/q4/q41/any-explicit/n}{49}
\pgfkeyssetvalue{/ed/q4/q41/any-explicit/pct}{87.5}
\pgfkeyssetvalue{/ed/q4/q42/mean-topics-per-responder}{3.4}
\pgfkeyssetvalue{/ed/q4/q42/allowed-prohibited/n}{48}
\pgfkeyssetvalue{/ed/q4/q42/allowed-prohibited/pct}{85.7}
\pgfkeyssetvalue{/ed/q4/q42/citation-disclosure/n}{31}
\pgfkeyssetvalue{/ed/q4/q42/citation-disclosure/pct}{55.4}
\pgfkeyssetvalue{/ed/q4/q42/assessment-rules/n}{21}
\pgfkeyssetvalue{/ed/q4/q42/assessment-rules/pct}{37.5}
\pgfkeyssetvalue{/ed/q4/q42/integrity/n}{38}
\pgfkeyssetvalue{/ed/q4/q42/integrity/pct}{67.9}
\pgfkeyssetvalue{/ed/q4/q42/consequences/n}{38}
\pgfkeyssetvalue{/ed/q4/q42/consequences/pct}{67.9}
\pgfkeyssetvalue{/ed/q4/q42/tool-guidance/n}{15}
\pgfkeyssetvalue{/ed/q4/q42/tool-guidance/pct}{26.8}
\pgfkeyssetvalue{/ed/q4/q42/top1/label}{Allowed vs prohibited uses}
\pgfkeyssetvalue{/ed/q4/q42/top1/n}{48}
\pgfkeyssetvalue{/ed/q4/q42/top1/pct}{85.7}
\pgfkeyssetvalue{/ed/q4/q42/top2/label}{Academic integrity}
\pgfkeyssetvalue{/ed/q4/q42/top2/n}{38}
\pgfkeyssetvalue{/ed/q4/q42/top2/pct}{67.9}
\pgfkeyssetvalue{/ed/q4/q42/top3/label}{Consequences of misuse}
\pgfkeyssetvalue{/ed/q4/q42/top3/n}{38}
\pgfkeyssetvalue{/ed/q4/q42/top3/pct}{67.9}
\pgfkeyssetvalue{/ed/q4/q44/yes-substantially/n}{22}
\pgfkeyssetvalue{/ed/q4/q44/yes-substantially/pct}{39.3}
\pgfkeyssetvalue{/ed/q4/q44/yes-slightly/n}{22}
\pgfkeyssetvalue{/ed/q4/q44/yes-slightly/pct}{39.3}
\pgfkeyssetvalue{/ed/q4/q44/no-considering/n}{8}
\pgfkeyssetvalue{/ed/q4/q44/no-considering/pct}{14.3}
\pgfkeyssetvalue{/ed/q4/q44/no-wont/n}{4}
\pgfkeyssetvalue{/ed/q4/q44/no-wont/pct}{7.1}
\pgfkeyssetvalue{/ed/q4/q44/yes-total/n}{44}
\pgfkeyssetvalue{/ed/q4/q44/yes-total/pct}{78.6}
\pgfkeyssetvalue{/ed/q4/q44/no-or-considering/n}{12}
\pgfkeyssetvalue{/ed/q4/q44/no-or-considering/pct}{21.4}
\pgfkeyssetvalue{/ed/q4/q47/n-responders}{49}
\pgfkeyssetvalue{/ed/q4/q47/oral/label}{Oral exams or interviews}
\pgfkeyssetvalue{/ed/q4/q47/oral/n}{38}
\pgfkeyssetvalue{/ed/q4/q47/oral/increased-pct}{84.2}
\pgfkeyssetvalue{/ed/q4/q47/oral/decreased-pct}{0.0}
\pgfkeyssetvalue{/ed/q4/q47/oral/no-change-pct}{15.8}
\pgfkeyssetvalue{/ed/q4/q47/oral/n-increased}{32}
\pgfkeyssetvalue{/ed/q4/q47/oral/n-decreased}{0}
\pgfkeyssetvalue{/ed/q4/q47/oral/n-no-change}{6}
\pgfkeyssetvalue{/ed/q4/q47/oral/substantive/n}{38}
\pgfkeyssetvalue{/ed/q4/q47/oral/substantive/n-increased}{32}
\pgfkeyssetvalue{/ed/q4/q47/oral/substantive/n-decreased}{0}
\pgfkeyssetvalue{/ed/q4/q47/oral/substantive/n-no-change}{6}
\pgfkeyssetvalue{/ed/q4/q47/oral/substantive/increased-pct}{84.2}
\pgfkeyssetvalue{/ed/q4/q47/oral/substantive/decreased-pct}{0.0}
\pgfkeyssetvalue{/ed/q4/q47/oral/substantive/no-change-pct}{15.8}
\pgfkeyssetvalue{/ed/q4/q47/oral/full-n/n}{56}
\pgfkeyssetvalue{/ed/q4/q47/oral/full-n/n-blank}{18}
\pgfkeyssetvalue{/ed/q4/q47/oral/full-n/blank-pct}{32.1}
\pgfkeyssetvalue{/ed/q4/q47/oral/full-n/increased-pct}{57.1}
\pgfkeyssetvalue{/ed/q4/q47/oral/full-n/decreased-pct}{0.0}
\pgfkeyssetvalue{/ed/q4/q47/oral/full-n/no-change-pct}{10.7}
\pgfkeyssetvalue{/ed/q4/q47/in-class/label}{In-class or supervised assessments}
\pgfkeyssetvalue{/ed/q4/q47/in-class/n}{44}
\pgfkeyssetvalue{/ed/q4/q47/in-class/increased-pct}{65.9}
\pgfkeyssetvalue{/ed/q4/q47/in-class/decreased-pct}{2.3}
\pgfkeyssetvalue{/ed/q4/q47/in-class/no-change-pct}{31.8}
\pgfkeyssetvalue{/ed/q4/q47/in-class/n-increased}{29}
\pgfkeyssetvalue{/ed/q4/q47/in-class/n-decreased}{1}
\pgfkeyssetvalue{/ed/q4/q47/in-class/n-no-change}{14}
\pgfkeyssetvalue{/ed/q4/q47/in-class/substantive/n}{44}
\pgfkeyssetvalue{/ed/q4/q47/in-class/substantive/n-increased}{29}
\pgfkeyssetvalue{/ed/q4/q47/in-class/substantive/n-decreased}{1}
\pgfkeyssetvalue{/ed/q4/q47/in-class/substantive/n-no-change}{14}
\pgfkeyssetvalue{/ed/q4/q47/in-class/substantive/increased-pct}{65.9}
\pgfkeyssetvalue{/ed/q4/q47/in-class/substantive/decreased-pct}{2.3}
\pgfkeyssetvalue{/ed/q4/q47/in-class/substantive/no-change-pct}{31.8}
\pgfkeyssetvalue{/ed/q4/q47/in-class/full-n/n}{56}
\pgfkeyssetvalue{/ed/q4/q47/in-class/full-n/n-blank}{12}
\pgfkeyssetvalue{/ed/q4/q47/in-class/full-n/blank-pct}{21.4}
\pgfkeyssetvalue{/ed/q4/q47/in-class/full-n/increased-pct}{51.8}
\pgfkeyssetvalue{/ed/q4/q47/in-class/full-n/decreased-pct}{1.8}
\pgfkeyssetvalue{/ed/q4/q47/in-class/full-n/no-change-pct}{25.0}
\pgfkeyssetvalue{/ed/q4/q47/reflective/label}{Reflective or justification-based components}
\pgfkeyssetvalue{/ed/q4/q47/reflective/n}{37}
\pgfkeyssetvalue{/ed/q4/q47/reflective/increased-pct}{64.9}
\pgfkeyssetvalue{/ed/q4/q47/reflective/decreased-pct}{18.9}
\pgfkeyssetvalue{/ed/q4/q47/reflective/no-change-pct}{16.2}
\pgfkeyssetvalue{/ed/q4/q47/reflective/n-increased}{24}
\pgfkeyssetvalue{/ed/q4/q47/reflective/n-decreased}{7}
\pgfkeyssetvalue{/ed/q4/q47/reflective/n-no-change}{6}
\pgfkeyssetvalue{/ed/q4/q47/reflective/substantive/n}{37}
\pgfkeyssetvalue{/ed/q4/q47/reflective/substantive/n-increased}{24}
\pgfkeyssetvalue{/ed/q4/q47/reflective/substantive/n-decreased}{7}
\pgfkeyssetvalue{/ed/q4/q47/reflective/substantive/n-no-change}{6}
\pgfkeyssetvalue{/ed/q4/q47/reflective/substantive/increased-pct}{64.9}
\pgfkeyssetvalue{/ed/q4/q47/reflective/substantive/decreased-pct}{18.9}
\pgfkeyssetvalue{/ed/q4/q47/reflective/substantive/no-change-pct}{16.2}
\pgfkeyssetvalue{/ed/q4/q47/reflective/full-n/n}{56}
\pgfkeyssetvalue{/ed/q4/q47/reflective/full-n/n-blank}{19}
\pgfkeyssetvalue{/ed/q4/q47/reflective/full-n/blank-pct}{33.9}
\pgfkeyssetvalue{/ed/q4/q47/reflective/full-n/increased-pct}{42.9}
\pgfkeyssetvalue{/ed/q4/q47/reflective/full-n/decreased-pct}{12.5}
\pgfkeyssetvalue{/ed/q4/q47/reflective/full-n/no-change-pct}{10.7}
\pgfkeyssetvalue{/ed/q4/q47/group-work/label}{Group work}
\pgfkeyssetvalue{/ed/q4/q47/group-work/n}{38}
\pgfkeyssetvalue{/ed/q4/q47/group-work/increased-pct}{36.8}
\pgfkeyssetvalue{/ed/q4/q47/group-work/decreased-pct}{13.2}
\pgfkeyssetvalue{/ed/q4/q47/group-work/no-change-pct}{50.0}
\pgfkeyssetvalue{/ed/q4/q47/group-work/n-increased}{14}
\pgfkeyssetvalue{/ed/q4/q47/group-work/n-decreased}{5}
\pgfkeyssetvalue{/ed/q4/q47/group-work/n-no-change}{19}
\pgfkeyssetvalue{/ed/q4/q47/group-work/substantive/n}{38}
\pgfkeyssetvalue{/ed/q4/q47/group-work/substantive/n-increased}{14}
\pgfkeyssetvalue{/ed/q4/q47/group-work/substantive/n-decreased}{5}
\pgfkeyssetvalue{/ed/q4/q47/group-work/substantive/n-no-change}{19}
\pgfkeyssetvalue{/ed/q4/q47/group-work/substantive/increased-pct}{36.8}
\pgfkeyssetvalue{/ed/q4/q47/group-work/substantive/decreased-pct}{13.2}
\pgfkeyssetvalue{/ed/q4/q47/group-work/substantive/no-change-pct}{50.0}
\pgfkeyssetvalue{/ed/q4/q47/group-work/full-n/n}{56}
\pgfkeyssetvalue{/ed/q4/q47/group-work/full-n/n-blank}{18}
\pgfkeyssetvalue{/ed/q4/q47/group-work/full-n/blank-pct}{32.1}
\pgfkeyssetvalue{/ed/q4/q47/group-work/full-n/increased-pct}{25.0}
\pgfkeyssetvalue{/ed/q4/q47/group-work/full-n/decreased-pct}{8.9}
\pgfkeyssetvalue{/ed/q4/q47/group-work/full-n/no-change-pct}{33.9}
\pgfkeyssetvalue{/ed/q4/q47/process-based/label}{Process-based assessment}
\pgfkeyssetvalue{/ed/q4/q47/process-based/n}{33}
\pgfkeyssetvalue{/ed/q4/q47/process-based/increased-pct}{42.4}
\pgfkeyssetvalue{/ed/q4/q47/process-based/decreased-pct}{3.0}
\pgfkeyssetvalue{/ed/q4/q47/process-based/no-change-pct}{54.5}
\pgfkeyssetvalue{/ed/q4/q47/process-based/n-increased}{14}
\pgfkeyssetvalue{/ed/q4/q47/process-based/n-decreased}{1}
\pgfkeyssetvalue{/ed/q4/q47/process-based/n-no-change}{18}
\pgfkeyssetvalue{/ed/q4/q47/process-based/substantive/n}{33}
\pgfkeyssetvalue{/ed/q4/q47/process-based/substantive/n-increased}{14}
\pgfkeyssetvalue{/ed/q4/q47/process-based/substantive/n-decreased}{1}
\pgfkeyssetvalue{/ed/q4/q47/process-based/substantive/n-no-change}{18}
\pgfkeyssetvalue{/ed/q4/q47/process-based/substantive/increased-pct}{42.4}
\pgfkeyssetvalue{/ed/q4/q47/process-based/substantive/decreased-pct}{3.0}
\pgfkeyssetvalue{/ed/q4/q47/process-based/substantive/no-change-pct}{54.5}
\pgfkeyssetvalue{/ed/q4/q47/process-based/full-n/n}{56}
\pgfkeyssetvalue{/ed/q4/q47/process-based/full-n/n-blank}{23}
\pgfkeyssetvalue{/ed/q4/q47/process-based/full-n/blank-pct}{41.1}
\pgfkeyssetvalue{/ed/q4/q47/process-based/full-n/increased-pct}{25.0}
\pgfkeyssetvalue{/ed/q4/q47/process-based/full-n/decreased-pct}{1.8}
\pgfkeyssetvalue{/ed/q4/q47/process-based/full-n/no-change-pct}{32.1}
\pgfkeyssetvalue{/ed/q4/q47/top1/label}{Oral exams or interviews}
\pgfkeyssetvalue{/ed/q4/q47/top1/increased-pct}{84.2}
\pgfkeyssetvalue{/ed/q4/q47/top2/label}{In-class or supervised assessments}
\pgfkeyssetvalue{/ed/q4/q47/top2/increased-pct}{65.9}
\pgfkeyssetvalue{/ed/q4/q47/top3/label}{Reflective or justification-based components}
\pgfkeyssetvalue{/ed/q4/q47/top3/increased-pct}{64.9}
\pgfkeyssetvalue{/ed/q4/q47/bottom1/label}{Group work}
\pgfkeyssetvalue{/ed/q4/q47/bottom1/increased-pct}{36.8}
\pgfkeyssetvalue{/ed/q4/q47/bottom1/no-change-pct}{50.0}
\pgfkeyssetvalue{/ed/q4/q47/bottom2/label}{Process-based assessment}
\pgfkeyssetvalue{/ed/q4/q47/bottom2/increased-pct}{42.4}
\pgfkeyssetvalue{/ed/q4/q47/bottom2/no-change-pct}{54.5}
\pgfkeyssetvalue{/ed/q4/q47/bottom3/label}{Reflective or justification-based components}
\pgfkeyssetvalue{/ed/q4/q47/bottom3/increased-pct}{64.9}
\pgfkeyssetvalue{/ed/q4/q47/bottom3/no-change-pct}{16.2}
\pgfkeyssetvalue{/ed/q4/q49/yes/n}{28}
\pgfkeyssetvalue{/ed/q4/q49/yes/pct}{50.0}
\pgfkeyssetvalue{/ed/q4/q49/no-considering/n}{15}
\pgfkeyssetvalue{/ed/q4/q49/no-considering/pct}{26.8}
\pgfkeyssetvalue{/ed/q4/q49/no-wont/n}{13}
\pgfkeyssetvalue{/ed/q4/q49/no-wont/pct}{23.2}
\pgfkeyssetvalue{/ed/q4/q410/n}{56}
\pgfkeyssetvalue{/ed/q4/q410/positive/n}{40}
\pgfkeyssetvalue{/ed/q4/q410/positive/pct}{71.4}
\pgfkeyssetvalue{/ed/q4/q410/positive/extreme/n}{12}
\pgfkeyssetvalue{/ed/q4/q410/positive/extreme/pct}{21.4}
\pgfkeyssetvalue{/ed/q4/q410/positive/base/n}{28}
\pgfkeyssetvalue{/ed/q4/q410/positive/base/pct}{50.0}
\pgfkeyssetvalue{/ed/q4/q410/neutral/n}{8}
\pgfkeyssetvalue{/ed/q4/q410/neutral/pct}{14.3}
\pgfkeyssetvalue{/ed/q4/q410/negative/n}{8}
\pgfkeyssetvalue{/ed/q4/q410/negative/pct}{14.3}
\pgfkeyssetvalue{/ed/q4/q410/negative/extreme/n}{3}
\pgfkeyssetvalue{/ed/q4/q410/negative/extreme/pct}{5.4}
\pgfkeyssetvalue{/ed/q4/q410/negative/base/n}{5}
\pgfkeyssetvalue{/ed/q4/q410/negative/base/pct}{8.9}
\pgfkeyssetvalue{/ed/q4/q410/strongly-disagree/n}{3}
\pgfkeyssetvalue{/ed/q4/q410/strongly-disagree/pct}{5.4}
\pgfkeyssetvalue{/ed/q4/q410/disagree/n}{5}
\pgfkeyssetvalue{/ed/q4/q410/disagree/pct}{8.9}
\pgfkeyssetvalue{/ed/q4/q410/neither-agree-nor-disagree/n}{8}
\pgfkeyssetvalue{/ed/q4/q410/neither-agree-nor-disagree/pct}{14.3}
\pgfkeyssetvalue{/ed/q4/q410/agree/n}{28}
\pgfkeyssetvalue{/ed/q4/q410/agree/pct}{50.0}
\pgfkeyssetvalue{/ed/q4/q410/strongly-agree/n}{12}
\pgfkeyssetvalue{/ed/q4/q410/strongly-agree/pct}{21.4}
\pgfkeyssetvalue{/ed/q4/q411/n}{56}
\pgfkeyssetvalue{/ed/q4/q411/positive/n}{20}
\pgfkeyssetvalue{/ed/q4/q411/positive/pct}{35.7}
\pgfkeyssetvalue{/ed/q4/q411/positive/extreme/n}{6}
\pgfkeyssetvalue{/ed/q4/q411/positive/extreme/pct}{10.7}
\pgfkeyssetvalue{/ed/q4/q411/positive/base/n}{14}
\pgfkeyssetvalue{/ed/q4/q411/positive/base/pct}{25.0}
\pgfkeyssetvalue{/ed/q4/q411/neutral/n}{16}
\pgfkeyssetvalue{/ed/q4/q411/neutral/pct}{28.6}
\pgfkeyssetvalue{/ed/q4/q411/negative/n}{20}
\pgfkeyssetvalue{/ed/q4/q411/negative/pct}{35.7}
\pgfkeyssetvalue{/ed/q4/q411/negative/extreme/n}{11}
\pgfkeyssetvalue{/ed/q4/q411/negative/extreme/pct}{19.6}
\pgfkeyssetvalue{/ed/q4/q411/negative/base/n}{9}
\pgfkeyssetvalue{/ed/q4/q411/negative/base/pct}{16.1}
\pgfkeyssetvalue{/ed/q4/q411/never/n}{11}
\pgfkeyssetvalue{/ed/q4/q411/never/pct}{19.6}
\pgfkeyssetvalue{/ed/q4/q411/rarely/n}{9}
\pgfkeyssetvalue{/ed/q4/q411/rarely/pct}{16.1}
\pgfkeyssetvalue{/ed/q4/q411/sometimes/n}{16}
\pgfkeyssetvalue{/ed/q4/q411/sometimes/pct}{28.6}
\pgfkeyssetvalue{/ed/q4/q411/often/n}{14}
\pgfkeyssetvalue{/ed/q4/q411/often/pct}{25.0}
\pgfkeyssetvalue{/ed/q4/q411/always/n}{6}
\pgfkeyssetvalue{/ed/q4/q411/always/pct}{10.7}
\pgfkeyssetvalue{/ed/q4/q412/n}{56}
\pgfkeyssetvalue{/ed/q4/q412/positive/n}{17}
\pgfkeyssetvalue{/ed/q4/q412/positive/pct}{30.4}
\pgfkeyssetvalue{/ed/q4/q412/positive/extreme/n}{3}
\pgfkeyssetvalue{/ed/q4/q412/positive/extreme/pct}{5.4}
\pgfkeyssetvalue{/ed/q4/q412/positive/base/n}{14}
\pgfkeyssetvalue{/ed/q4/q412/positive/base/pct}{25.0}
\pgfkeyssetvalue{/ed/q4/q412/neutral/n}{23}
\pgfkeyssetvalue{/ed/q4/q412/neutral/pct}{41.1}
\pgfkeyssetvalue{/ed/q4/q412/negative/n}{16}
\pgfkeyssetvalue{/ed/q4/q412/negative/pct}{28.6}
\pgfkeyssetvalue{/ed/q4/q412/negative/extreme/n}{7}
\pgfkeyssetvalue{/ed/q4/q412/negative/extreme/pct}{12.5}
\pgfkeyssetvalue{/ed/q4/q412/negative/base/n}{9}
\pgfkeyssetvalue{/ed/q4/q412/negative/base/pct}{16.1}
\pgfkeyssetvalue{/ed/q4/q412/never/n}{7}
\pgfkeyssetvalue{/ed/q4/q412/never/pct}{12.5}
\pgfkeyssetvalue{/ed/q4/q412/rarely/n}{9}
\pgfkeyssetvalue{/ed/q4/q412/rarely/pct}{16.1}
\pgfkeyssetvalue{/ed/q4/q412/sometimes/n}{23}
\pgfkeyssetvalue{/ed/q4/q412/sometimes/pct}{41.1}
\pgfkeyssetvalue{/ed/q4/q412/often/n}{14}
\pgfkeyssetvalue{/ed/q4/q412/often/pct}{25.0}
\pgfkeyssetvalue{/ed/q4/q412/always/n}{3}
\pgfkeyssetvalue{/ed/q4/q412/always/pct}{5.4}
\pgfkeyssetvalue{/ed/q4/detection-mismatch/check-pct}{35.7}
\pgfkeyssetvalue{/ed/q4/detection-mismatch/detect-pct}{30.4}
\pgfkeyssetvalue{/ed/q4/detection-mismatch/gap-pct}{5.4}

%% file: values/educator-q3-7-values.tex
\pgfkeyssetvalue{/ed/q37/n}{54}
\pgfkeyssetvalue{/ed/q37/n-full-sample}{54}
\pgfkeyssetvalue{/ed/q37/avg-codes-per-respondent}{1.39}
\pgfkeyssetvalue{/ed/q37/total-code-applications}{75}
\pgfkeyssetvalue{/ed/q37/code/time-saving/n}{14}
\pgfkeyssetvalue{/ed/q37/code/time-saving/pct}{25.9}
\pgfkeyssetvalue{/ed/q37/code/time-saving/label}{Time saving / efficiency}
\pgfkeyssetvalue{/ed/q37/code/content-generation/n}{19}
\pgfkeyssetvalue{/ed/q37/code/content-generation/pct}{35.2}
\pgfkeyssetvalue{/ed/q37/code/content-generation/label}{Content generation support}
\pgfkeyssetvalue{/ed/q37/code/admin-doc/n}{4}
\pgfkeyssetvalue{/ed/q37/code/admin-doc/pct}{7.4}
\pgfkeyssetvalue{/ed/q37/code/admin-doc/label}{Admin/documentation support}
\pgfkeyssetvalue{/ed/q37/code/changes-content/n}{8}
\pgfkeyssetvalue{/ed/q37/code/changes-content/pct}{14.8}
\pgfkeyssetvalue{/ed/q37/code/changes-content/label}{Changes to teaching content}
\pgfkeyssetvalue{/ed/q37/code/changes-assessment/n}{7}
\pgfkeyssetvalue{/ed/q37/code/changes-assessment/pct}{13.0}
\pgfkeyssetvalue{/ed/q37/code/changes-assessment/label}{Changes to assessment}
\pgfkeyssetvalue{/ed/q37/code/student-facing/n}{8}
\pgfkeyssetvalue{/ed/q37/code/student-facing/pct}{14.8}
\pgfkeyssetvalue{/ed/q37/code/student-facing/label}{Student-facing use affecting prep}
\pgfkeyssetvalue{/ed/q37/code/no-effect/n}{11}
\pgfkeyssetvalue{/ed/q37/code/no-effect/pct}{20.4}
\pgfkeyssetvalue{/ed/q37/code/no-effect/label}{No effect / not adopted}
\pgfkeyssetvalue{/ed/q37/code/uncodable/n}{4}
\pgfkeyssetvalue{/ed/q37/code/uncodable/pct}{7.4}
\pgfkeyssetvalue{/ed/q37/code/uncodable/label}{Uncodable}
\pgfkeyssetvalue{/ed/q37/top1/label}{Content generation support}
\pgfkeyssetvalue{/ed/q37/top1/slug}{content-generation}
\pgfkeyssetvalue{/ed/q37/top1/n}{19}
\pgfkeyssetvalue{/ed/q37/top1/pct}{35.2}
\pgfkeyssetvalue{/ed/q37/top2/label}{Time saving / efficiency}
\pgfkeyssetvalue{/ed/q37/top2/slug}{time-saving}
\pgfkeyssetvalue{/ed/q37/top2/n}{14}
\pgfkeyssetvalue{/ed/q37/top2/pct}{25.9}
\pgfkeyssetvalue{/ed/q37/top3/label}{No effect / not adopted}
\pgfkeyssetvalue{/ed/q37/top3/slug}{no-effect}
\pgfkeyssetvalue{/ed/q37/top3/n}{11}
\pgfkeyssetvalue{/ed/q37/top3/pct}{20.4}

%% file: values/educator-q4-3-values.tex
\pgfkeyssetvalue{/ed/q43/non-codeable}{4}
\pgfkeyssetvalue{/ed/q43/n}{23}
\pgfkeyssetvalue{/ed/q43/n-full-sample}{56}
\pgfkeyssetvalue{/ed/q43/avg-codes-per-respondent}{2.91}
\pgfkeyssetvalue{/ed/q43/total-code-applications}{67}
\pgfkeyssetvalue{/ed/q43/code/permission-scope/n}{19}
\pgfkeyssetvalue{/ed/q43/code/permission-scope/pct}{82.6}
\pgfkeyssetvalue{/ed/q43/code/permission-scope/label}{Permission scope}
\pgfkeyssetvalue{/ed/q43/code/disclosure/n}{13}
\pgfkeyssetvalue{/ed/q43/code/disclosure/pct}{56.5}
\pgfkeyssetvalue{/ed/q43/code/disclosure/label}{Disclosure requirement}
\pgfkeyssetvalue{/ed/q43/code/citation-format/n}{2}
\pgfkeyssetvalue{/ed/q43/code/citation-format/pct}{8.7}
\pgfkeyssetvalue{/ed/q43/code/citation-format/label}{Citation format specified}
\pgfkeyssetvalue{/ed/q43/code/accountability/n}{11}
\pgfkeyssetvalue{/ed/q43/code/accountability/pct}{47.8}
\pgfkeyssetvalue{/ed/q43/code/accountability/label}{Student accountability}
\pgfkeyssetvalue{/ed/q43/code/educative/n}{9}
\pgfkeyssetvalue{/ed/q43/code/educative/pct}{39.1}
\pgfkeyssetvalue{/ed/q43/code/educative/label}{Educative framing}
\pgfkeyssetvalue{/ed/q43/code/worked-examples/n}{3}
\pgfkeyssetvalue{/ed/q43/code/worked-examples/pct}{13.0}
\pgfkeyssetvalue{/ed/q43/code/worked-examples/label}{Worked examples}
\pgfkeyssetvalue{/ed/q43/code/integrity-framing/n}{10}
\pgfkeyssetvalue{/ed/q43/code/integrity-framing/pct}{43.5}
\pgfkeyssetvalue{/ed/q43/code/integrity-framing/label}{Academic integrity framing}
\pgfkeyssetvalue{/ed/q43/code/uncodable/n}{4}
\pgfkeyssetvalue{/ed/q43/code/uncodable/pct}{17.4}
\pgfkeyssetvalue{/ed/q43/code/uncodable/label}{Uncodable}
\pgfkeyssetvalue{/ed/q43/top1/label}{Permission scope}
\pgfkeyssetvalue{/ed/q43/top1/slug}{permission-scope}
\pgfkeyssetvalue{/ed/q43/top1/n}{19}
\pgfkeyssetvalue{/ed/q43/top1/pct}{82.6}
\pgfkeyssetvalue{/ed/q43/top2/label}{Disclosure requirement}
\pgfkeyssetvalue{/ed/q43/top2/slug}{disclosure}
\pgfkeyssetvalue{/ed/q43/top2/n}{13}
\pgfkeyssetvalue{/ed/q43/top2/pct}{56.5}
\pgfkeyssetvalue{/ed/q43/top3/label}{Student accountability}
\pgfkeyssetvalue{/ed/q43/top3/slug}{accountability}
\pgfkeyssetvalue{/ed/q43/top3/n}{11}
\pgfkeyssetvalue{/ed/q43/top3/pct}{47.8}

%% file: values/hiring-q3-values.tex
\pgfkeyssetvalue{/hp/q3/n}{24}
\pgfkeyssetvalue{/hp/q3/q31/other-only/n}{1}
\pgfkeyssetvalue{/hp/q3/q31/other-only/pct}{4.2}
\pgfkeyssetvalue{/hp/q3/q31/mean-tools-per-respondent}{4.2}
\pgfkeyssetvalue{/hp/q3/q31/median-tools-per-respondent}{4.0}
\pgfkeyssetvalue{/hp/q3/q31/min-tools-per-respondent}{0}
\pgfkeyssetvalue{/hp/q3/q31/max-tools-per-respondent}{6}
\pgfkeyssetvalue{/hp/q3/q31/chatgpt/n}{21}
\pgfkeyssetvalue{/hp/q3/q31/chatgpt/pct}{87.5}
\pgfkeyssetvalue{/hp/q3/q31/copilot/n}{16}
\pgfkeyssetvalue{/hp/q3/q31/copilot/pct}{66.7}
\pgfkeyssetvalue{/hp/q3/q31/gemini/n}{21}
\pgfkeyssetvalue{/hp/q3/q31/gemini/pct}{87.5}
\pgfkeyssetvalue{/hp/q3/q31/claude/n}{20}
\pgfkeyssetvalue{/hp/q3/q31/claude/pct}{83.3}
\pgfkeyssetvalue{/hp/q3/q31/cursor/n}{14}
\pgfkeyssetvalue{/hp/q3/q31/cursor/pct}{58.3}
\pgfkeyssetvalue{/hp/q3/q31/perplexity/n}{9}
\pgfkeyssetvalue{/hp/q3/q31/perplexity/pct}{37.5}
\pgfkeyssetvalue{/hp/q3/q31/other/n}{3}
\pgfkeyssetvalue{/hp/q3/q31/other/pct}{12.5}
\pgfkeyssetvalue{/hp/q3/q32/top-box/n}{20}
\pgfkeyssetvalue{/hp/q3/q32/top-box/pct}{83.3}
\pgfkeyssetvalue{/hp/q3/q32/middle/n}{4}
\pgfkeyssetvalue{/hp/q3/q32/middle/pct}{16.7}
\pgfkeyssetvalue{/hp/q3/q32/bottom-box/n}{0}
\pgfkeyssetvalue{/hp/q3/q32/bottom-box/pct}{0.0}
\pgfkeyssetvalue{/hp/q3/q32/never/n}{0}
\pgfkeyssetvalue{/hp/q3/q32/never/pct}{0.0}
\pgfkeyssetvalue{/hp/q3/q32/rarely/n}{0}
\pgfkeyssetvalue{/hp/q3/q32/rarely/pct}{0.0}
\pgfkeyssetvalue{/hp/q3/q32/sometimes/n}{4}
\pgfkeyssetvalue{/hp/q3/q32/sometimes/pct}{16.7}
\pgfkeyssetvalue{/hp/q3/q32/often/n}{9}
\pgfkeyssetvalue{/hp/q3/q32/often/pct}{37.5}
\pgfkeyssetvalue{/hp/q3/q32/always/n}{11}
\pgfkeyssetvalue{/hp/q3/q32/always/pct}{45.8}
\pgfkeyssetvalue{/hp/q3/q33/claude/n}{15}
\pgfkeyssetvalue{/hp/q3/q33/claude/pct}{62.5}
\pgfkeyssetvalue{/hp/q3/q33/copilot/n}{3}
\pgfkeyssetvalue{/hp/q3/q33/copilot/pct}{12.5}
\pgfkeyssetvalue{/hp/q3/q33/gemini/n}{2}
\pgfkeyssetvalue{/hp/q3/q33/gemini/pct}{8.3}
\pgfkeyssetvalue{/hp/q3/q33/other/n}{2}
\pgfkeyssetvalue{/hp/q3/q33/other/pct}{8.3}
\pgfkeyssetvalue{/hp/q3/q33/chatgpt/n}{2}
\pgfkeyssetvalue{/hp/q3/q33/chatgpt/pct}{8.3}
\pgfkeyssetvalue{/hp/q3/q33/top1/label}{Claude}
\pgfkeyssetvalue{/hp/q3/q33/top1/n}{15}
\pgfkeyssetvalue{/hp/q3/q33/top1/pct}{62.5}
\pgfkeyssetvalue{/hp/q3/q33/top2/label}{Copilot}
\pgfkeyssetvalue{/hp/q3/q33/top2/n}{3}
\pgfkeyssetvalue{/hp/q3/q33/top2/pct}{12.5}
\pgfkeyssetvalue{/hp/q3/q33/top3/label}{Gemini}
\pgfkeyssetvalue{/hp/q3/q33/top3/n}{2}
\pgfkeyssetvalue{/hp/q3/q33/top3/pct}{8.3}
\pgfkeyssetvalue{/hp/q3/q33/no-response/n}{0}
\pgfkeyssetvalue{/hp/q3/q33/no-response/pct}{0.0}
\pgfkeyssetvalue{/hp/q3/q34/any-use/n}{23}
\pgfkeyssetvalue{/hp/q3/q34/any-use/pct}{95.8}
\pgfkeyssetvalue{/hp/q3/q34/high-use/n}{17}
\pgfkeyssetvalue{/hp/q3/q34/high-use/pct}{70.8}
\pgfkeyssetvalue{/hp/q3/q34/limited-or-some/n}{6}
\pgfkeyssetvalue{/hp/q3/q34/limited-or-some/pct}{25.0}
\pgfkeyssetvalue{/hp/q3/q34/not-used/n}{1}
\pgfkeyssetvalue{/hp/q3/q34/not-used/pct}{4.2}
\pgfkeyssetvalue{/hp/q3/q34/limited/n}{2}
\pgfkeyssetvalue{/hp/q3/q34/limited/pct}{8.3}
\pgfkeyssetvalue{/hp/q3/q34/some-teams/n}{4}
\pgfkeyssetvalue{/hp/q3/q34/some-teams/pct}{16.7}
\pgfkeyssetvalue{/hp/q3/q34/widely-used/n}{10}
\pgfkeyssetvalue{/hp/q3/q34/widely-used/pct}{41.7}
\pgfkeyssetvalue{/hp/q3/q34/standard-practice/n}{7}
\pgfkeyssetvalue{/hp/q3/q34/standard-practice/pct}{29.2}
\pgfkeyssetvalue{/hp/q3/q35/any-guidance/n}{22}
\pgfkeyssetvalue{/hp/q3/q35/any-guidance/pct}{91.7}
\pgfkeyssetvalue{/hp/q3/q35/any-explicit/n}{12}
\pgfkeyssetvalue{/hp/q3/q35/any-explicit/pct}{50.0}
\pgfkeyssetvalue{/hp/q3/q35/none/n}{2}
\pgfkeyssetvalue{/hp/q3/q35/none/pct}{8.3}
\pgfkeyssetvalue{/hp/q3/q35/informal/n}{5}
\pgfkeyssetvalue{/hp/q3/q35/informal/pct}{20.8}
\pgfkeyssetvalue{/hp/q3/q35/in-dev/n}{5}
\pgfkeyssetvalue{/hp/q3/q35/in-dev/pct}{20.8}
\pgfkeyssetvalue{/hp/q3/q35/clear-formal/n}{12}
\pgfkeyssetvalue{/hp/q3/q35/clear-formal/pct}{50.0}
\pgfkeyssetvalue{/hp/q3/q35/not-sure/n}{0}
\pgfkeyssetvalue{/hp/q3/q35/not-sure/pct}{0.0}

%% file: values/hiring-q4-values.tex
\pgfkeyssetvalue{/hp/q4/n}{24}
\pgfkeyssetvalue{/hp/q4/q41/yes/n}{8}
\pgfkeyssetvalue{/hp/q4/q41/yes/pct}{33.3}
\pgfkeyssetvalue{/hp/q4/q41/no/n}{16}
\pgfkeyssetvalue{/hp/q4/q41/no/pct}{66.7}
\pgfkeyssetvalue{/hp/q4/q43/yes/n}{14}
\pgfkeyssetvalue{/hp/q4/q43/yes/pct}{58.3}
\pgfkeyssetvalue{/hp/q4/q43/no/n}{10}
\pgfkeyssetvalue{/hp/q4/q43/no/pct}{41.7}
\pgfkeyssetvalue{/hp/q4/q45/yes/n}{12}
\pgfkeyssetvalue{/hp/q4/q45/yes/pct}{50.0}
\pgfkeyssetvalue{/hp/q4/q45/no/n}{12}
\pgfkeyssetvalue{/hp/q4/q45/no/pct}{50.0}
\pgfkeyssetvalue{/hp/q4/q411/yes/n}{18}
\pgfkeyssetvalue{/hp/q4/q411/yes/pct}{75.0}
\pgfkeyssetvalue{/hp/q4/q411/no/n}{6}
\pgfkeyssetvalue{/hp/q4/q411/no/pct}{25.0}
\pgfkeyssetvalue{/hp/q4/q47/problem-decomposition/label}{Problem decomposition}
\pgfkeyssetvalue{/hp/q4/q47/problem-decomposition/n-substantive}{24}
\pgfkeyssetvalue{/hp/q4/q47/problem-decomposition/n-na}{0}
\pgfkeyssetvalue{/hp/q4/q47/problem-decomposition/positive/pct}{79.2}
\pgfkeyssetvalue{/hp/q4/q47/problem-decomposition/positive/n}{19}
\pgfkeyssetvalue{/hp/q4/q47/problem-decomposition/neutral/pct}{8.3}
\pgfkeyssetvalue{/hp/q4/q47/problem-decomposition/neutral/n}{2}
\pgfkeyssetvalue{/hp/q4/q47/problem-decomposition/negative/pct}{12.5}
\pgfkeyssetvalue{/hp/q4/q47/problem-decomposition/negative/n}{3}
\pgfkeyssetvalue{/hp/q4/q47/code-comprehension/label}{Code comprehension}
\pgfkeyssetvalue{/hp/q4/q47/code-comprehension/n-substantive}{24}
\pgfkeyssetvalue{/hp/q4/q47/code-comprehension/n-na}{0}
\pgfkeyssetvalue{/hp/q4/q47/code-comprehension/positive/pct}{66.7}
\pgfkeyssetvalue{/hp/q4/q47/code-comprehension/positive/n}{16}
\pgfkeyssetvalue{/hp/q4/q47/code-comprehension/neutral/pct}{20.8}
\pgfkeyssetvalue{/hp/q4/q47/code-comprehension/neutral/n}{5}
\pgfkeyssetvalue{/hp/q4/q47/code-comprehension/negative/pct}{12.5}
\pgfkeyssetvalue{/hp/q4/q47/code-comprehension/negative/n}{3}
\pgfkeyssetvalue{/hp/q4/q47/code-development/label}{Code development}
\pgfkeyssetvalue{/hp/q4/q47/code-development/n-substantive}{24}
\pgfkeyssetvalue{/hp/q4/q47/code-development/n-na}{0}
\pgfkeyssetvalue{/hp/q4/q47/code-development/positive/pct}{54.2}
\pgfkeyssetvalue{/hp/q4/q47/code-development/positive/n}{13}
\pgfkeyssetvalue{/hp/q4/q47/code-development/neutral/pct}{12.5}
\pgfkeyssetvalue{/hp/q4/q47/code-development/neutral/n}{3}
\pgfkeyssetvalue{/hp/q4/q47/code-development/negative/pct}{33.3}
\pgfkeyssetvalue{/hp/q4/q47/code-development/negative/n}{8}
\pgfkeyssetvalue{/hp/q4/q47/code-review/label}{Code review}
\pgfkeyssetvalue{/hp/q4/q47/code-review/n-substantive}{24}
\pgfkeyssetvalue{/hp/q4/q47/code-review/n-na}{0}
\pgfkeyssetvalue{/hp/q4/q47/code-review/positive/pct}{75.0}
\pgfkeyssetvalue{/hp/q4/q47/code-review/positive/n}{18}
\pgfkeyssetvalue{/hp/q4/q47/code-review/neutral/pct}{20.8}
\pgfkeyssetvalue{/hp/q4/q47/code-review/neutral/n}{5}
\pgfkeyssetvalue{/hp/q4/q47/code-review/negative/pct}{4.2}
\pgfkeyssetvalue{/hp/q4/q47/code-review/negative/n}{1}
\pgfkeyssetvalue{/hp/q4/q47/debugging/label}{Debugging}
\pgfkeyssetvalue{/hp/q4/q47/debugging/n-substantive}{24}
\pgfkeyssetvalue{/hp/q4/q47/debugging/n-na}{0}
\pgfkeyssetvalue{/hp/q4/q47/debugging/positive/pct}{66.7}
\pgfkeyssetvalue{/hp/q4/q47/debugging/positive/n}{16}
\pgfkeyssetvalue{/hp/q4/q47/debugging/neutral/pct}{29.2}
\pgfkeyssetvalue{/hp/q4/q47/debugging/neutral/n}{7}
\pgfkeyssetvalue{/hp/q4/q47/debugging/negative/pct}{4.2}
\pgfkeyssetvalue{/hp/q4/q47/debugging/negative/n}{1}
\pgfkeyssetvalue{/hp/q4/q47/documentation/label}{Documentation}
\pgfkeyssetvalue{/hp/q4/q47/documentation/n-substantive}{23}
\pgfkeyssetvalue{/hp/q4/q47/documentation/n-na}{1}
\pgfkeyssetvalue{/hp/q4/q47/documentation/positive/pct}{47.8}
\pgfkeyssetvalue{/hp/q4/q47/documentation/positive/n}{11}
\pgfkeyssetvalue{/hp/q4/q47/documentation/neutral/pct}{34.8}
\pgfkeyssetvalue{/hp/q4/q47/documentation/neutral/n}{8}
\pgfkeyssetvalue{/hp/q4/q47/documentation/negative/pct}{17.4}
\pgfkeyssetvalue{/hp/q4/q47/documentation/negative/n}{4}
\pgfkeyssetvalue{/hp/q4/q47/system-design/label}{System-level design}
\pgfkeyssetvalue{/hp/q4/q47/system-design/n-substantive}{24}
\pgfkeyssetvalue{/hp/q4/q47/system-design/n-na}{0}
\pgfkeyssetvalue{/hp/q4/q47/system-design/positive/pct}{62.5}
\pgfkeyssetvalue{/hp/q4/q47/system-design/positive/n}{15}
\pgfkeyssetvalue{/hp/q4/q47/system-design/neutral/pct}{25.0}
\pgfkeyssetvalue{/hp/q4/q47/system-design/neutral/n}{6}
\pgfkeyssetvalue{/hp/q4/q47/system-design/negative/pct}{12.5}
\pgfkeyssetvalue{/hp/q4/q47/system-design/negative/n}{3}
\pgfkeyssetvalue{/hp/q4/q47/algorithmic-thinking/label}{Algorithmic thinking}
\pgfkeyssetvalue{/hp/q4/q47/algorithmic-thinking/n-substantive}{24}
\pgfkeyssetvalue{/hp/q4/q47/algorithmic-thinking/n-na}{0}
\pgfkeyssetvalue{/hp/q4/q47/algorithmic-thinking/positive/pct}{66.7}
\pgfkeyssetvalue{/hp/q4/q47/algorithmic-thinking/positive/n}{16}
\pgfkeyssetvalue{/hp/q4/q47/algorithmic-thinking/neutral/pct}{25.0}
\pgfkeyssetvalue{/hp/q4/q47/algorithmic-thinking/neutral/n}{6}
\pgfkeyssetvalue{/hp/q4/q47/algorithmic-thinking/negative/pct}{8.3}
\pgfkeyssetvalue{/hp/q4/q47/algorithmic-thinking/negative/n}{2}
\pgfkeyssetvalue{/hp/q4/q47/critical-eval-ai/label}{Critical evaluation of AI-generated output}
\pgfkeyssetvalue{/hp/q4/q47/critical-eval-ai/n-substantive}{24}
\pgfkeyssetvalue{/hp/q4/q47/critical-eval-ai/n-na}{0}
\pgfkeyssetvalue{/hp/q4/q47/critical-eval-ai/positive/pct}{100.0}
\pgfkeyssetvalue{/hp/q4/q47/critical-eval-ai/positive/n}{24}
\pgfkeyssetvalue{/hp/q4/q47/critical-eval-ai/neutral/pct}{0.0}
\pgfkeyssetvalue{/hp/q4/q47/critical-eval-ai/neutral/n}{0}
\pgfkeyssetvalue{/hp/q4/q47/critical-eval-ai/negative/pct}{0.0}
\pgfkeyssetvalue{/hp/q4/q47/critical-eval-ai/negative/n}{0}
\pgfkeyssetvalue{/hp/q4/q47/learn-adapt/label}{Ability to learn and adapt independently}
\pgfkeyssetvalue{/hp/q4/q47/learn-adapt/n-substantive}{24}
\pgfkeyssetvalue{/hp/q4/q47/learn-adapt/n-na}{0}
\pgfkeyssetvalue{/hp/q4/q47/learn-adapt/positive/pct}{91.7}
\pgfkeyssetvalue{/hp/q4/q47/learn-adapt/positive/n}{22}
\pgfkeyssetvalue{/hp/q4/q47/learn-adapt/neutral/pct}{8.3}
\pgfkeyssetvalue{/hp/q4/q47/learn-adapt/neutral/n}{2}
\pgfkeyssetvalue{/hp/q4/q47/learn-adapt/negative/pct}{0.0}
\pgfkeyssetvalue{/hp/q4/q47/learn-adapt/negative/n}{0}
\pgfkeyssetvalue{/hp/q4/q47/collaboration/label}{Collaboration and communication}
\pgfkeyssetvalue{/hp/q4/q47/collaboration/n-substantive}{24}
\pgfkeyssetvalue{/hp/q4/q47/collaboration/n-na}{0}
\pgfkeyssetvalue{/hp/q4/q47/collaboration/positive/pct}{83.3}
\pgfkeyssetvalue{/hp/q4/q47/collaboration/positive/n}{20}
\pgfkeyssetvalue{/hp/q4/q47/collaboration/neutral/pct}{12.5}
\pgfkeyssetvalue{/hp/q4/q47/collaboration/neutral/n}{3}
\pgfkeyssetvalue{/hp/q4/q47/collaboration/negative/pct}{4.2}
\pgfkeyssetvalue{/hp/q4/q47/collaboration/negative/n}{1}
\pgfkeyssetvalue{/hp/q4/q47/responsible-genai/label}{Responsible and effective use of GenAI tools}
\pgfkeyssetvalue{/hp/q4/q47/responsible-genai/n-substantive}{24}
\pgfkeyssetvalue{/hp/q4/q47/responsible-genai/n-na}{0}
\pgfkeyssetvalue{/hp/q4/q47/responsible-genai/positive/pct}{95.8}
\pgfkeyssetvalue{/hp/q4/q47/responsible-genai/positive/n}{23}
\pgfkeyssetvalue{/hp/q4/q47/responsible-genai/neutral/pct}{4.2}
\pgfkeyssetvalue{/hp/q4/q47/responsible-genai/neutral/n}{1}
\pgfkeyssetvalue{/hp/q4/q47/responsible-genai/negative/pct}{0.0}
\pgfkeyssetvalue{/hp/q4/q47/responsible-genai/negative/n}{0}
\pgfkeyssetvalue{/hp/q4/q47/top1/label}{Critical evaluation of AI-generated output}
\pgfkeyssetvalue{/hp/q4/q47/top1/pct}{100.0}
\pgfkeyssetvalue{/hp/q4/q47/top1/n}{24}
\pgfkeyssetvalue{/hp/q4/q47/top1/skill-num}{9}
\pgfkeyssetvalue{/hp/q4/q47/top2/label}{Responsible and effective use of GenAI tools}
\pgfkeyssetvalue{/hp/q4/q47/top2/pct}{95.8}
\pgfkeyssetvalue{/hp/q4/q47/top2/n}{23}
\pgfkeyssetvalue{/hp/q4/q47/top2/skill-num}{12}
\pgfkeyssetvalue{/hp/q4/q47/top3/label}{Ability to learn and adapt independently}
\pgfkeyssetvalue{/hp/q4/q47/top3/pct}{91.7}
\pgfkeyssetvalue{/hp/q4/q47/top3/n}{22}
\pgfkeyssetvalue{/hp/q4/q47/top3/skill-num}{10}
\pgfkeyssetvalue{/hp/q4/q47/bottom1/label}{Documentation}
\pgfkeyssetvalue{/hp/q4/q47/bottom1/pct}{47.8}
\pgfkeyssetvalue{/hp/q4/q47/bottom1/n}{11}
\pgfkeyssetvalue{/hp/q4/q47/bottom1/skill-num}{6}
\pgfkeyssetvalue{/hp/q4/q47/bottom2/label}{Code development}
\pgfkeyssetvalue{/hp/q4/q47/bottom2/pct}{54.2}
\pgfkeyssetvalue{/hp/q4/q47/bottom2/n}{13}
\pgfkeyssetvalue{/hp/q4/q47/bottom2/skill-num}{3}
\pgfkeyssetvalue{/hp/q4/q47/bottom3/label}{System-level design}
\pgfkeyssetvalue{/hp/q4/q47/bottom3/pct}{62.5}
\pgfkeyssetvalue{/hp/q4/q47/bottom3/n}{15}
\pgfkeyssetvalue{/hp/q4/q47/bottom3/skill-num}{7}
\pgfkeyssetvalue{/hp/q4/q48/n}{24}
\pgfkeyssetvalue{/hp/q4/q48/n-substantive}{24}
\pgfkeyssetvalue{/hp/q4/q48/n-na}{0}
\pgfkeyssetvalue{/hp/q4/q48/positive/n}{14}
\pgfkeyssetvalue{/hp/q4/q48/positive/pct}{58.3}
\pgfkeyssetvalue{/hp/q4/q48/positive/extreme/n}{12}
\pgfkeyssetvalue{/hp/q4/q48/positive/extreme/pct}{50.0}
\pgfkeyssetvalue{/hp/q4/q48/positive/base/n}{2}
\pgfkeyssetvalue{/hp/q4/q48/positive/base/pct}{8.3}
\pgfkeyssetvalue{/hp/q4/q48/neutral/n}{6}
\pgfkeyssetvalue{/hp/q4/q48/neutral/pct}{25.0}
\pgfkeyssetvalue{/hp/q4/q48/negative/n}{4}
\pgfkeyssetvalue{/hp/q4/q48/negative/pct}{16.7}
\pgfkeyssetvalue{/hp/q4/q48/negative/extreme/n}{0}
\pgfkeyssetvalue{/hp/q4/q48/negative/extreme/pct}{0.0}
\pgfkeyssetvalue{/hp/q4/q48/negative/base/n}{4}
\pgfkeyssetvalue{/hp/q4/q48/negative/base/pct}{16.7}
\pgfkeyssetvalue{/hp/q4/q49/n}{24}
\pgfkeyssetvalue{/hp/q4/q49/n-substantive}{22}
\pgfkeyssetvalue{/hp/q4/q49/n-na}{2}
\pgfkeyssetvalue{/hp/q4/q49/positive/n}{6}
\pgfkeyssetvalue{/hp/q4/q49/positive/pct}{27.3}
\pgfkeyssetvalue{/hp/q4/q49/positive/extreme/n}{2}
\pgfkeyssetvalue{/hp/q4/q49/positive/extreme/pct}{9.1}
\pgfkeyssetvalue{/hp/q4/q49/positive/base/n}{4}
\pgfkeyssetvalue{/hp/q4/q49/positive/base/pct}{18.2}
\pgfkeyssetvalue{/hp/q4/q49/neutral/n}{8}
\pgfkeyssetvalue{/hp/q4/q49/neutral/pct}{36.4}
\pgfkeyssetvalue{/hp/q4/q49/negative/n}{8}
\pgfkeyssetvalue{/hp/q4/q49/negative/pct}{36.4}
\pgfkeyssetvalue{/hp/q4/q49/negative/extreme/n}{4}
\pgfkeyssetvalue{/hp/q4/q49/negative/extreme/pct}{18.2}
\pgfkeyssetvalue{/hp/q4/q49/negative/base/n}{4}
\pgfkeyssetvalue{/hp/q4/q49/negative/base/pct}{18.2}
\pgfkeyssetvalue{/hp/q4/q410/n}{24}
\pgfkeyssetvalue{/hp/q4/q410/n-substantive}{24}
\pgfkeyssetvalue{/hp/q4/q410/n-na}{0}
\pgfkeyssetvalue{/hp/q4/q410/positive/n}{12}
\pgfkeyssetvalue{/hp/q4/q410/positive/pct}{50.0}
\pgfkeyssetvalue{/hp/q4/q410/positive/extreme/n}{3}
\pgfkeyssetvalue{/hp/q4/q410/positive/extreme/pct}{12.5}
\pgfkeyssetvalue{/hp/q4/q410/positive/base/n}{9}
\pgfkeyssetvalue{/hp/q4/q410/positive/base/pct}{37.5}
\pgfkeyssetvalue{/hp/q4/q410/neutral/n}{4}
\pgfkeyssetvalue{/hp/q4/q410/neutral/pct}{16.7}
\pgfkeyssetvalue{/hp/q4/q410/negative/n}{8}
\pgfkeyssetvalue{/hp/q4/q410/negative/pct}{33.3}
\pgfkeyssetvalue{/hp/q4/q410/negative/extreme/n}{5}
\pgfkeyssetvalue{/hp/q4/q410/negative/extreme/pct}{20.8}
\pgfkeyssetvalue{/hp/q4/q410/negative/base/n}{3}
\pgfkeyssetvalue{/hp/q4/q410/negative/base/pct}{12.5}

%% file: values/educator-q5-values.tex
\pgfkeyssetvalue{/ed/q5/n}{56}
\pgfkeyssetvalue{/ed/q5/q51/n}{56}
\pgfkeyssetvalue{/ed/q5/q51/n-substantive}{56}
\pgfkeyssetvalue{/ed/q5/q51/n-na}{0}
\pgfkeyssetvalue{/ed/q5/q51/positive/n}{53}
\pgfkeyssetvalue{/ed/q5/q51/positive/pct}{94.6}
\pgfkeyssetvalue{/ed/q5/q51/positive/extreme/n}{33}
\pgfkeyssetvalue{/ed/q5/q51/positive/extreme/pct}{58.9}
\pgfkeyssetvalue{/ed/q5/q51/positive/base/n}{20}
\pgfkeyssetvalue{/ed/q5/q51/positive/base/pct}{35.7}
\pgfkeyssetvalue{/ed/q5/q51/neutral/n}{2}
\pgfkeyssetvalue{/ed/q5/q51/neutral/pct}{3.6}
\pgfkeyssetvalue{/ed/q5/q51/negative/n}{1}
\pgfkeyssetvalue{/ed/q5/q51/negative/pct}{1.8}
\pgfkeyssetvalue{/ed/q5/q51/negative/extreme/n}{0}
\pgfkeyssetvalue{/ed/q5/q51/negative/extreme/pct}{0.0}
\pgfkeyssetvalue{/ed/q5/q51/negative/base/n}{1}
\pgfkeyssetvalue{/ed/q5/q51/negative/base/pct}{1.8}
\pgfkeyssetvalue{/ed/q5/q52/n}{56}
\pgfkeyssetvalue{/ed/q5/q52/n-substantive}{54}
\pgfkeyssetvalue{/ed/q5/q52/n-na}{2}
\pgfkeyssetvalue{/ed/q5/q52/positive/n}{41}
\pgfkeyssetvalue{/ed/q5/q52/positive/pct}{75.9}
\pgfkeyssetvalue{/ed/q5/q52/positive/extreme/n}{15}
\pgfkeyssetvalue{/ed/q5/q52/positive/extreme/pct}{27.8}
\pgfkeyssetvalue{/ed/q5/q52/positive/base/n}{26}
\pgfkeyssetvalue{/ed/q5/q52/positive/base/pct}{48.1}
\pgfkeyssetvalue{/ed/q5/q52/neutral/n}{8}
\pgfkeyssetvalue{/ed/q5/q52/neutral/pct}{14.8}
\pgfkeyssetvalue{/ed/q5/q52/negative/n}{5}
\pgfkeyssetvalue{/ed/q5/q52/negative/pct}{9.3}
\pgfkeyssetvalue{/ed/q5/q52/negative/extreme/n}{0}
\pgfkeyssetvalue{/ed/q5/q52/negative/extreme/pct}{0.0}
\pgfkeyssetvalue{/ed/q5/q52/negative/base/n}{5}
\pgfkeyssetvalue{/ed/q5/q52/negative/base/pct}{9.3}
\pgfkeyssetvalue{/ed/q5/q53/n}{56}
\pgfkeyssetvalue{/ed/q5/q53/n-substantive}{56}
\pgfkeyssetvalue{/ed/q5/q53/n-na}{0}
\pgfkeyssetvalue{/ed/q5/q53/positive/n}{2}
\pgfkeyssetvalue{/ed/q5/q53/positive/pct}{3.6}
\pgfkeyssetvalue{/ed/q5/q53/positive/extreme/n}{1}
\pgfkeyssetvalue{/ed/q5/q53/positive/extreme/pct}{1.8}
\pgfkeyssetvalue{/ed/q5/q53/positive/base/n}{1}
\pgfkeyssetvalue{/ed/q5/q53/positive/base/pct}{1.8}
\pgfkeyssetvalue{/ed/q5/q53/neutral/n}{19}
\pgfkeyssetvalue{/ed/q5/q53/neutral/pct}{33.9}
\pgfkeyssetvalue{/ed/q5/q53/negative/n}{35}
\pgfkeyssetvalue{/ed/q5/q53/negative/pct}{62.5}
\pgfkeyssetvalue{/ed/q5/q53/negative/extreme/n}{13}
\pgfkeyssetvalue{/ed/q5/q53/negative/extreme/pct}{23.2}
\pgfkeyssetvalue{/ed/q5/q53/negative/base/n}{22}
\pgfkeyssetvalue{/ed/q5/q53/negative/base/pct}{39.3}
\pgfkeyssetvalue{/ed/q5/q54/problem-decomposition/label}{Problem decomposition}
\pgfkeyssetvalue{/ed/q5/q54/problem-decomposition/n}{56}
\pgfkeyssetvalue{/ed/q5/q54/problem-decomposition/n-substantive}{56}
\pgfkeyssetvalue{/ed/q5/q54/problem-decomposition/n-na}{0}
\pgfkeyssetvalue{/ed/q5/q54/problem-decomposition/positive/n}{50}
\pgfkeyssetvalue{/ed/q5/q54/problem-decomposition/positive/pct}{89.3}
\pgfkeyssetvalue{/ed/q5/q54/problem-decomposition/positive/extreme/n}{34}
\pgfkeyssetvalue{/ed/q5/q54/problem-decomposition/positive/extreme/pct}{60.7}
\pgfkeyssetvalue{/ed/q5/q54/problem-decomposition/positive/base/n}{16}
\pgfkeyssetvalue{/ed/q5/q54/problem-decomposition/positive/base/pct}{28.6}
\pgfkeyssetvalue{/ed/q5/q54/problem-decomposition/neutral/n}{3}
\pgfkeyssetvalue{/ed/q5/q54/problem-decomposition/neutral/pct}{5.4}
\pgfkeyssetvalue{/ed/q5/q54/problem-decomposition/negative/n}{3}
\pgfkeyssetvalue{/ed/q5/q54/problem-decomposition/negative/pct}{5.4}
\pgfkeyssetvalue{/ed/q5/q54/problem-decomposition/negative/extreme/n}{0}
\pgfkeyssetvalue{/ed/q5/q54/problem-decomposition/negative/extreme/pct}{0.0}
\pgfkeyssetvalue{/ed/q5/q54/problem-decomposition/negative/base/n}{3}
\pgfkeyssetvalue{/ed/q5/q54/problem-decomposition/negative/base/pct}{5.4}
\pgfkeyssetvalue{/ed/q5/q54/code-comprehension/label}{Code comprehension}
\pgfkeyssetvalue{/ed/q5/q54/code-comprehension/n}{56}
\pgfkeyssetvalue{/ed/q5/q54/code-comprehension/n-substantive}{56}
\pgfkeyssetvalue{/ed/q5/q54/code-comprehension/n-na}{0}
\pgfkeyssetvalue{/ed/q5/q54/code-comprehension/positive/n}{47}
\pgfkeyssetvalue{/ed/q5/q54/code-comprehension/positive/pct}{83.9}
\pgfkeyssetvalue{/ed/q5/q54/code-comprehension/positive/extreme/n}{23}
\pgfkeyssetvalue{/ed/q5/q54/code-comprehension/positive/extreme/pct}{41.1}
\pgfkeyssetvalue{/ed/q5/q54/code-comprehension/positive/base/n}{24}
\pgfkeyssetvalue{/ed/q5/q54/code-comprehension/positive/base/pct}{42.9}
\pgfkeyssetvalue{/ed/q5/q54/code-comprehension/neutral/n}{7}
\pgfkeyssetvalue{/ed/q5/q54/code-comprehension/neutral/pct}{12.5}
\pgfkeyssetvalue{/ed/q5/q54/code-comprehension/negative/n}{2}
\pgfkeyssetvalue{/ed/q5/q54/code-comprehension/negative/pct}{3.6}
\pgfkeyssetvalue{/ed/q5/q54/code-comprehension/negative/extreme/n}{0}
\pgfkeyssetvalue{/ed/q5/q54/code-comprehension/negative/extreme/pct}{0.0}
\pgfkeyssetvalue{/ed/q5/q54/code-comprehension/negative/base/n}{2}
\pgfkeyssetvalue{/ed/q5/q54/code-comprehension/negative/base/pct}{3.6}
\pgfkeyssetvalue{/ed/q5/q54/code-development/label}{Code development}
\pgfkeyssetvalue{/ed/q5/q54/code-development/n}{56}
\pgfkeyssetvalue{/ed/q5/q54/code-development/n-substantive}{56}
\pgfkeyssetvalue{/ed/q5/q54/code-development/n-na}{0}
\pgfkeyssetvalue{/ed/q5/q54/code-development/positive/n}{22}
\pgfkeyssetvalue{/ed/q5/q54/code-development/positive/pct}{39.3}
\pgfkeyssetvalue{/ed/q5/q54/code-development/positive/extreme/n}{7}
\pgfkeyssetvalue{/ed/q5/q54/code-development/positive/extreme/pct}{12.5}
\pgfkeyssetvalue{/ed/q5/q54/code-development/positive/base/n}{15}
\pgfkeyssetvalue{/ed/q5/q54/code-development/positive/base/pct}{26.8}
\pgfkeyssetvalue{/ed/q5/q54/code-development/neutral/n}{20}
\pgfkeyssetvalue{/ed/q5/q54/code-development/neutral/pct}{35.7}
\pgfkeyssetvalue{/ed/q5/q54/code-development/negative/n}{14}
\pgfkeyssetvalue{/ed/q5/q54/code-development/negative/pct}{25.0}
\pgfkeyssetvalue{/ed/q5/q54/code-development/negative/extreme/n}{1}
\pgfkeyssetvalue{/ed/q5/q54/code-development/negative/extreme/pct}{1.8}
\pgfkeyssetvalue{/ed/q5/q54/code-development/negative/base/n}{13}
\pgfkeyssetvalue{/ed/q5/q54/code-development/negative/base/pct}{23.2}
\pgfkeyssetvalue{/ed/q5/q54/code-review/label}{Code review}
\pgfkeyssetvalue{/ed/q5/q54/code-review/n}{56}
\pgfkeyssetvalue{/ed/q5/q54/code-review/n-substantive}{56}
\pgfkeyssetvalue{/ed/q5/q54/code-review/n-na}{0}
\pgfkeyssetvalue{/ed/q5/q54/code-review/positive/n}{48}
\pgfkeyssetvalue{/ed/q5/q54/code-review/positive/pct}{85.7}
\pgfkeyssetvalue{/ed/q5/q54/code-review/positive/extreme/n}{28}
\pgfkeyssetvalue{/ed/q5/q54/code-review/positive/extreme/pct}{50.0}
\pgfkeyssetvalue{/ed/q5/q54/code-review/positive/base/n}{20}
\pgfkeyssetvalue{/ed/q5/q54/code-review/positive/base/pct}{35.7}
\pgfkeyssetvalue{/ed/q5/q54/code-review/neutral/n}{7}
\pgfkeyssetvalue{/ed/q5/q54/code-review/neutral/pct}{12.5}
\pgfkeyssetvalue{/ed/q5/q54/code-review/negative/n}{1}
\pgfkeyssetvalue{/ed/q5/q54/code-review/negative/pct}{1.8}
\pgfkeyssetvalue{/ed/q5/q54/code-review/negative/extreme/n}{0}
\pgfkeyssetvalue{/ed/q5/q54/code-review/negative/extreme/pct}{0.0}
\pgfkeyssetvalue{/ed/q5/q54/code-review/negative/base/n}{1}
\pgfkeyssetvalue{/ed/q5/q54/code-review/negative/base/pct}{1.8}
\pgfkeyssetvalue{/ed/q5/q54/debugging/label}{Debugging}
\pgfkeyssetvalue{/ed/q5/q54/debugging/n}{56}
\pgfkeyssetvalue{/ed/q5/q54/debugging/n-substantive}{56}
\pgfkeyssetvalue{/ed/q5/q54/debugging/n-na}{0}
\pgfkeyssetvalue{/ed/q5/q54/debugging/positive/n}{44}
\pgfkeyssetvalue{/ed/q5/q54/debugging/positive/pct}{78.6}
\pgfkeyssetvalue{/ed/q5/q54/debugging/positive/extreme/n}{24}
\pgfkeyssetvalue{/ed/q5/q54/debugging/positive/extreme/pct}{42.9}
\pgfkeyssetvalue{/ed/q5/q54/debugging/positive/base/n}{20}
\pgfkeyssetvalue{/ed/q5/q54/debugging/positive/base/pct}{35.7}
\pgfkeyssetvalue{/ed/q5/q54/debugging/neutral/n}{10}
\pgfkeyssetvalue{/ed/q5/q54/debugging/neutral/pct}{17.9}
\pgfkeyssetvalue{/ed/q5/q54/debugging/negative/n}{2}
\pgfkeyssetvalue{/ed/q5/q54/debugging/negative/pct}{3.6}
\pgfkeyssetvalue{/ed/q5/q54/debugging/negative/extreme/n}{0}
\pgfkeyssetvalue{/ed/q5/q54/debugging/negative/extreme/pct}{0.0}
\pgfkeyssetvalue{/ed/q5/q54/debugging/negative/base/n}{2}
\pgfkeyssetvalue{/ed/q5/q54/debugging/negative/base/pct}{3.6}
\pgfkeyssetvalue{/ed/q5/q54/documentation/label}{Documentation}
\pgfkeyssetvalue{/ed/q5/q54/documentation/n}{56}
\pgfkeyssetvalue{/ed/q5/q54/documentation/n-substantive}{56}
\pgfkeyssetvalue{/ed/q5/q54/documentation/n-na}{0}
\pgfkeyssetvalue{/ed/q5/q54/documentation/positive/n}{22}
\pgfkeyssetvalue{/ed/q5/q54/documentation/positive/pct}{39.3}
\pgfkeyssetvalue{/ed/q5/q54/documentation/positive/extreme/n}{7}
\pgfkeyssetvalue{/ed/q5/q54/documentation/positive/extreme/pct}{12.5}
\pgfkeyssetvalue{/ed/q5/q54/documentation/positive/base/n}{15}
\pgfkeyssetvalue{/ed/q5/q54/documentation/positive/base/pct}{26.8}
\pgfkeyssetvalue{/ed/q5/q54/documentation/neutral/n}{19}
\pgfkeyssetvalue{/ed/q5/q54/documentation/neutral/pct}{33.9}
\pgfkeyssetvalue{/ed/q5/q54/documentation/negative/n}{15}
\pgfkeyssetvalue{/ed/q5/q54/documentation/negative/pct}{26.8}
\pgfkeyssetvalue{/ed/q5/q54/documentation/negative/extreme/n}{2}
\pgfkeyssetvalue{/ed/q5/q54/documentation/negative/extreme/pct}{3.6}
\pgfkeyssetvalue{/ed/q5/q54/documentation/negative/base/n}{13}
\pgfkeyssetvalue{/ed/q5/q54/documentation/negative/base/pct}{23.2}
\pgfkeyssetvalue{/ed/q5/q54/system-design/label}{System-level design}
\pgfkeyssetvalue{/ed/q5/q54/system-design/n}{56}
\pgfkeyssetvalue{/ed/q5/q54/system-design/n-substantive}{55}
\pgfkeyssetvalue{/ed/q5/q54/system-design/n-na}{1}
\pgfkeyssetvalue{/ed/q5/q54/system-design/positive/n}{47}
\pgfkeyssetvalue{/ed/q5/q54/system-design/positive/pct}{85.5}
\pgfkeyssetvalue{/ed/q5/q54/system-design/positive/extreme/n}{29}
\pgfkeyssetvalue{/ed/q5/q54/system-design/positive/extreme/pct}{52.7}
\pgfkeyssetvalue{/ed/q5/q54/system-design/positive/base/n}{18}
\pgfkeyssetvalue{/ed/q5/q54/system-design/positive/base/pct}{32.7}
\pgfkeyssetvalue{/ed/q5/q54/system-design/neutral/n}{6}
\pgfkeyssetvalue{/ed/q5/q54/system-design/neutral/pct}{10.9}
\pgfkeyssetvalue{/ed/q5/q54/system-design/negative/n}{2}
\pgfkeyssetvalue{/ed/q5/q54/system-design/negative/pct}{3.6}
\pgfkeyssetvalue{/ed/q5/q54/system-design/negative/extreme/n}{0}
\pgfkeyssetvalue{/ed/q5/q54/system-design/negative/extreme/pct}{0.0}
\pgfkeyssetvalue{/ed/q5/q54/system-design/negative/base/n}{2}
\pgfkeyssetvalue{/ed/q5/q54/system-design/negative/base/pct}{3.6}
\pgfkeyssetvalue{/ed/q5/q54/algorithmic-thinking/label}{Algorithmic thinking}
\pgfkeyssetvalue{/ed/q5/q54/algorithmic-thinking/n}{56}
\pgfkeyssetvalue{/ed/q5/q54/algorithmic-thinking/n-substantive}{55}
\pgfkeyssetvalue{/ed/q5/q54/algorithmic-thinking/n-na}{1}
\pgfkeyssetvalue{/ed/q5/q54/algorithmic-thinking/positive/n}{44}
\pgfkeyssetvalue{/ed/q5/q54/algorithmic-thinking/positive/pct}{80.0}
\pgfkeyssetvalue{/ed/q5/q54/algorithmic-thinking/positive/extreme/n}{30}
\pgfkeyssetvalue{/ed/q5/q54/algorithmic-thinking/positive/extreme/pct}{54.5}
\pgfkeyssetvalue{/ed/q5/q54/algorithmic-thinking/positive/base/n}{14}
\pgfkeyssetvalue{/ed/q5/q54/algorithmic-thinking/positive/base/pct}{25.5}
\pgfkeyssetvalue{/ed/q5/q54/algorithmic-thinking/neutral/n}{8}
\pgfkeyssetvalue{/ed/q5/q54/algorithmic-thinking/neutral/pct}{14.5}
\pgfkeyssetvalue{/ed/q5/q54/algorithmic-thinking/negative/n}{3}
\pgfkeyssetvalue{/ed/q5/q54/algorithmic-thinking/negative/pct}{5.5}
\pgfkeyssetvalue{/ed/q5/q54/algorithmic-thinking/negative/extreme/n}{0}
\pgfkeyssetvalue{/ed/q5/q54/algorithmic-thinking/negative/extreme/pct}{0.0}
\pgfkeyssetvalue{/ed/q5/q54/algorithmic-thinking/negative/base/n}{3}
\pgfkeyssetvalue{/ed/q5/q54/algorithmic-thinking/negative/base/pct}{5.5}
\pgfkeyssetvalue{/ed/q5/q54/critical-eval-ai/label}{Critical evaluation of AI-generated output}
\pgfkeyssetvalue{/ed/q5/q54/critical-eval-ai/n}{56}
\pgfkeyssetvalue{/ed/q5/q54/critical-eval-ai/n-substantive}{56}
\pgfkeyssetvalue{/ed/q5/q54/critical-eval-ai/n-na}{0}
\pgfkeyssetvalue{/ed/q5/q54/critical-eval-ai/positive/n}{52}
\pgfkeyssetvalue{/ed/q5/q54/critical-eval-ai/positive/pct}{92.9}
\pgfkeyssetvalue{/ed/q5/q54/critical-eval-ai/positive/extreme/n}{42}
\pgfkeyssetvalue{/ed/q5/q54/critical-eval-ai/positive/extreme/pct}{75.0}
\pgfkeyssetvalue{/ed/q5/q54/critical-eval-ai/positive/base/n}{10}
\pgfkeyssetvalue{/ed/q5/q54/critical-eval-ai/positive/base/pct}{17.9}
\pgfkeyssetvalue{/ed/q5/q54/critical-eval-ai/neutral/n}{1}
\pgfkeyssetvalue{/ed/q5/q54/critical-eval-ai/neutral/pct}{1.8}
\pgfkeyssetvalue{/ed/q5/q54/critical-eval-ai/negative/n}{3}
\pgfkeyssetvalue{/ed/q5/q54/critical-eval-ai/negative/pct}{5.4}
\pgfkeyssetvalue{/ed/q5/q54/critical-eval-ai/negative/extreme/n}{1}
\pgfkeyssetvalue{/ed/q5/q54/critical-eval-ai/negative/extreme/pct}{1.8}
\pgfkeyssetvalue{/ed/q5/q54/critical-eval-ai/negative/base/n}{2}
\pgfkeyssetvalue{/ed/q5/q54/critical-eval-ai/negative/base/pct}{3.6}
\pgfkeyssetvalue{/ed/q5/q54/learn-adapt/label}{Ability to learn and adapt independently}
\pgfkeyssetvalue{/ed/q5/q54/learn-adapt/n}{56}
\pgfkeyssetvalue{/ed/q5/q54/learn-adapt/n-substantive}{55}
\pgfkeyssetvalue{/ed/q5/q54/learn-adapt/n-na}{1}
\pgfkeyssetvalue{/ed/q5/q54/learn-adapt/positive/n}{52}
\pgfkeyssetvalue{/ed/q5/q54/learn-adapt/positive/pct}{94.5}
\pgfkeyssetvalue{/ed/q5/q54/learn-adapt/positive/extreme/n}{40}
\pgfkeyssetvalue{/ed/q5/q54/learn-adapt/positive/extreme/pct}{72.7}
\pgfkeyssetvalue{/ed/q5/q54/learn-adapt/positive/base/n}{12}
\pgfkeyssetvalue{/ed/q5/q54/learn-adapt/positive/base/pct}{21.8}
\pgfkeyssetvalue{/ed/q5/q54/learn-adapt/neutral/n}{2}
\pgfkeyssetvalue{/ed/q5/q54/learn-adapt/neutral/pct}{3.6}
\pgfkeyssetvalue{/ed/q5/q54/learn-adapt/negative/n}{1}
\pgfkeyssetvalue{/ed/q5/q54/learn-adapt/negative/pct}{1.8}
\pgfkeyssetvalue{/ed/q5/q54/learn-adapt/negative/extreme/n}{0}
\pgfkeyssetvalue{/ed/q5/q54/learn-adapt/negative/extreme/pct}{0.0}
\pgfkeyssetvalue{/ed/q5/q54/learn-adapt/negative/base/n}{1}
\pgfkeyssetvalue{/ed/q5/q54/learn-adapt/negative/base/pct}{1.8}
\pgfkeyssetvalue{/ed/q5/q54/collaboration/label}{Collaboration and communication}
\pgfkeyssetvalue{/ed/q5/q54/collaboration/n}{56}
\pgfkeyssetvalue{/ed/q5/q54/collaboration/n-substantive}{55}
\pgfkeyssetvalue{/ed/q5/q54/collaboration/n-na}{1}
\pgfkeyssetvalue{/ed/q5/q54/collaboration/positive/n}{53}
\pgfkeyssetvalue{/ed/q5/q54/collaboration/positive/pct}{96.4}
\pgfkeyssetvalue{/ed/q5/q54/collaboration/positive/extreme/n}{33}
\pgfkeyssetvalue{/ed/q5/q54/collaboration/positive/extreme/pct}{60.0}
\pgfkeyssetvalue{/ed/q5/q54/collaboration/positive/base/n}{20}
\pgfkeyssetvalue{/ed/q5/q54/collaboration/positive/base/pct}{36.4}
\pgfkeyssetvalue{/ed/q5/q54/collaboration/neutral/n}{0}
\pgfkeyssetvalue{/ed/q5/q54/collaboration/neutral/pct}{0.0}
\pgfkeyssetvalue{/ed/q5/q54/collaboration/negative/n}{2}
\pgfkeyssetvalue{/ed/q5/q54/collaboration/negative/pct}{3.6}
\pgfkeyssetvalue{/ed/q5/q54/collaboration/negative/extreme/n}{0}
\pgfkeyssetvalue{/ed/q5/q54/collaboration/negative/extreme/pct}{0.0}
\pgfkeyssetvalue{/ed/q5/q54/collaboration/negative/base/n}{2}
\pgfkeyssetvalue{/ed/q5/q54/collaboration/negative/base/pct}{3.6}
\pgfkeyssetvalue{/ed/q5/q54/responsible-genai/label}{Responsible and effective use of GenAI tools}
\pgfkeyssetvalue{/ed/q5/q54/responsible-genai/n}{56}
\pgfkeyssetvalue{/ed/q5/q54/responsible-genai/n-substantive}{55}
\pgfkeyssetvalue{/ed/q5/q54/responsible-genai/n-na}{1}
\pgfkeyssetvalue{/ed/q5/q54/responsible-genai/positive/n}{48}
\pgfkeyssetvalue{/ed/q5/q54/responsible-genai/positive/pct}{87.3}
\pgfkeyssetvalue{/ed/q5/q54/responsible-genai/positive/extreme/n}{33}
\pgfkeyssetvalue{/ed/q5/q54/responsible-genai/positive/extreme/pct}{60.0}
\pgfkeyssetvalue{/ed/q5/q54/responsible-genai/positive/base/n}{15}
\pgfkeyssetvalue{/ed/q5/q54/responsible-genai/positive/base/pct}{27.3}
\pgfkeyssetvalue{/ed/q5/q54/responsible-genai/neutral/n}{2}
\pgfkeyssetvalue{/ed/q5/q54/responsible-genai/neutral/pct}{3.6}
\pgfkeyssetvalue{/ed/q5/q54/responsible-genai/negative/n}{5}
\pgfkeyssetvalue{/ed/q5/q54/responsible-genai/negative/pct}{9.1}
\pgfkeyssetvalue{/ed/q5/q54/responsible-genai/negative/extreme/n}{2}
\pgfkeyssetvalue{/ed/q5/q54/responsible-genai/negative/extreme/pct}{3.6}
\pgfkeyssetvalue{/ed/q5/q54/responsible-genai/negative/base/n}{3}
\pgfkeyssetvalue{/ed/q5/q54/responsible-genai/negative/base/pct}{5.5}
\pgfkeyssetvalue{/ed/q5/q54/top1/label}{Collaboration and communication}
\pgfkeyssetvalue{/ed/q5/q54/top1/pct}{96.4}
\pgfkeyssetvalue{/ed/q5/q54/top1/skill-num}{11}
\pgfkeyssetvalue{/ed/q5/q54/top2/label}{Ability to learn and adapt independently}
\pgfkeyssetvalue{/ed/q5/q54/top2/pct}{94.5}
\pgfkeyssetvalue{/ed/q5/q54/top2/skill-num}{10}
\pgfkeyssetvalue{/ed/q5/q54/top3/label}{Critical evaluation of AI-generated output}
\pgfkeyssetvalue{/ed/q5/q54/top3/pct}{92.9}
\pgfkeyssetvalue{/ed/q5/q54/top3/skill-num}{9}
\pgfkeyssetvalue{/ed/q5/q54/bottom1/label}{Documentation}
\pgfkeyssetvalue{/ed/q5/q54/bottom1/pct}{39.3}
\pgfkeyssetvalue{/ed/q5/q54/bottom1/skill-num}{6}
\pgfkeyssetvalue{/ed/q5/q54/bottom2/label}{Code development}
\pgfkeyssetvalue{/ed/q5/q54/bottom2/pct}{39.3}
\pgfkeyssetvalue{/ed/q5/q54/bottom2/skill-num}{3}
\pgfkeyssetvalue{/ed/q5/q54/bottom3/label}{Debugging}
\pgfkeyssetvalue{/ed/q5/q54/bottom3/pct}{78.6}
\pgfkeyssetvalue{/ed/q5/q54/bottom3/skill-num}{5}
\pgfkeyssetvalue{/ed/q5/q55/problem-decomposition/label}{Problem decomposition}
\pgfkeyssetvalue{/ed/q5/q55/problem-decomposition/n}{56}
\pgfkeyssetvalue{/ed/q5/q55/problem-decomposition/n-substantive}{55}
\pgfkeyssetvalue{/ed/q5/q55/problem-decomposition/n-na}{1}
\pgfkeyssetvalue{/ed/q5/q55/problem-decomposition/positive/n}{25}
\pgfkeyssetvalue{/ed/q5/q55/problem-decomposition/positive/pct}{45.5}
\pgfkeyssetvalue{/ed/q5/q55/problem-decomposition/positive/extreme/n}{7}
\pgfkeyssetvalue{/ed/q5/q55/problem-decomposition/positive/extreme/pct}{12.7}
\pgfkeyssetvalue{/ed/q5/q55/problem-decomposition/positive/base/n}{18}
\pgfkeyssetvalue{/ed/q5/q55/problem-decomposition/positive/base/pct}{32.7}
\pgfkeyssetvalue{/ed/q5/q55/problem-decomposition/neutral/n}{25}
\pgfkeyssetvalue{/ed/q5/q55/problem-decomposition/neutral/pct}{45.5}
\pgfkeyssetvalue{/ed/q5/q55/problem-decomposition/negative/n}{5}
\pgfkeyssetvalue{/ed/q5/q55/problem-decomposition/negative/pct}{9.1}
\pgfkeyssetvalue{/ed/q5/q55/problem-decomposition/negative/extreme/n}{0}
\pgfkeyssetvalue{/ed/q5/q55/problem-decomposition/negative/extreme/pct}{0.0}
\pgfkeyssetvalue{/ed/q5/q55/problem-decomposition/negative/base/n}{5}
\pgfkeyssetvalue{/ed/q5/q55/problem-decomposition/negative/base/pct}{9.1}
\pgfkeyssetvalue{/ed/q5/q55/problem-decomposition/weakness-pct}{9.1}
\pgfkeyssetvalue{/ed/q5/q55/problem-decomposition/weakness-n}{5}
\pgfkeyssetvalue{/ed/q5/q55/code-comprehension/label}{Code comprehension}
\pgfkeyssetvalue{/ed/q5/q55/code-comprehension/n}{56}
\pgfkeyssetvalue{/ed/q5/q55/code-comprehension/n-substantive}{54}
\pgfkeyssetvalue{/ed/q5/q55/code-comprehension/n-na}{2}
\pgfkeyssetvalue{/ed/q5/q55/code-comprehension/positive/n}{29}
\pgfkeyssetvalue{/ed/q5/q55/code-comprehension/positive/pct}{53.7}
\pgfkeyssetvalue{/ed/q5/q55/code-comprehension/positive/extreme/n}{2}
\pgfkeyssetvalue{/ed/q5/q55/code-comprehension/positive/extreme/pct}{3.7}
\pgfkeyssetvalue{/ed/q5/q55/code-comprehension/positive/base/n}{27}
\pgfkeyssetvalue{/ed/q5/q55/code-comprehension/positive/base/pct}{50.0}
\pgfkeyssetvalue{/ed/q5/q55/code-comprehension/neutral/n}{16}
\pgfkeyssetvalue{/ed/q5/q55/code-comprehension/neutral/pct}{29.6}
\pgfkeyssetvalue{/ed/q5/q55/code-comprehension/negative/n}{9}
\pgfkeyssetvalue{/ed/q5/q55/code-comprehension/negative/pct}{16.7}
\pgfkeyssetvalue{/ed/q5/q55/code-comprehension/negative/extreme/n}{1}
\pgfkeyssetvalue{/ed/q5/q55/code-comprehension/negative/extreme/pct}{1.9}
\pgfkeyssetvalue{/ed/q5/q55/code-comprehension/negative/base/n}{8}
\pgfkeyssetvalue{/ed/q5/q55/code-comprehension/negative/base/pct}{14.8}
\pgfkeyssetvalue{/ed/q5/q55/code-comprehension/weakness-pct}{16.7}
\pgfkeyssetvalue{/ed/q5/q55/code-comprehension/weakness-n}{9}
\pgfkeyssetvalue{/ed/q5/q55/code-development/label}{Code development}
\pgfkeyssetvalue{/ed/q5/q55/code-development/n}{56}
\pgfkeyssetvalue{/ed/q5/q55/code-development/n-substantive}{54}
\pgfkeyssetvalue{/ed/q5/q55/code-development/n-na}{2}
\pgfkeyssetvalue{/ed/q5/q55/code-development/positive/n}{31}
\pgfkeyssetvalue{/ed/q5/q55/code-development/positive/pct}{57.4}
\pgfkeyssetvalue{/ed/q5/q55/code-development/positive/extreme/n}{7}
\pgfkeyssetvalue{/ed/q5/q55/code-development/positive/extreme/pct}{13.0}
\pgfkeyssetvalue{/ed/q5/q55/code-development/positive/base/n}{24}
\pgfkeyssetvalue{/ed/q5/q55/code-development/positive/base/pct}{44.4}
\pgfkeyssetvalue{/ed/q5/q55/code-development/neutral/n}{15}
\pgfkeyssetvalue{/ed/q5/q55/code-development/neutral/pct}{27.8}
\pgfkeyssetvalue{/ed/q5/q55/code-development/negative/n}{8}
\pgfkeyssetvalue{/ed/q5/q55/code-development/negative/pct}{14.8}
\pgfkeyssetvalue{/ed/q5/q55/code-development/negative/extreme/n}{2}
\pgfkeyssetvalue{/ed/q5/q55/code-development/negative/extreme/pct}{3.7}
\pgfkeyssetvalue{/ed/q5/q55/code-development/negative/base/n}{6}
\pgfkeyssetvalue{/ed/q5/q55/code-development/negative/base/pct}{11.1}
\pgfkeyssetvalue{/ed/q5/q55/code-development/weakness-pct}{14.8}
\pgfkeyssetvalue{/ed/q5/q55/code-development/weakness-n}{8}
\pgfkeyssetvalue{/ed/q5/q55/code-review/label}{Code review}
\pgfkeyssetvalue{/ed/q5/q55/code-review/n}{56}
\pgfkeyssetvalue{/ed/q5/q55/code-review/n-substantive}{53}
\pgfkeyssetvalue{/ed/q5/q55/code-review/n-na}{3}
\pgfkeyssetvalue{/ed/q5/q55/code-review/positive/n}{20}
\pgfkeyssetvalue{/ed/q5/q55/code-review/positive/pct}{37.7}
\pgfkeyssetvalue{/ed/q5/q55/code-review/positive/extreme/n}{3}
\pgfkeyssetvalue{/ed/q5/q55/code-review/positive/extreme/pct}{5.7}
\pgfkeyssetvalue{/ed/q5/q55/code-review/positive/base/n}{17}
\pgfkeyssetvalue{/ed/q5/q55/code-review/positive/base/pct}{32.1}
\pgfkeyssetvalue{/ed/q5/q55/code-review/neutral/n}{21}
\pgfkeyssetvalue{/ed/q5/q55/code-review/neutral/pct}{39.6}
\pgfkeyssetvalue{/ed/q5/q55/code-review/negative/n}{12}
\pgfkeyssetvalue{/ed/q5/q55/code-review/negative/pct}{22.6}
\pgfkeyssetvalue{/ed/q5/q55/code-review/negative/extreme/n}{2}
\pgfkeyssetvalue{/ed/q5/q55/code-review/negative/extreme/pct}{3.8}
\pgfkeyssetvalue{/ed/q5/q55/code-review/negative/base/n}{10}
\pgfkeyssetvalue{/ed/q5/q55/code-review/negative/base/pct}{18.9}
\pgfkeyssetvalue{/ed/q5/q55/code-review/weakness-pct}{22.6}
\pgfkeyssetvalue{/ed/q5/q55/code-review/weakness-n}{12}
\pgfkeyssetvalue{/ed/q5/q55/debugging/label}{Debugging}
\pgfkeyssetvalue{/ed/q5/q55/debugging/n}{56}
\pgfkeyssetvalue{/ed/q5/q55/debugging/n-substantive}{54}
\pgfkeyssetvalue{/ed/q5/q55/debugging/n-na}{2}
\pgfkeyssetvalue{/ed/q5/q55/debugging/positive/n}{21}
\pgfkeyssetvalue{/ed/q5/q55/debugging/positive/pct}{38.9}
\pgfkeyssetvalue{/ed/q5/q55/debugging/positive/extreme/n}{3}
\pgfkeyssetvalue{/ed/q5/q55/debugging/positive/extreme/pct}{5.6}
\pgfkeyssetvalue{/ed/q5/q55/debugging/positive/base/n}{18}
\pgfkeyssetvalue{/ed/q5/q55/debugging/positive/base/pct}{33.3}
\pgfkeyssetvalue{/ed/q5/q55/debugging/neutral/n}{22}
\pgfkeyssetvalue{/ed/q5/q55/debugging/neutral/pct}{40.7}
\pgfkeyssetvalue{/ed/q5/q55/debugging/negative/n}{11}
\pgfkeyssetvalue{/ed/q5/q55/debugging/negative/pct}{20.4}
\pgfkeyssetvalue{/ed/q5/q55/debugging/negative/extreme/n}{2}
\pgfkeyssetvalue{/ed/q5/q55/debugging/negative/extreme/pct}{3.7}
\pgfkeyssetvalue{/ed/q5/q55/debugging/negative/base/n}{9}
\pgfkeyssetvalue{/ed/q5/q55/debugging/negative/base/pct}{16.7}
\pgfkeyssetvalue{/ed/q5/q55/debugging/weakness-pct}{20.4}
\pgfkeyssetvalue{/ed/q5/q55/debugging/weakness-n}{11}
\pgfkeyssetvalue{/ed/q5/q55/documentation/label}{Documentation}
\pgfkeyssetvalue{/ed/q5/q55/documentation/n}{56}
\pgfkeyssetvalue{/ed/q5/q55/documentation/n-substantive}{50}
\pgfkeyssetvalue{/ed/q5/q55/documentation/n-na}{6}
\pgfkeyssetvalue{/ed/q5/q55/documentation/positive/n}{8}
\pgfkeyssetvalue{/ed/q5/q55/documentation/positive/pct}{16.0}
\pgfkeyssetvalue{/ed/q5/q55/documentation/positive/extreme/n}{4}
\pgfkeyssetvalue{/ed/q5/q55/documentation/positive/extreme/pct}{8.0}
\pgfkeyssetvalue{/ed/q5/q55/documentation/positive/base/n}{4}
\pgfkeyssetvalue{/ed/q5/q55/documentation/positive/base/pct}{8.0}
\pgfkeyssetvalue{/ed/q5/q55/documentation/neutral/n}{23}
\pgfkeyssetvalue{/ed/q5/q55/documentation/neutral/pct}{46.0}
\pgfkeyssetvalue{/ed/q5/q55/documentation/negative/n}{19}
\pgfkeyssetvalue{/ed/q5/q55/documentation/negative/pct}{38.0}
\pgfkeyssetvalue{/ed/q5/q55/documentation/negative/extreme/n}{2}
\pgfkeyssetvalue{/ed/q5/q55/documentation/negative/extreme/pct}{4.0}
\pgfkeyssetvalue{/ed/q5/q55/documentation/negative/base/n}{17}
\pgfkeyssetvalue{/ed/q5/q55/documentation/negative/base/pct}{34.0}
\pgfkeyssetvalue{/ed/q5/q55/documentation/weakness-pct}{38.0}
\pgfkeyssetvalue{/ed/q5/q55/documentation/weakness-n}{19}
\pgfkeyssetvalue{/ed/q5/q55/system-design/label}{System-level design}
\pgfkeyssetvalue{/ed/q5/q55/system-design/n}{56}
\pgfkeyssetvalue{/ed/q5/q55/system-design/n-substantive}{53}
\pgfkeyssetvalue{/ed/q5/q55/system-design/n-na}{3}
\pgfkeyssetvalue{/ed/q5/q55/system-design/positive/n}{15}
\pgfkeyssetvalue{/ed/q5/q55/system-design/positive/pct}{28.3}
\pgfkeyssetvalue{/ed/q5/q55/system-design/positive/extreme/n}{2}
\pgfkeyssetvalue{/ed/q5/q55/system-design/positive/extreme/pct}{3.8}
\pgfkeyssetvalue{/ed/q5/q55/system-design/positive/base/n}{13}
\pgfkeyssetvalue{/ed/q5/q55/system-design/positive/base/pct}{24.5}
\pgfkeyssetvalue{/ed/q5/q55/system-design/neutral/n}{24}
\pgfkeyssetvalue{/ed/q5/q55/system-design/neutral/pct}{45.3}
\pgfkeyssetvalue{/ed/q5/q55/system-design/negative/n}{14}
\pgfkeyssetvalue{/ed/q5/q55/system-design/negative/pct}{26.4}
\pgfkeyssetvalue{/ed/q5/q55/system-design/negative/extreme/n}{1}
\pgfkeyssetvalue{/ed/q5/q55/system-design/negative/extreme/pct}{1.9}
\pgfkeyssetvalue{/ed/q5/q55/system-design/negative/base/n}{13}
\pgfkeyssetvalue{/ed/q5/q55/system-design/negative/base/pct}{24.5}
\pgfkeyssetvalue{/ed/q5/q55/system-design/weakness-pct}{26.4}
\pgfkeyssetvalue{/ed/q5/q55/system-design/weakness-n}{14}
\pgfkeyssetvalue{/ed/q5/q55/algorithmic-thinking/label}{Algorithmic thinking}
\pgfkeyssetvalue{/ed/q5/q55/algorithmic-thinking/n}{56}
\pgfkeyssetvalue{/ed/q5/q55/algorithmic-thinking/n-substantive}{52}
\pgfkeyssetvalue{/ed/q5/q55/algorithmic-thinking/n-na}{4}
\pgfkeyssetvalue{/ed/q5/q55/algorithmic-thinking/positive/n}{21}
\pgfkeyssetvalue{/ed/q5/q55/algorithmic-thinking/positive/pct}{40.4}
\pgfkeyssetvalue{/ed/q5/q55/algorithmic-thinking/positive/extreme/n}{7}
\pgfkeyssetvalue{/ed/q5/q55/algorithmic-thinking/positive/extreme/pct}{13.5}
\pgfkeyssetvalue{/ed/q5/q55/algorithmic-thinking/positive/base/n}{14}
\pgfkeyssetvalue{/ed/q5/q55/algorithmic-thinking/positive/base/pct}{26.9}
\pgfkeyssetvalue{/ed/q5/q55/algorithmic-thinking/neutral/n}{23}
\pgfkeyssetvalue{/ed/q5/q55/algorithmic-thinking/neutral/pct}{44.2}
\pgfkeyssetvalue{/ed/q5/q55/algorithmic-thinking/negative/n}{8}
\pgfkeyssetvalue{/ed/q5/q55/algorithmic-thinking/negative/pct}{15.4}
\pgfkeyssetvalue{/ed/q5/q55/algorithmic-thinking/negative/extreme/n}{0}
\pgfkeyssetvalue{/ed/q5/q55/algorithmic-thinking/negative/extreme/pct}{0.0}
\pgfkeyssetvalue{/ed/q5/q55/algorithmic-thinking/negative/base/n}{8}
\pgfkeyssetvalue{/ed/q5/q55/algorithmic-thinking/negative/base/pct}{15.4}
\pgfkeyssetvalue{/ed/q5/q55/algorithmic-thinking/weakness-pct}{15.4}
\pgfkeyssetvalue{/ed/q5/q55/algorithmic-thinking/weakness-n}{8}
\pgfkeyssetvalue{/ed/q5/q55/critical-eval-ai/label}{Critical evaluation of AI-generated output}
\pgfkeyssetvalue{/ed/q5/q55/critical-eval-ai/n}{56}
\pgfkeyssetvalue{/ed/q5/q55/critical-eval-ai/n-substantive}{52}
\pgfkeyssetvalue{/ed/q5/q55/critical-eval-ai/n-na}{4}
\pgfkeyssetvalue{/ed/q5/q55/critical-eval-ai/positive/n}{12}
\pgfkeyssetvalue{/ed/q5/q55/critical-eval-ai/positive/pct}{23.1}
\pgfkeyssetvalue{/ed/q5/q55/critical-eval-ai/positive/extreme/n}{2}
\pgfkeyssetvalue{/ed/q5/q55/critical-eval-ai/positive/extreme/pct}{3.8}
\pgfkeyssetvalue{/ed/q5/q55/critical-eval-ai/positive/base/n}{10}
\pgfkeyssetvalue{/ed/q5/q55/critical-eval-ai/positive/base/pct}{19.2}
\pgfkeyssetvalue{/ed/q5/q55/critical-eval-ai/neutral/n}{20}
\pgfkeyssetvalue{/ed/q5/q55/critical-eval-ai/neutral/pct}{38.5}
\pgfkeyssetvalue{/ed/q5/q55/critical-eval-ai/negative/n}{20}
\pgfkeyssetvalue{/ed/q5/q55/critical-eval-ai/negative/pct}{38.5}
\pgfkeyssetvalue{/ed/q5/q55/critical-eval-ai/negative/extreme/n}{4}
\pgfkeyssetvalue{/ed/q5/q55/critical-eval-ai/negative/extreme/pct}{7.7}
\pgfkeyssetvalue{/ed/q5/q55/critical-eval-ai/negative/base/n}{16}
\pgfkeyssetvalue{/ed/q5/q55/critical-eval-ai/negative/base/pct}{30.8}
\pgfkeyssetvalue{/ed/q5/q55/critical-eval-ai/weakness-pct}{38.5}
\pgfkeyssetvalue{/ed/q5/q55/critical-eval-ai/weakness-n}{20}
\pgfkeyssetvalue{/ed/q5/q55/learn-adapt/label}{Ability to learn and adapt independently}
\pgfkeyssetvalue{/ed/q5/q55/learn-adapt/n}{56}
\pgfkeyssetvalue{/ed/q5/q55/learn-adapt/n-substantive}{54}
\pgfkeyssetvalue{/ed/q5/q55/learn-adapt/n-na}{2}
\pgfkeyssetvalue{/ed/q5/q55/learn-adapt/positive/n}{23}
\pgfkeyssetvalue{/ed/q5/q55/learn-adapt/positive/pct}{42.6}
\pgfkeyssetvalue{/ed/q5/q55/learn-adapt/positive/extreme/n}{7}
\pgfkeyssetvalue{/ed/q5/q55/learn-adapt/positive/extreme/pct}{13.0}
\pgfkeyssetvalue{/ed/q5/q55/learn-adapt/positive/base/n}{16}
\pgfkeyssetvalue{/ed/q5/q55/learn-adapt/positive/base/pct}{29.6}
\pgfkeyssetvalue{/ed/q5/q55/learn-adapt/neutral/n}{19}
\pgfkeyssetvalue{/ed/q5/q55/learn-adapt/neutral/pct}{35.2}
\pgfkeyssetvalue{/ed/q5/q55/learn-adapt/negative/n}{12}
\pgfkeyssetvalue{/ed/q5/q55/learn-adapt/negative/pct}{22.2}
\pgfkeyssetvalue{/ed/q5/q55/learn-adapt/negative/extreme/n}{2}
\pgfkeyssetvalue{/ed/q5/q55/learn-adapt/negative/extreme/pct}{3.7}
\pgfkeyssetvalue{/ed/q5/q55/learn-adapt/negative/base/n}{10}
\pgfkeyssetvalue{/ed/q5/q55/learn-adapt/negative/base/pct}{18.5}
\pgfkeyssetvalue{/ed/q5/q55/learn-adapt/weakness-pct}{22.2}
\pgfkeyssetvalue{/ed/q5/q55/learn-adapt/weakness-n}{12}
\pgfkeyssetvalue{/ed/q5/q55/collaboration/label}{Collaboration and communication}
\pgfkeyssetvalue{/ed/q5/q55/collaboration/n}{56}
\pgfkeyssetvalue{/ed/q5/q55/collaboration/n-substantive}{55}
\pgfkeyssetvalue{/ed/q5/q55/collaboration/n-na}{1}
\pgfkeyssetvalue{/ed/q5/q55/collaboration/positive/n}{23}
\pgfkeyssetvalue{/ed/q5/q55/collaboration/positive/pct}{41.8}
\pgfkeyssetvalue{/ed/q5/q55/collaboration/positive/extreme/n}{5}
\pgfkeyssetvalue{/ed/q5/q55/collaboration/positive/extreme/pct}{9.1}
\pgfkeyssetvalue{/ed/q5/q55/collaboration/positive/base/n}{18}
\pgfkeyssetvalue{/ed/q5/q55/collaboration/positive/base/pct}{32.7}
\pgfkeyssetvalue{/ed/q5/q55/collaboration/neutral/n}{21}
\pgfkeyssetvalue{/ed/q5/q55/collaboration/neutral/pct}{38.2}
\pgfkeyssetvalue{/ed/q5/q55/collaboration/negative/n}{11}
\pgfkeyssetvalue{/ed/q5/q55/collaboration/negative/pct}{20.0}
\pgfkeyssetvalue{/ed/q5/q55/collaboration/negative/extreme/n}{1}
\pgfkeyssetvalue{/ed/q5/q55/collaboration/negative/extreme/pct}{1.8}
\pgfkeyssetvalue{/ed/q5/q55/collaboration/negative/base/n}{10}
\pgfkeyssetvalue{/ed/q5/q55/collaboration/negative/base/pct}{18.2}
\pgfkeyssetvalue{/ed/q5/q55/collaboration/weakness-pct}{20.0}
\pgfkeyssetvalue{/ed/q5/q55/collaboration/weakness-n}{11}
\pgfkeyssetvalue{/ed/q5/q55/responsible-genai/label}{Responsible and effective use of GenAI tools}
\pgfkeyssetvalue{/ed/q5/q55/responsible-genai/n}{56}
\pgfkeyssetvalue{/ed/q5/q55/responsible-genai/n-substantive}{51}
\pgfkeyssetvalue{/ed/q5/q55/responsible-genai/n-na}{5}
\pgfkeyssetvalue{/ed/q5/q55/responsible-genai/positive/n}{11}
\pgfkeyssetvalue{/ed/q5/q55/responsible-genai/positive/pct}{21.6}
\pgfkeyssetvalue{/ed/q5/q55/responsible-genai/positive/extreme/n}{2}
\pgfkeyssetvalue{/ed/q5/q55/responsible-genai/positive/extreme/pct}{3.9}
\pgfkeyssetvalue{/ed/q5/q55/responsible-genai/positive/base/n}{9}
\pgfkeyssetvalue{/ed/q5/q55/responsible-genai/positive/base/pct}{17.6}
\pgfkeyssetvalue{/ed/q5/q55/responsible-genai/neutral/n}{18}
\pgfkeyssetvalue{/ed/q5/q55/responsible-genai/neutral/pct}{35.3}
\pgfkeyssetvalue{/ed/q5/q55/responsible-genai/negative/n}{22}
\pgfkeyssetvalue{/ed/q5/q55/responsible-genai/negative/pct}{43.1}
\pgfkeyssetvalue{/ed/q5/q55/responsible-genai/negative/extreme/n}{8}
\pgfkeyssetvalue{/ed/q5/q55/responsible-genai/negative/extreme/pct}{15.7}
\pgfkeyssetvalue{/ed/q5/q55/responsible-genai/negative/base/n}{14}
\pgfkeyssetvalue{/ed/q5/q55/responsible-genai/negative/base/pct}{27.5}
\pgfkeyssetvalue{/ed/q5/q55/responsible-genai/weakness-pct}{43.1}
\pgfkeyssetvalue{/ed/q5/q55/responsible-genai/weakness-n}{22}
\pgfkeyssetvalue{/ed/q5/q55/weakest1/label}{Responsible and effective use of GenAI tools}
\pgfkeyssetvalue{/ed/q5/q55/weakest1/pct}{43.1}
\pgfkeyssetvalue{/ed/q5/q55/weakest1/skill-num}{12}
\pgfkeyssetvalue{/ed/q5/q55/weakest2/label}{Critical evaluation of AI-generated output}
\pgfkeyssetvalue{/ed/q5/q55/weakest2/pct}{38.5}
\pgfkeyssetvalue{/ed/q5/q55/weakest2/skill-num}{9}
\pgfkeyssetvalue{/ed/q5/q55/weakest3/label}{Documentation}
\pgfkeyssetvalue{/ed/q5/q55/weakest3/pct}{38.0}
\pgfkeyssetvalue{/ed/q5/q55/weakest3/skill-num}{6}
\pgfkeyssetvalue{/ed/q5/q56/s1/label}{I am concerned that students will become overly reliant on GenAI to solve their assignments}
\pgfkeyssetvalue{/ed/q5/q56/s1/framing}{concern}
\pgfkeyssetvalue{/ed/q5/q56/s1/n}{56}
\pgfkeyssetvalue{/ed/q5/q56/s1/n-substantive}{56}
\pgfkeyssetvalue{/ed/q5/q56/s1/n-na}{0}
\pgfkeyssetvalue{/ed/q5/q56/s1/positive/n}{47}
\pgfkeyssetvalue{/ed/q5/q56/s1/positive/pct}{83.9}
\pgfkeyssetvalue{/ed/q5/q56/s1/positive/extreme/n}{24}
\pgfkeyssetvalue{/ed/q5/q56/s1/positive/extreme/pct}{42.9}
\pgfkeyssetvalue{/ed/q5/q56/s1/positive/base/n}{23}
\pgfkeyssetvalue{/ed/q5/q56/s1/positive/base/pct}{41.1}
\pgfkeyssetvalue{/ed/q5/q56/s1/neutral/n}{8}
\pgfkeyssetvalue{/ed/q5/q56/s1/neutral/pct}{14.3}
\pgfkeyssetvalue{/ed/q5/q56/s1/negative/n}{1}
\pgfkeyssetvalue{/ed/q5/q56/s1/negative/pct}{1.8}
\pgfkeyssetvalue{/ed/q5/q56/s1/negative/extreme/n}{0}
\pgfkeyssetvalue{/ed/q5/q56/s1/negative/extreme/pct}{0.0}
\pgfkeyssetvalue{/ed/q5/q56/s1/negative/base/n}{1}
\pgfkeyssetvalue{/ed/q5/q56/s1/negative/base/pct}{1.8}
\pgfkeyssetvalue{/ed/q5/q56/s2/label}{I believe GenAI can help students explore topics more deeply and independently}
\pgfkeyssetvalue{/ed/q5/q56/s2/framing}{benefit}
\pgfkeyssetvalue{/ed/q5/q56/s2/n}{56}
\pgfkeyssetvalue{/ed/q5/q56/s2/n-substantive}{56}
\pgfkeyssetvalue{/ed/q5/q56/s2/n-na}{0}
\pgfkeyssetvalue{/ed/q5/q56/s2/positive/n}{38}
\pgfkeyssetvalue{/ed/q5/q56/s2/positive/pct}{67.9}
\pgfkeyssetvalue{/ed/q5/q56/s2/positive/extreme/n}{8}
\pgfkeyssetvalue{/ed/q5/q56/s2/positive/extreme/pct}{14.3}
\pgfkeyssetvalue{/ed/q5/q56/s2/positive/base/n}{30}
\pgfkeyssetvalue{/ed/q5/q56/s2/positive/base/pct}{53.6}
\pgfkeyssetvalue{/ed/q5/q56/s2/neutral/n}{14}
\pgfkeyssetvalue{/ed/q5/q56/s2/neutral/pct}{25.0}
\pgfkeyssetvalue{/ed/q5/q56/s2/negative/n}{4}
\pgfkeyssetvalue{/ed/q5/q56/s2/negative/pct}{7.1}
\pgfkeyssetvalue{/ed/q5/q56/s2/negative/extreme/n}{1}
\pgfkeyssetvalue{/ed/q5/q56/s2/negative/extreme/pct}{1.8}
\pgfkeyssetvalue{/ed/q5/q56/s2/negative/base/n}{3}
\pgfkeyssetvalue{/ed/q5/q56/s2/negative/base/pct}{5.4}
\pgfkeyssetvalue{/ed/q5/q56/s3/label}{I am concerned that students submit GenAI-generated work without understanding it}
\pgfkeyssetvalue{/ed/q5/q56/s3/framing}{concern}
\pgfkeyssetvalue{/ed/q5/q56/s3/n}{56}
\pgfkeyssetvalue{/ed/q5/q56/s3/n-substantive}{56}
\pgfkeyssetvalue{/ed/q5/q56/s3/n-na}{0}
\pgfkeyssetvalue{/ed/q5/q56/s3/positive/n}{50}
\pgfkeyssetvalue{/ed/q5/q56/s3/positive/pct}{89.3}
\pgfkeyssetvalue{/ed/q5/q56/s3/positive/extreme/n}{36}
\pgfkeyssetvalue{/ed/q5/q56/s3/positive/extreme/pct}{64.3}
\pgfkeyssetvalue{/ed/q5/q56/s3/positive/base/n}{14}
\pgfkeyssetvalue{/ed/q5/q56/s3/positive/base/pct}{25.0}
\pgfkeyssetvalue{/ed/q5/q56/s3/neutral/n}{6}
\pgfkeyssetvalue{/ed/q5/q56/s3/neutral/pct}{10.7}
\pgfkeyssetvalue{/ed/q5/q56/s3/negative/n}{0}
\pgfkeyssetvalue{/ed/q5/q56/s3/negative/pct}{0.0}
\pgfkeyssetvalue{/ed/q5/q56/s3/negative/extreme/n}{0}
\pgfkeyssetvalue{/ed/q5/q56/s3/negative/extreme/pct}{0.0}
\pgfkeyssetvalue{/ed/q5/q56/s3/negative/base/n}{0}
\pgfkeyssetvalue{/ed/q5/q56/s3/negative/base/pct}{0.0}
\pgfkeyssetvalue{/ed/q5/q56/s4/label}{I believe GenAI can help students develop critical thinking skills when used appropriately}
\pgfkeyssetvalue{/ed/q5/q56/s4/framing}{benefit}
\pgfkeyssetvalue{/ed/q5/q56/s4/n}{56}
\pgfkeyssetvalue{/ed/q5/q56/s4/n-substantive}{56}
\pgfkeyssetvalue{/ed/q5/q56/s4/n-na}{0}
\pgfkeyssetvalue{/ed/q5/q56/s4/positive/n}{28}
\pgfkeyssetvalue{/ed/q5/q56/s4/positive/pct}{50.0}
\pgfkeyssetvalue{/ed/q5/q56/s4/positive/extreme/n}{6}
\pgfkeyssetvalue{/ed/q5/q56/s4/positive/extreme/pct}{10.7}
\pgfkeyssetvalue{/ed/q5/q56/s4/positive/base/n}{22}
\pgfkeyssetvalue{/ed/q5/q56/s4/positive/base/pct}{39.3}
\pgfkeyssetvalue{/ed/q5/q56/s4/neutral/n}{17}
\pgfkeyssetvalue{/ed/q5/q56/s4/neutral/pct}{30.4}
\pgfkeyssetvalue{/ed/q5/q56/s4/negative/n}{11}
\pgfkeyssetvalue{/ed/q5/q56/s4/negative/pct}{19.6}
\pgfkeyssetvalue{/ed/q5/q56/s4/negative/extreme/n}{3}
\pgfkeyssetvalue{/ed/q5/q56/s4/negative/extreme/pct}{5.4}
\pgfkeyssetvalue{/ed/q5/q56/s4/negative/base/n}{8}
\pgfkeyssetvalue{/ed/q5/q56/s4/negative/base/pct}{14.3}
\pgfkeyssetvalue{/ed/q5/q56/s5/label}{I think that using GenAI for learning undermines students' mastery of foundational computing skills}
\pgfkeyssetvalue{/ed/q5/q56/s5/framing}{concern}
\pgfkeyssetvalue{/ed/q5/q56/s5/n}{56}
\pgfkeyssetvalue{/ed/q5/q56/s5/n-substantive}{56}
\pgfkeyssetvalue{/ed/q5/q56/s5/n-na}{0}
\pgfkeyssetvalue{/ed/q5/q56/s5/positive/n}{33}
\pgfkeyssetvalue{/ed/q5/q56/s5/positive/pct}{58.9}
\pgfkeyssetvalue{/ed/q5/q56/s5/positive/extreme/n}{13}
\pgfkeyssetvalue{/ed/q5/q56/s5/positive/extreme/pct}{23.2}
\pgfkeyssetvalue{/ed/q5/q56/s5/positive/base/n}{20}
\pgfkeyssetvalue{/ed/q5/q56/s5/positive/base/pct}{35.7}
\pgfkeyssetvalue{/ed/q5/q56/s5/neutral/n}{13}
\pgfkeyssetvalue{/ed/q5/q56/s5/neutral/pct}{23.2}
\pgfkeyssetvalue{/ed/q5/q56/s5/negative/n}{10}
\pgfkeyssetvalue{/ed/q5/q56/s5/negative/pct}{17.9}
\pgfkeyssetvalue{/ed/q5/q56/s5/negative/extreme/n}{0}
\pgfkeyssetvalue{/ed/q5/q56/s5/negative/extreme/pct}{0.0}
\pgfkeyssetvalue{/ed/q5/q56/s5/negative/base/n}{10}
\pgfkeyssetvalue{/ed/q5/q56/s5/negative/base/pct}{17.9}
\pgfkeyssetvalue{/ed/q5/q56/s6/label}{I believe GenAI has increased the frequency of academic integrity violations}
\pgfkeyssetvalue{/ed/q5/q56/s6/framing}{concern}
\pgfkeyssetvalue{/ed/q5/q56/s6/n}{56}
\pgfkeyssetvalue{/ed/q5/q56/s6/n-substantive}{56}
\pgfkeyssetvalue{/ed/q5/q56/s6/n-na}{0}
\pgfkeyssetvalue{/ed/q5/q56/s6/positive/n}{39}
\pgfkeyssetvalue{/ed/q5/q56/s6/positive/pct}{69.6}
\pgfkeyssetvalue{/ed/q5/q56/s6/positive/extreme/n}{20}
\pgfkeyssetvalue{/ed/q5/q56/s6/positive/extreme/pct}{35.7}
\pgfkeyssetvalue{/ed/q5/q56/s6/positive/base/n}{19}
\pgfkeyssetvalue{/ed/q5/q56/s6/positive/base/pct}{33.9}
\pgfkeyssetvalue{/ed/q5/q56/s6/neutral/n}{12}
\pgfkeyssetvalue{/ed/q5/q56/s6/neutral/pct}{21.4}
\pgfkeyssetvalue{/ed/q5/q56/s6/negative/n}{5}
\pgfkeyssetvalue{/ed/q5/q56/s6/negative/pct}{8.9}
\pgfkeyssetvalue{/ed/q5/q56/s6/negative/extreme/n}{0}
\pgfkeyssetvalue{/ed/q5/q56/s6/negative/extreme/pct}{0.0}
\pgfkeyssetvalue{/ed/q5/q56/s6/negative/base/n}{5}
\pgfkeyssetvalue{/ed/q5/q56/s6/negative/base/pct}{8.9}
\pgfkeyssetvalue{/ed/q5/q56/s7/label}{I am concerned that students cannot distinguish correct from incorrect GenAI-generated answers}
\pgfkeyssetvalue{/ed/q5/q56/s7/framing}{concern}
\pgfkeyssetvalue{/ed/q5/q56/s7/n}{56}
\pgfkeyssetvalue{/ed/q5/q56/s7/n-substantive}{56}
\pgfkeyssetvalue{/ed/q5/q56/s7/n-na}{0}
\pgfkeyssetvalue{/ed/q5/q56/s7/positive/n}{41}
\pgfkeyssetvalue{/ed/q5/q56/s7/positive/pct}{73.2}
\pgfkeyssetvalue{/ed/q5/q56/s7/positive/extreme/n}{18}
\pgfkeyssetvalue{/ed/q5/q56/s7/positive/extreme/pct}{32.1}
\pgfkeyssetvalue{/ed/q5/q56/s7/positive/base/n}{23}
\pgfkeyssetvalue{/ed/q5/q56/s7/positive/base/pct}{41.1}
\pgfkeyssetvalue{/ed/q5/q56/s7/neutral/n}{14}
\pgfkeyssetvalue{/ed/q5/q56/s7/neutral/pct}{25.0}
\pgfkeyssetvalue{/ed/q5/q56/s7/negative/n}{1}
\pgfkeyssetvalue{/ed/q5/q56/s7/negative/pct}{1.8}
\pgfkeyssetvalue{/ed/q5/q56/s7/negative/extreme/n}{0}
\pgfkeyssetvalue{/ed/q5/q56/s7/negative/extreme/pct}{0.0}
\pgfkeyssetvalue{/ed/q5/q56/s7/negative/base/n}{1}
\pgfkeyssetvalue{/ed/q5/q56/s7/negative/base/pct}{1.8}
\pgfkeyssetvalue{/ed/q5/q56/s8/label}{I think GenAI tools can enhance student learning by providing personalized feedback}
\pgfkeyssetvalue{/ed/q5/q56/s8/framing}{benefit}
\pgfkeyssetvalue{/ed/q5/q56/s8/n}{56}
\pgfkeyssetvalue{/ed/q5/q56/s8/n-substantive}{56}
\pgfkeyssetvalue{/ed/q5/q56/s8/n-na}{0}
\pgfkeyssetvalue{/ed/q5/q56/s8/positive/n}{31}
\pgfkeyssetvalue{/ed/q5/q56/s8/positive/pct}{55.4}
\pgfkeyssetvalue{/ed/q5/q56/s8/positive/extreme/n}{6}
\pgfkeyssetvalue{/ed/q5/q56/s8/positive/extreme/pct}{10.7}
\pgfkeyssetvalue{/ed/q5/q56/s8/positive/base/n}{25}
\pgfkeyssetvalue{/ed/q5/q56/s8/positive/base/pct}{44.6}
\pgfkeyssetvalue{/ed/q5/q56/s8/neutral/n}{16}
\pgfkeyssetvalue{/ed/q5/q56/s8/neutral/pct}{28.6}
\pgfkeyssetvalue{/ed/q5/q56/s8/negative/n}{9}
\pgfkeyssetvalue{/ed/q5/q56/s8/negative/pct}{16.1}
\pgfkeyssetvalue{/ed/q5/q56/s8/negative/extreme/n}{3}
\pgfkeyssetvalue{/ed/q5/q56/s8/negative/extreme/pct}{5.4}
\pgfkeyssetvalue{/ed/q5/q56/s8/negative/base/n}{6}
\pgfkeyssetvalue{/ed/q5/q56/s8/negative/base/pct}{10.7}
\pgfkeyssetvalue{/ed/q5/q56/s9/label}{I am concerned that students use GenAI more than they report}
\pgfkeyssetvalue{/ed/q5/q56/s9/framing}{concern}
\pgfkeyssetvalue{/ed/q5/q56/s9/n}{56}
\pgfkeyssetvalue{/ed/q5/q56/s9/n-substantive}{56}
\pgfkeyssetvalue{/ed/q5/q56/s9/n-na}{0}
\pgfkeyssetvalue{/ed/q5/q56/s9/positive/n}{36}
\pgfkeyssetvalue{/ed/q5/q56/s9/positive/pct}{64.3}
\pgfkeyssetvalue{/ed/q5/q56/s9/positive/extreme/n}{15}
\pgfkeyssetvalue{/ed/q5/q56/s9/positive/extreme/pct}{26.8}
\pgfkeyssetvalue{/ed/q5/q56/s9/positive/base/n}{21}
\pgfkeyssetvalue{/ed/q5/q56/s9/positive/base/pct}{37.5}
\pgfkeyssetvalue{/ed/q5/q56/s9/neutral/n}{17}
\pgfkeyssetvalue{/ed/q5/q56/s9/neutral/pct}{30.4}
\pgfkeyssetvalue{/ed/q5/q56/s9/negative/n}{3}
\pgfkeyssetvalue{/ed/q5/q56/s9/negative/pct}{5.4}
\pgfkeyssetvalue{/ed/q5/q56/s9/negative/extreme/n}{1}
\pgfkeyssetvalue{/ed/q5/q56/s9/negative/extreme/pct}{1.8}
\pgfkeyssetvalue{/ed/q5/q56/s9/negative/base/n}{2}
\pgfkeyssetvalue{/ed/q5/q56/s9/negative/base/pct}{3.6}

%% file: values/hiring-q5-values.tex
\pgfkeyssetvalue{/hp/q5/n}{24}
\pgfkeyssetvalue{/hp/q5/q51/n}{24}
\pgfkeyssetvalue{/hp/q5/q51/n-substantive}{21}
\pgfkeyssetvalue{/hp/q5/q51/n-na}{3}
\pgfkeyssetvalue{/hp/q5/q51/positive/n}{3}
\pgfkeyssetvalue{/hp/q5/q51/positive/pct}{14.3}
\pgfkeyssetvalue{/hp/q5/q51/positive/extreme/n}{2}
\pgfkeyssetvalue{/hp/q5/q51/positive/extreme/pct}{9.5}
\pgfkeyssetvalue{/hp/q5/q51/positive/base/n}{1}
\pgfkeyssetvalue{/hp/q5/q51/positive/base/pct}{4.8}
\pgfkeyssetvalue{/hp/q5/q51/neutral/n}{6}
\pgfkeyssetvalue{/hp/q5/q51/neutral/pct}{28.6}
\pgfkeyssetvalue{/hp/q5/q51/negative/n}{12}
\pgfkeyssetvalue{/hp/q5/q51/negative/pct}{57.1}
\pgfkeyssetvalue{/hp/q5/q51/negative/extreme/n}{3}
\pgfkeyssetvalue{/hp/q5/q51/negative/extreme/pct}{14.3}
\pgfkeyssetvalue{/hp/q5/q51/negative/base/n}{9}
\pgfkeyssetvalue{/hp/q5/q51/negative/base/pct}{42.9}
\pgfkeyssetvalue{/hp/q5/q52/n}{24}
\pgfkeyssetvalue{/hp/q5/q52/n-substantive}{23}
\pgfkeyssetvalue{/hp/q5/q52/n-na}{1}
\pgfkeyssetvalue{/hp/q5/q52/positive/n}{16}
\pgfkeyssetvalue{/hp/q5/q52/positive/pct}{69.6}
\pgfkeyssetvalue{/hp/q5/q52/positive/extreme/n}{11}
\pgfkeyssetvalue{/hp/q5/q52/positive/extreme/pct}{47.8}
\pgfkeyssetvalue{/hp/q5/q52/positive/base/n}{5}
\pgfkeyssetvalue{/hp/q5/q52/positive/base/pct}{21.7}
\pgfkeyssetvalue{/hp/q5/q52/neutral/n}{5}
\pgfkeyssetvalue{/hp/q5/q52/neutral/pct}{21.7}
\pgfkeyssetvalue{/hp/q5/q52/negative/n}{2}
\pgfkeyssetvalue{/hp/q5/q52/negative/pct}{8.7}
\pgfkeyssetvalue{/hp/q5/q52/negative/extreme/n}{0}
\pgfkeyssetvalue{/hp/q5/q52/negative/extreme/pct}{0.0}
\pgfkeyssetvalue{/hp/q5/q52/negative/base/n}{2}
\pgfkeyssetvalue{/hp/q5/q52/negative/base/pct}{8.7}
\pgfkeyssetvalue{/hp/q5/q53/problem-decomposition/label}{Problem decomposition}
\pgfkeyssetvalue{/hp/q5/q53/problem-decomposition/n}{24}
\pgfkeyssetvalue{/hp/q5/q53/problem-decomposition/n-substantive}{20}
\pgfkeyssetvalue{/hp/q5/q53/problem-decomposition/n-na}{4}
\pgfkeyssetvalue{/hp/q5/q53/problem-decomposition/positive/n}{10}
\pgfkeyssetvalue{/hp/q5/q53/problem-decomposition/positive/pct}{50.0}
\pgfkeyssetvalue{/hp/q5/q53/problem-decomposition/positive/extreme/n}{4}
\pgfkeyssetvalue{/hp/q5/q53/problem-decomposition/positive/extreme/pct}{20.0}
\pgfkeyssetvalue{/hp/q5/q53/problem-decomposition/positive/base/n}{6}
\pgfkeyssetvalue{/hp/q5/q53/problem-decomposition/positive/base/pct}{30.0}
\pgfkeyssetvalue{/hp/q5/q53/problem-decomposition/neutral/n}{8}
\pgfkeyssetvalue{/hp/q5/q53/problem-decomposition/neutral/pct}{40.0}
\pgfkeyssetvalue{/hp/q5/q53/problem-decomposition/negative/n}{2}
\pgfkeyssetvalue{/hp/q5/q53/problem-decomposition/negative/pct}{10.0}
\pgfkeyssetvalue{/hp/q5/q53/problem-decomposition/negative/extreme/n}{1}
\pgfkeyssetvalue{/hp/q5/q53/problem-decomposition/negative/extreme/pct}{5.0}
\pgfkeyssetvalue{/hp/q5/q53/problem-decomposition/negative/base/n}{1}
\pgfkeyssetvalue{/hp/q5/q53/problem-decomposition/negative/base/pct}{5.0}
\pgfkeyssetvalue{/hp/q5/q53/problem-decomposition/gap-pct}{50.0}
\pgfkeyssetvalue{/hp/q5/q53/problem-decomposition/gap-n}{10}
\pgfkeyssetvalue{/hp/q5/q53/code-comprehension/label}{Code comprehension}
\pgfkeyssetvalue{/hp/q5/q53/code-comprehension/n}{24}
\pgfkeyssetvalue{/hp/q5/q53/code-comprehension/n-substantive}{18}
\pgfkeyssetvalue{/hp/q5/q53/code-comprehension/n-na}{6}
\pgfkeyssetvalue{/hp/q5/q53/code-comprehension/positive/n}{9}
\pgfkeyssetvalue{/hp/q5/q53/code-comprehension/positive/pct}{50.0}
\pgfkeyssetvalue{/hp/q5/q53/code-comprehension/positive/extreme/n}{2}
\pgfkeyssetvalue{/hp/q5/q53/code-comprehension/positive/extreme/pct}{11.1}
\pgfkeyssetvalue{/hp/q5/q53/code-comprehension/positive/base/n}{7}
\pgfkeyssetvalue{/hp/q5/q53/code-comprehension/positive/base/pct}{38.9}
\pgfkeyssetvalue{/hp/q5/q53/code-comprehension/neutral/n}{6}
\pgfkeyssetvalue{/hp/q5/q53/code-comprehension/neutral/pct}{33.3}
\pgfkeyssetvalue{/hp/q5/q53/code-comprehension/negative/n}{3}
\pgfkeyssetvalue{/hp/q5/q53/code-comprehension/negative/pct}{16.7}
\pgfkeyssetvalue{/hp/q5/q53/code-comprehension/negative/extreme/n}{0}
\pgfkeyssetvalue{/hp/q5/q53/code-comprehension/negative/extreme/pct}{0.0}
\pgfkeyssetvalue{/hp/q5/q53/code-comprehension/negative/base/n}{3}
\pgfkeyssetvalue{/hp/q5/q53/code-comprehension/negative/base/pct}{16.7}
\pgfkeyssetvalue{/hp/q5/q53/code-comprehension/gap-pct}{50.0}
\pgfkeyssetvalue{/hp/q5/q53/code-comprehension/gap-n}{9}
\pgfkeyssetvalue{/hp/q5/q53/code-development/label}{Code development}
\pgfkeyssetvalue{/hp/q5/q53/code-development/n}{24}
\pgfkeyssetvalue{/hp/q5/q53/code-development/n-substantive}{18}
\pgfkeyssetvalue{/hp/q5/q53/code-development/n-na}{6}
\pgfkeyssetvalue{/hp/q5/q53/code-development/positive/n}{6}
\pgfkeyssetvalue{/hp/q5/q53/code-development/positive/pct}{33.3}
\pgfkeyssetvalue{/hp/q5/q53/code-development/positive/extreme/n}{1}
\pgfkeyssetvalue{/hp/q5/q53/code-development/positive/extreme/pct}{5.6}
\pgfkeyssetvalue{/hp/q5/q53/code-development/positive/base/n}{5}
\pgfkeyssetvalue{/hp/q5/q53/code-development/positive/base/pct}{27.8}
\pgfkeyssetvalue{/hp/q5/q53/code-development/neutral/n}{8}
\pgfkeyssetvalue{/hp/q5/q53/code-development/neutral/pct}{44.4}
\pgfkeyssetvalue{/hp/q5/q53/code-development/negative/n}{4}
\pgfkeyssetvalue{/hp/q5/q53/code-development/negative/pct}{22.2}
\pgfkeyssetvalue{/hp/q5/q53/code-development/negative/extreme/n}{0}
\pgfkeyssetvalue{/hp/q5/q53/code-development/negative/extreme/pct}{0.0}
\pgfkeyssetvalue{/hp/q5/q53/code-development/negative/base/n}{4}
\pgfkeyssetvalue{/hp/q5/q53/code-development/negative/base/pct}{22.2}
\pgfkeyssetvalue{/hp/q5/q53/code-development/gap-pct}{33.3}
\pgfkeyssetvalue{/hp/q5/q53/code-development/gap-n}{6}
\pgfkeyssetvalue{/hp/q5/q53/code-review/label}{Code review}
\pgfkeyssetvalue{/hp/q5/q53/code-review/n}{24}
\pgfkeyssetvalue{/hp/q5/q53/code-review/n-substantive}{18}
\pgfkeyssetvalue{/hp/q5/q53/code-review/n-na}{6}
\pgfkeyssetvalue{/hp/q5/q53/code-review/positive/n}{12}
\pgfkeyssetvalue{/hp/q5/q53/code-review/positive/pct}{66.7}
\pgfkeyssetvalue{/hp/q5/q53/code-review/positive/extreme/n}{3}
\pgfkeyssetvalue{/hp/q5/q53/code-review/positive/extreme/pct}{16.7}
\pgfkeyssetvalue{/hp/q5/q53/code-review/positive/base/n}{9}
\pgfkeyssetvalue{/hp/q5/q53/code-review/positive/base/pct}{50.0}
\pgfkeyssetvalue{/hp/q5/q53/code-review/neutral/n}{2}
\pgfkeyssetvalue{/hp/q5/q53/code-review/neutral/pct}{11.1}
\pgfkeyssetvalue{/hp/q5/q53/code-review/negative/n}{4}
\pgfkeyssetvalue{/hp/q5/q53/code-review/negative/pct}{22.2}
\pgfkeyssetvalue{/hp/q5/q53/code-review/negative/extreme/n}{0}
\pgfkeyssetvalue{/hp/q5/q53/code-review/negative/extreme/pct}{0.0}
\pgfkeyssetvalue{/hp/q5/q53/code-review/negative/base/n}{4}
\pgfkeyssetvalue{/hp/q5/q53/code-review/negative/base/pct}{22.2}
\pgfkeyssetvalue{/hp/q5/q53/code-review/gap-pct}{66.7}
\pgfkeyssetvalue{/hp/q5/q53/code-review/gap-n}{12}
\pgfkeyssetvalue{/hp/q5/q53/debugging/label}{Debugging}
\pgfkeyssetvalue{/hp/q5/q53/debugging/n}{24}
\pgfkeyssetvalue{/hp/q5/q53/debugging/n-substantive}{18}
\pgfkeyssetvalue{/hp/q5/q53/debugging/n-na}{6}
\pgfkeyssetvalue{/hp/q5/q53/debugging/positive/n}{11}
\pgfkeyssetvalue{/hp/q5/q53/debugging/positive/pct}{61.1}
\pgfkeyssetvalue{/hp/q5/q53/debugging/positive/extreme/n}{5}
\pgfkeyssetvalue{/hp/q5/q53/debugging/positive/extreme/pct}{27.8}
\pgfkeyssetvalue{/hp/q5/q53/debugging/positive/base/n}{6}
\pgfkeyssetvalue{/hp/q5/q53/debugging/positive/base/pct}{33.3}
\pgfkeyssetvalue{/hp/q5/q53/debugging/neutral/n}{4}
\pgfkeyssetvalue{/hp/q5/q53/debugging/neutral/pct}{22.2}
\pgfkeyssetvalue{/hp/q5/q53/debugging/negative/n}{3}
\pgfkeyssetvalue{/hp/q5/q53/debugging/negative/pct}{16.7}
\pgfkeyssetvalue{/hp/q5/q53/debugging/negative/extreme/n}{0}
\pgfkeyssetvalue{/hp/q5/q53/debugging/negative/extreme/pct}{0.0}
\pgfkeyssetvalue{/hp/q5/q53/debugging/negative/base/n}{3}
\pgfkeyssetvalue{/hp/q5/q53/debugging/negative/base/pct}{16.7}
\pgfkeyssetvalue{/hp/q5/q53/debugging/gap-pct}{61.1}
\pgfkeyssetvalue{/hp/q5/q53/debugging/gap-n}{11}
\pgfkeyssetvalue{/hp/q5/q53/documentation/label}{Documentation}
\pgfkeyssetvalue{/hp/q5/q53/documentation/n}{24}
\pgfkeyssetvalue{/hp/q5/q53/documentation/n-substantive}{17}
\pgfkeyssetvalue{/hp/q5/q53/documentation/n-na}{7}
\pgfkeyssetvalue{/hp/q5/q53/documentation/positive/n}{5}
\pgfkeyssetvalue{/hp/q5/q53/documentation/positive/pct}{29.4}
\pgfkeyssetvalue{/hp/q5/q53/documentation/positive/extreme/n}{2}
\pgfkeyssetvalue{/hp/q5/q53/documentation/positive/extreme/pct}{11.8}
\pgfkeyssetvalue{/hp/q5/q53/documentation/positive/base/n}{3}
\pgfkeyssetvalue{/hp/q5/q53/documentation/positive/base/pct}{17.6}
\pgfkeyssetvalue{/hp/q5/q53/documentation/neutral/n}{7}
\pgfkeyssetvalue{/hp/q5/q53/documentation/neutral/pct}{41.2}
\pgfkeyssetvalue{/hp/q5/q53/documentation/negative/n}{5}
\pgfkeyssetvalue{/hp/q5/q53/documentation/negative/pct}{29.4}
\pgfkeyssetvalue{/hp/q5/q53/documentation/negative/extreme/n}{1}
\pgfkeyssetvalue{/hp/q5/q53/documentation/negative/extreme/pct}{5.9}
\pgfkeyssetvalue{/hp/q5/q53/documentation/negative/base/n}{4}
\pgfkeyssetvalue{/hp/q5/q53/documentation/negative/base/pct}{23.5}
\pgfkeyssetvalue{/hp/q5/q53/documentation/gap-pct}{29.4}
\pgfkeyssetvalue{/hp/q5/q53/documentation/gap-n}{5}
\pgfkeyssetvalue{/hp/q5/q53/system-design/label}{System-level design}
\pgfkeyssetvalue{/hp/q5/q53/system-design/n}{24}
\pgfkeyssetvalue{/hp/q5/q53/system-design/n-substantive}{18}
\pgfkeyssetvalue{/hp/q5/q53/system-design/n-na}{6}
\pgfkeyssetvalue{/hp/q5/q53/system-design/positive/n}{10}
\pgfkeyssetvalue{/hp/q5/q53/system-design/positive/pct}{55.6}
\pgfkeyssetvalue{/hp/q5/q53/system-design/positive/extreme/n}{5}
\pgfkeyssetvalue{/hp/q5/q53/system-design/positive/extreme/pct}{27.8}
\pgfkeyssetvalue{/hp/q5/q53/system-design/positive/base/n}{5}
\pgfkeyssetvalue{/hp/q5/q53/system-design/positive/base/pct}{27.8}
\pgfkeyssetvalue{/hp/q5/q53/system-design/neutral/n}{7}
\pgfkeyssetvalue{/hp/q5/q53/system-design/neutral/pct}{38.9}
\pgfkeyssetvalue{/hp/q5/q53/system-design/negative/n}{1}
\pgfkeyssetvalue{/hp/q5/q53/system-design/negative/pct}{5.6}
\pgfkeyssetvalue{/hp/q5/q53/system-design/negative/extreme/n}{0}
\pgfkeyssetvalue{/hp/q5/q53/system-design/negative/extreme/pct}{0.0}
\pgfkeyssetvalue{/hp/q5/q53/system-design/negative/base/n}{1}
\pgfkeyssetvalue{/hp/q5/q53/system-design/negative/base/pct}{5.6}
\pgfkeyssetvalue{/hp/q5/q53/system-design/gap-pct}{55.6}
\pgfkeyssetvalue{/hp/q5/q53/system-design/gap-n}{10}
\pgfkeyssetvalue{/hp/q5/q53/algorithmic-thinking/label}{Algorithmic thinking}
\pgfkeyssetvalue{/hp/q5/q53/algorithmic-thinking/n}{24}
\pgfkeyssetvalue{/hp/q5/q53/algorithmic-thinking/n-substantive}{17}
\pgfkeyssetvalue{/hp/q5/q53/algorithmic-thinking/n-na}{7}
\pgfkeyssetvalue{/hp/q5/q53/algorithmic-thinking/positive/n}{7}
\pgfkeyssetvalue{/hp/q5/q53/algorithmic-thinking/positive/pct}{41.2}
\pgfkeyssetvalue{/hp/q5/q53/algorithmic-thinking/positive/extreme/n}{1}
\pgfkeyssetvalue{/hp/q5/q53/algorithmic-thinking/positive/extreme/pct}{5.9}
\pgfkeyssetvalue{/hp/q5/q53/algorithmic-thinking/positive/base/n}{6}
\pgfkeyssetvalue{/hp/q5/q53/algorithmic-thinking/positive/base/pct}{35.3}
\pgfkeyssetvalue{/hp/q5/q53/algorithmic-thinking/neutral/n}{2}
\pgfkeyssetvalue{/hp/q5/q53/algorithmic-thinking/neutral/pct}{11.8}
\pgfkeyssetvalue{/hp/q5/q53/algorithmic-thinking/negative/n}{8}
\pgfkeyssetvalue{/hp/q5/q53/algorithmic-thinking/negative/pct}{47.1}
\pgfkeyssetvalue{/hp/q5/q53/algorithmic-thinking/negative/extreme/n}{0}
\pgfkeyssetvalue{/hp/q5/q53/algorithmic-thinking/negative/extreme/pct}{0.0}
\pgfkeyssetvalue{/hp/q5/q53/algorithmic-thinking/negative/base/n}{8}
\pgfkeyssetvalue{/hp/q5/q53/algorithmic-thinking/negative/base/pct}{47.1}
\pgfkeyssetvalue{/hp/q5/q53/algorithmic-thinking/gap-pct}{41.2}
\pgfkeyssetvalue{/hp/q5/q53/algorithmic-thinking/gap-n}{7}
\pgfkeyssetvalue{/hp/q5/q53/critical-eval-ai/label}{Critical evaluation of AI-generated output}
\pgfkeyssetvalue{/hp/q5/q53/critical-eval-ai/n}{24}
\pgfkeyssetvalue{/hp/q5/q53/critical-eval-ai/n-substantive}{17}
\pgfkeyssetvalue{/hp/q5/q53/critical-eval-ai/n-na}{7}
\pgfkeyssetvalue{/hp/q5/q53/critical-eval-ai/positive/n}{10}
\pgfkeyssetvalue{/hp/q5/q53/critical-eval-ai/positive/pct}{58.8}
\pgfkeyssetvalue{/hp/q5/q53/critical-eval-ai/positive/extreme/n}{3}
\pgfkeyssetvalue{/hp/q5/q53/critical-eval-ai/positive/extreme/pct}{17.6}
\pgfkeyssetvalue{/hp/q5/q53/critical-eval-ai/positive/base/n}{7}
\pgfkeyssetvalue{/hp/q5/q53/critical-eval-ai/positive/base/pct}{41.2}
\pgfkeyssetvalue{/hp/q5/q53/critical-eval-ai/neutral/n}{6}
\pgfkeyssetvalue{/hp/q5/q53/critical-eval-ai/neutral/pct}{35.3}
\pgfkeyssetvalue{/hp/q5/q53/critical-eval-ai/negative/n}{1}
\pgfkeyssetvalue{/hp/q5/q53/critical-eval-ai/negative/pct}{5.9}
\pgfkeyssetvalue{/hp/q5/q53/critical-eval-ai/negative/extreme/n}{1}
\pgfkeyssetvalue{/hp/q5/q53/critical-eval-ai/negative/extreme/pct}{5.9}
\pgfkeyssetvalue{/hp/q5/q53/critical-eval-ai/negative/base/n}{0}
\pgfkeyssetvalue{/hp/q5/q53/critical-eval-ai/negative/base/pct}{0.0}
\pgfkeyssetvalue{/hp/q5/q53/critical-eval-ai/gap-pct}{58.8}
\pgfkeyssetvalue{/hp/q5/q53/critical-eval-ai/gap-n}{10}
\pgfkeyssetvalue{/hp/q5/q53/learn-adapt/label}{Ability to learn and adapt independently}
\pgfkeyssetvalue{/hp/q5/q53/learn-adapt/n}{24}
\pgfkeyssetvalue{/hp/q5/q53/learn-adapt/n-substantive}{20}
\pgfkeyssetvalue{/hp/q5/q53/learn-adapt/n-na}{4}
\pgfkeyssetvalue{/hp/q5/q53/learn-adapt/positive/n}{8}
\pgfkeyssetvalue{/hp/q5/q53/learn-adapt/positive/pct}{40.0}
\pgfkeyssetvalue{/hp/q5/q53/learn-adapt/positive/extreme/n}{2}
\pgfkeyssetvalue{/hp/q5/q53/learn-adapt/positive/extreme/pct}{10.0}
\pgfkeyssetvalue{/hp/q5/q53/learn-adapt/positive/base/n}{6}
\pgfkeyssetvalue{/hp/q5/q53/learn-adapt/positive/base/pct}{30.0}
\pgfkeyssetvalue{/hp/q5/q53/learn-adapt/neutral/n}{6}
\pgfkeyssetvalue{/hp/q5/q53/learn-adapt/neutral/pct}{30.0}
\pgfkeyssetvalue{/hp/q5/q53/learn-adapt/negative/n}{6}
\pgfkeyssetvalue{/hp/q5/q53/learn-adapt/negative/pct}{30.0}
\pgfkeyssetvalue{/hp/q5/q53/learn-adapt/negative/extreme/n}{2}
\pgfkeyssetvalue{/hp/q5/q53/learn-adapt/negative/extreme/pct}{10.0}
\pgfkeyssetvalue{/hp/q5/q53/learn-adapt/negative/base/n}{4}
\pgfkeyssetvalue{/hp/q5/q53/learn-adapt/negative/base/pct}{20.0}
\pgfkeyssetvalue{/hp/q5/q53/learn-adapt/gap-pct}{40.0}
\pgfkeyssetvalue{/hp/q5/q53/learn-adapt/gap-n}{8}
\pgfkeyssetvalue{/hp/q5/q53/collaboration/label}{Collaboration and communication}
\pgfkeyssetvalue{/hp/q5/q53/collaboration/n}{24}
\pgfkeyssetvalue{/hp/q5/q53/collaboration/n-substantive}{20}
\pgfkeyssetvalue{/hp/q5/q53/collaboration/n-na}{4}
\pgfkeyssetvalue{/hp/q5/q53/collaboration/positive/n}{10}
\pgfkeyssetvalue{/hp/q5/q53/collaboration/positive/pct}{50.0}
\pgfkeyssetvalue{/hp/q5/q53/collaboration/positive/extreme/n}{5}
\pgfkeyssetvalue{/hp/q5/q53/collaboration/positive/extreme/pct}{25.0}
\pgfkeyssetvalue{/hp/q5/q53/collaboration/positive/base/n}{5}
\pgfkeyssetvalue{/hp/q5/q53/collaboration/positive/base/pct}{25.0}
\pgfkeyssetvalue{/hp/q5/q53/collaboration/neutral/n}{3}
\pgfkeyssetvalue{/hp/q5/q53/collaboration/neutral/pct}{15.0}
\pgfkeyssetvalue{/hp/q5/q53/collaboration/negative/n}{7}
\pgfkeyssetvalue{/hp/q5/q53/collaboration/negative/pct}{35.0}
\pgfkeyssetvalue{/hp/q5/q53/collaboration/negative/extreme/n}{2}
\pgfkeyssetvalue{/hp/q5/q53/collaboration/negative/extreme/pct}{10.0}
\pgfkeyssetvalue{/hp/q5/q53/collaboration/negative/base/n}{5}
\pgfkeyssetvalue{/hp/q5/q53/collaboration/negative/base/pct}{25.0}
\pgfkeyssetvalue{/hp/q5/q53/collaboration/gap-pct}{50.0}
\pgfkeyssetvalue{/hp/q5/q53/collaboration/gap-n}{10}
\pgfkeyssetvalue{/hp/q5/q53/responsible-genai/label}{Responsible and effective use of GenAI tools}
\pgfkeyssetvalue{/hp/q5/q53/responsible-genai/n}{24}
\pgfkeyssetvalue{/hp/q5/q53/responsible-genai/n-substantive}{15}
\pgfkeyssetvalue{/hp/q5/q53/responsible-genai/n-na}{9}
\pgfkeyssetvalue{/hp/q5/q53/responsible-genai/positive/n}{6}
\pgfkeyssetvalue{/hp/q5/q53/responsible-genai/positive/pct}{40.0}
\pgfkeyssetvalue{/hp/q5/q53/responsible-genai/positive/extreme/n}{3}
\pgfkeyssetvalue{/hp/q5/q53/responsible-genai/positive/extreme/pct}{20.0}
\pgfkeyssetvalue{/hp/q5/q53/responsible-genai/positive/base/n}{3}
\pgfkeyssetvalue{/hp/q5/q53/responsible-genai/positive/base/pct}{20.0}
\pgfkeyssetvalue{/hp/q5/q53/responsible-genai/neutral/n}{2}
\pgfkeyssetvalue{/hp/q5/q53/responsible-genai/neutral/pct}{13.3}
\pgfkeyssetvalue{/hp/q5/q53/responsible-genai/negative/n}{7}
\pgfkeyssetvalue{/hp/q5/q53/responsible-genai/negative/pct}{46.7}
\pgfkeyssetvalue{/hp/q5/q53/responsible-genai/negative/extreme/n}{1}
\pgfkeyssetvalue{/hp/q5/q53/responsible-genai/negative/extreme/pct}{6.7}
\pgfkeyssetvalue{/hp/q5/q53/responsible-genai/negative/base/n}{6}
\pgfkeyssetvalue{/hp/q5/q53/responsible-genai/negative/base/pct}{40.0}
\pgfkeyssetvalue{/hp/q5/q53/responsible-genai/gap-pct}{40.0}
\pgfkeyssetvalue{/hp/q5/q53/responsible-genai/gap-n}{6}
\pgfkeyssetvalue{/hp/q5/q53/top-gap1/label}{Code review}
\pgfkeyssetvalue{/hp/q5/q53/top-gap1/pct}{66.7}
\pgfkeyssetvalue{/hp/q5/q53/top-gap1/n}{12}
\pgfkeyssetvalue{/hp/q5/q53/top-gap1/skill-num}{4}
\pgfkeyssetvalue{/hp/q5/q53/top-gap2/label}{Debugging}
\pgfkeyssetvalue{/hp/q5/q53/top-gap2/pct}{61.1}
\pgfkeyssetvalue{/hp/q5/q53/top-gap2/n}{11}
\pgfkeyssetvalue{/hp/q5/q53/top-gap2/skill-num}{5}
\pgfkeyssetvalue{/hp/q5/q53/top-gap3/label}{Critical evaluation of AI-generated output}
\pgfkeyssetvalue{/hp/q5/q53/top-gap3/pct}{58.8}
\pgfkeyssetvalue{/hp/q5/q53/top-gap3/n}{10}
\pgfkeyssetvalue{/hp/q5/q53/top-gap3/skill-num}{9}

%% file: values/skill-comparison-values.tex
\pgfkeyssetvalue{/cmp/skill/ed-n}{56}
\pgfkeyssetvalue{/cmp/skill/hp-n}{24}
\pgfkeyssetvalue{/cmp/skill/importance/problem-decomposition/label}{Problem decomposition}
\pgfkeyssetvalue{/cmp/skill/importance/problem-decomposition/ed-pct}{89.3}
\pgfkeyssetvalue{/cmp/skill/importance/problem-decomposition/hp-pct}{79.2}
\pgfkeyssetvalue{/cmp/skill/importance/problem-decomposition/gap}{10.1}
\pgfkeyssetvalue{/cmp/skill/importance/problem-decomposition/abs-gap}{10.1}
\pgfkeyssetvalue{/cmp/skill/importance/code-comprehension/label}{Code comprehension}
\pgfkeyssetvalue{/cmp/skill/importance/code-comprehension/ed-pct}{83.9}
\pgfkeyssetvalue{/cmp/skill/importance/code-comprehension/hp-pct}{66.7}
\pgfkeyssetvalue{/cmp/skill/importance/code-comprehension/gap}{17.3}
\pgfkeyssetvalue{/cmp/skill/importance/code-comprehension/abs-gap}{17.3}
\pgfkeyssetvalue{/cmp/skill/importance/code-development/label}{Code development}
\pgfkeyssetvalue{/cmp/skill/importance/code-development/ed-pct}{39.3}
\pgfkeyssetvalue{/cmp/skill/importance/code-development/hp-pct}{54.2}
\pgfkeyssetvalue{/cmp/skill/importance/code-development/gap}{-14.9}
\pgfkeyssetvalue{/cmp/skill/importance/code-development/abs-gap}{14.9}
\pgfkeyssetvalue{/cmp/skill/importance/code-review/label}{Code review}
\pgfkeyssetvalue{/cmp/skill/importance/code-review/ed-pct}{85.7}
\pgfkeyssetvalue{/cmp/skill/importance/code-review/hp-pct}{75.0}
\pgfkeyssetvalue{/cmp/skill/importance/code-review/gap}{10.7}
\pgfkeyssetvalue{/cmp/skill/importance/code-review/abs-gap}{10.7}
\pgfkeyssetvalue{/cmp/skill/importance/debugging/label}{Debugging}
\pgfkeyssetvalue{/cmp/skill/importance/debugging/ed-pct}{78.6}
\pgfkeyssetvalue{/cmp/skill/importance/debugging/hp-pct}{66.7}
\pgfkeyssetvalue{/cmp/skill/importance/debugging/gap}{11.9}
\pgfkeyssetvalue{/cmp/skill/importance/debugging/abs-gap}{11.9}
\pgfkeyssetvalue{/cmp/skill/importance/documentation/label}{Documentation}
\pgfkeyssetvalue{/cmp/skill/importance/documentation/ed-pct}{39.3}
\pgfkeyssetvalue{/cmp/skill/importance/documentation/hp-pct}{47.8}
\pgfkeyssetvalue{/cmp/skill/importance/documentation/gap}{-8.5}
\pgfkeyssetvalue{/cmp/skill/importance/documentation/abs-gap}{8.5}
\pgfkeyssetvalue{/cmp/skill/importance/system-design/label}{System-level design}
\pgfkeyssetvalue{/cmp/skill/importance/system-design/ed-pct}{85.5}
\pgfkeyssetvalue{/cmp/skill/importance/system-design/hp-pct}{62.5}
\pgfkeyssetvalue{/cmp/skill/importance/system-design/gap}{23.0}
\pgfkeyssetvalue{/cmp/skill/importance/system-design/abs-gap}{23.0}
\pgfkeyssetvalue{/cmp/skill/importance/algorithmic-thinking/label}{Algorithmic thinking}
\pgfkeyssetvalue{/cmp/skill/importance/algorithmic-thinking/ed-pct}{80.0}
\pgfkeyssetvalue{/cmp/skill/importance/algorithmic-thinking/hp-pct}{66.7}
\pgfkeyssetvalue{/cmp/skill/importance/algorithmic-thinking/gap}{13.3}
\pgfkeyssetvalue{/cmp/skill/importance/algorithmic-thinking/abs-gap}{13.3}
\pgfkeyssetvalue{/cmp/skill/importance/critical-eval-ai/label}{Critical evaluation of AI-generated output}
\pgfkeyssetvalue{/cmp/skill/importance/critical-eval-ai/ed-pct}{92.9}
\pgfkeyssetvalue{/cmp/skill/importance/critical-eval-ai/hp-pct}{100.0}
\pgfkeyssetvalue{/cmp/skill/importance/critical-eval-ai/gap}{-7.1}
\pgfkeyssetvalue{/cmp/skill/importance/critical-eval-ai/abs-gap}{7.1}
\pgfkeyssetvalue{/cmp/skill/importance/learn-adapt/label}{Ability to learn and adapt independently}
\pgfkeyssetvalue{/cmp/skill/importance/learn-adapt/ed-pct}{94.5}
\pgfkeyssetvalue{/cmp/skill/importance/learn-adapt/hp-pct}{91.7}
\pgfkeyssetvalue{/cmp/skill/importance/learn-adapt/gap}{2.9}
\pgfkeyssetvalue{/cmp/skill/importance/learn-adapt/abs-gap}{2.9}
\pgfkeyssetvalue{/cmp/skill/importance/collaboration/label}{Collaboration and communication}
\pgfkeyssetvalue{/cmp/skill/importance/collaboration/ed-pct}{96.4}
\pgfkeyssetvalue{/cmp/skill/importance/collaboration/hp-pct}{83.3}
\pgfkeyssetvalue{/cmp/skill/importance/collaboration/gap}{13.0}
\pgfkeyssetvalue{/cmp/skill/importance/collaboration/abs-gap}{13.0}
\pgfkeyssetvalue{/cmp/skill/importance/responsible-genai/label}{Responsible and effective use of GenAI tools}
\pgfkeyssetvalue{/cmp/skill/importance/responsible-genai/ed-pct}{87.3}
\pgfkeyssetvalue{/cmp/skill/importance/responsible-genai/hp-pct}{95.8}
\pgfkeyssetvalue{/cmp/skill/importance/responsible-genai/gap}{-8.6}
\pgfkeyssetvalue{/cmp/skill/importance/responsible-genai/abs-gap}{8.6}
\pgfkeyssetvalue{/cmp/skill/importance/mean-abs-gap}{11.8}
\pgfkeyssetvalue{/cmp/skill/importance/max-abs-gap}{23.0}
\pgfkeyssetvalue{/cmp/skill/importance/n-converge-10pp}{4}
\pgfkeyssetvalue{/cmp/skill/importance/n-diverge-20pp}{1}
\pgfkeyssetvalue{/cmp/skill/importance/n-ed-higher}{7}
\pgfkeyssetvalue{/cmp/skill/importance/n-hp-higher}{1}
\pgfkeyssetvalue{/cmp/skill/importance/max-div/label}{System-level design}
\pgfkeyssetvalue{/cmp/skill/importance/max-div/abs-gap}{23.0}
\pgfkeyssetvalue{/cmp/skill/readiness/problem-decomposition/label}{Problem decomposition}
\pgfkeyssetvalue{/cmp/skill/readiness/problem-decomposition/ed-pct}{9.1}
\pgfkeyssetvalue{/cmp/skill/readiness/problem-decomposition/hp-pct}{50.0}
\pgfkeyssetvalue{/cmp/skill/readiness/problem-decomposition/hp-n-sub}{20}
\pgfkeyssetvalue{/cmp/skill/readiness/problem-decomposition/gap}{-40.9}
\pgfkeyssetvalue{/cmp/skill/readiness/problem-decomposition/abs-gap}{40.9}
\pgfkeyssetvalue{/cmp/skill/readiness/code-comprehension/label}{Code comprehension}
\pgfkeyssetvalue{/cmp/skill/readiness/code-comprehension/ed-pct}{16.7}
\pgfkeyssetvalue{/cmp/skill/readiness/code-comprehension/hp-pct}{50.0}
\pgfkeyssetvalue{/cmp/skill/readiness/code-comprehension/hp-n-sub}{18}
\pgfkeyssetvalue{/cmp/skill/readiness/code-comprehension/gap}{-33.3}
\pgfkeyssetvalue{/cmp/skill/readiness/code-comprehension/abs-gap}{33.3}
\pgfkeyssetvalue{/cmp/skill/readiness/code-development/label}{Code development}
\pgfkeyssetvalue{/cmp/skill/readiness/code-development/ed-pct}{14.8}
\pgfkeyssetvalue{/cmp/skill/readiness/code-development/hp-pct}{33.3}
\pgfkeyssetvalue{/cmp/skill/readiness/code-development/hp-n-sub}{18}
\pgfkeyssetvalue{/cmp/skill/readiness/code-development/gap}{-18.5}
\pgfkeyssetvalue{/cmp/skill/readiness/code-development/abs-gap}{18.5}
\pgfkeyssetvalue{/cmp/skill/readiness/code-review/label}{Code review}
\pgfkeyssetvalue{/cmp/skill/readiness/code-review/ed-pct}{22.6}
\pgfkeyssetvalue{/cmp/skill/readiness/code-review/hp-pct}{66.7}
\pgfkeyssetvalue{/cmp/skill/readiness/code-review/hp-n-sub}{18}
\pgfkeyssetvalue{/cmp/skill/readiness/code-review/gap}{-44.0}
\pgfkeyssetvalue{/cmp/skill/readiness/code-review/abs-gap}{44.0}
\pgfkeyssetvalue{/cmp/skill/readiness/debugging/label}{Debugging}
\pgfkeyssetvalue{/cmp/skill/readiness/debugging/ed-pct}{20.4}
\pgfkeyssetvalue{/cmp/skill/readiness/debugging/hp-pct}{61.1}
\pgfkeyssetvalue{/cmp/skill/readiness/debugging/hp-n-sub}{18}
\pgfkeyssetvalue{/cmp/skill/readiness/debugging/gap}{-40.7}
\pgfkeyssetvalue{/cmp/skill/readiness/debugging/abs-gap}{40.7}
\pgfkeyssetvalue{/cmp/skill/readiness/documentation/label}{Documentation}
\pgfkeyssetvalue{/cmp/skill/readiness/documentation/ed-pct}{38.0}
\pgfkeyssetvalue{/cmp/skill/readiness/documentation/hp-pct}{29.4}
\pgfkeyssetvalue{/cmp/skill/readiness/documentation/hp-n-sub}{17}
\pgfkeyssetvalue{/cmp/skill/readiness/documentation/gap}{8.6}
\pgfkeyssetvalue{/cmp/skill/readiness/documentation/abs-gap}{8.6}
\pgfkeyssetvalue{/cmp/skill/readiness/system-design/label}{System-level design}
\pgfkeyssetvalue{/cmp/skill/readiness/system-design/ed-pct}{26.4}
\pgfkeyssetvalue{/cmp/skill/readiness/system-design/hp-pct}{55.6}
\pgfkeyssetvalue{/cmp/skill/readiness/system-design/hp-n-sub}{18}
\pgfkeyssetvalue{/cmp/skill/readiness/system-design/gap}{-29.1}
\pgfkeyssetvalue{/cmp/skill/readiness/system-design/abs-gap}{29.1}
\pgfkeyssetvalue{/cmp/skill/readiness/algorithmic-thinking/label}{Algorithmic thinking}
\pgfkeyssetvalue{/cmp/skill/readiness/algorithmic-thinking/ed-pct}{15.4}
\pgfkeyssetvalue{/cmp/skill/readiness/algorithmic-thinking/hp-pct}{41.2}
\pgfkeyssetvalue{/cmp/skill/readiness/algorithmic-thinking/hp-n-sub}{17}
\pgfkeyssetvalue{/cmp/skill/readiness/algorithmic-thinking/gap}{-25.8}
\pgfkeyssetvalue{/cmp/skill/readiness/algorithmic-thinking/abs-gap}{25.8}
\pgfkeyssetvalue{/cmp/skill/readiness/critical-eval-ai/label}{Critical evaluation of AI-generated output}
\pgfkeyssetvalue{/cmp/skill/readiness/critical-eval-ai/ed-pct}{38.5}
\pgfkeyssetvalue{/cmp/skill/readiness/critical-eval-ai/hp-pct}{58.8}
\pgfkeyssetvalue{/cmp/skill/readiness/critical-eval-ai/hp-n-sub}{17}
\pgfkeyssetvalue{/cmp/skill/readiness/critical-eval-ai/gap}{-20.4}
\pgfkeyssetvalue{/cmp/skill/readiness/critical-eval-ai/abs-gap}{20.4}
\pgfkeyssetvalue{/cmp/skill/readiness/learn-adapt/label}{Ability to learn and adapt independently}
\pgfkeyssetvalue{/cmp/skill/readiness/learn-adapt/ed-pct}{22.2}
\pgfkeyssetvalue{/cmp/skill/readiness/learn-adapt/hp-pct}{40.0}
\pgfkeyssetvalue{/cmp/skill/readiness/learn-adapt/hp-n-sub}{20}
\pgfkeyssetvalue{/cmp/skill/readiness/learn-adapt/gap}{-17.8}
\pgfkeyssetvalue{/cmp/skill/readiness/learn-adapt/abs-gap}{17.8}
\pgfkeyssetvalue{/cmp/skill/readiness/collaboration/label}{Collaboration and communication}
\pgfkeyssetvalue{/cmp/skill/readiness/collaboration/ed-pct}{20.0}
\pgfkeyssetvalue{/cmp/skill/readiness/collaboration/hp-pct}{50.0}
\pgfkeyssetvalue{/cmp/skill/readiness/collaboration/hp-n-sub}{20}
\pgfkeyssetvalue{/cmp/skill/readiness/collaboration/gap}{-30.0}
\pgfkeyssetvalue{/cmp/skill/readiness/collaboration/abs-gap}{30.0}
\pgfkeyssetvalue{/cmp/skill/readiness/responsible-genai/label}{Responsible and effective use of GenAI tools}
\pgfkeyssetvalue{/cmp/skill/readiness/responsible-genai/ed-pct}{43.1}
\pgfkeyssetvalue{/cmp/skill/readiness/responsible-genai/hp-pct}{40.0}
\pgfkeyssetvalue{/cmp/skill/readiness/responsible-genai/hp-n-sub}{15}
\pgfkeyssetvalue{/cmp/skill/readiness/responsible-genai/gap}{3.1}
\pgfkeyssetvalue{/cmp/skill/readiness/responsible-genai/abs-gap}{3.1}
\pgfkeyssetvalue{/cmp/skill/readiness/mean-abs-gap}{26.0}
\pgfkeyssetvalue{/cmp/skill/readiness/max-abs-gap}{44.0}
\pgfkeyssetvalue{/cmp/skill/readiness/n-converge-10pp}{2}
\pgfkeyssetvalue{/cmp/skill/readiness/n-diverge-20pp}{8}
\pgfkeyssetvalue{/cmp/skill/readiness/n-ed-higher}{0}
\pgfkeyssetvalue{/cmp/skill/readiness/n-hp-higher}{10}
\pgfkeyssetvalue{/cmp/skill/readiness/max-div/label}{Code review}
\pgfkeyssetvalue{/cmp/skill/readiness/max-div/abs-gap}{44.0}
\pgfkeyssetvalue{/cmp/skill/readiness/top-hp-gap/label}{Code review}
\pgfkeyssetvalue{/cmp/skill/readiness/top-hp-gap/pct}{66.7}
\pgfkeyssetvalue{/cmp/skill/readiness/top-ed-weakness/label}{Responsible and effective use of GenAI tools}
\pgfkeyssetvalue{/cmp/skill/readiness/top-ed-weakness/pct}{43.1}

%% file: values/screening-values.tex
\pgfkeyssetvalue{/screening/hp/raw/n}{58}
\pgfkeyssetvalue{/screening/hp/attempted/n}{57}
\pgfkeyssetvalue{/screening/hp/excluded/n}{33}
\pgfkeyssetvalue{/screening/hp/final/n}{24}
\pgfkeyssetvalue{/screening/hp/excluded/survey-preview/n}{1}
\pgfkeyssetvalue{/screening/hp/excluded/survey-preview/n}{1}
\pgfkeyssetvalue{/screening/hp/excluded/progress-incomplete/n}{21}
\pgfkeyssetvalue{/screening/hp/excluded/not-finished/n}{0}
\pgfkeyssetvalue{/screening/hp/excluded/did-not-consent/n}{0}
\pgfkeyssetvalue{/screening/hp/excluded/q1-no/n}{12}

\pgfkeyssetvalue{/screening/ed/raw/n}{84}
\pgfkeyssetvalue{/screening/ed/attempted/n}{83}
\pgfkeyssetvalue{/screening/ed/excluded/n}{27}
\pgfkeyssetvalue{/screening/ed/final/n}{56}
\pgfkeyssetvalue{/screening/ed/excluded/survey-preview/n}{1}
\pgfkeyssetvalue{/screening/ed/excluded/survey-preview/n}{1}
\pgfkeyssetvalue{/screening/ed/excluded/progress-incomplete/n}{19}
\pgfkeyssetvalue{/screening/ed/excluded/not-finished/n}{0}
\pgfkeyssetvalue{/screening/ed/excluded/did-not-consent/n}{0}
\pgfkeyssetvalue{/screening/ed/excluded/q11-none-of-these/n}{2}
\pgfkeyssetvalue{/screening/ed/excluded/q12-no/n}{6}

%% file: values/hiring-q3-6-values.tex
\pgfkeyssetvalue{/hp/q36/n}{15}
\pgfkeyssetvalue{/hp/q36/n-full-sample}{24}
\pgfkeyssetvalue{/hp/q36/avg-codes-per-respondent}{1.80}
\pgfkeyssetvalue{/hp/q36/total-code-applications}{27}
\pgfkeyssetvalue{/hp/q36/code/required-use/n}{3}
\pgfkeyssetvalue{/hp/q36/code/required-use/pct}{20.0}
\pgfkeyssetvalue{/hp/q36/code/required-use/label}{Required use}
\pgfkeyssetvalue{/hp/q36/code/required-use/group}{Mandate}
\pgfkeyssetvalue{/hp/q36/code/encouraged-use/n}{5}
\pgfkeyssetvalue{/hp/q36/code/encouraged-use/pct}{33.3}
\pgfkeyssetvalue{/hp/q36/code/encouraged-use/label}{Encouraged use}
\pgfkeyssetvalue{/hp/q36/code/encouraged-use/group}{Mandate}
\pgfkeyssetvalue{/hp/q36/code/mandatory-training/n}{2}
\pgfkeyssetvalue{/hp/q36/code/mandatory-training/pct}{13.3}
\pgfkeyssetvalue{/hp/q36/code/mandatory-training/label}{Mandatory training/onboarding}
\pgfkeyssetvalue{/hp/q36/code/mandatory-training/group}{Mandate}
\pgfkeyssetvalue{/hp/q36/code/approved-tools/n}{3}
\pgfkeyssetvalue{/hp/q36/code/approved-tools/pct}{20.0}
\pgfkeyssetvalue{/hp/q36/code/approved-tools/label}{Approved-tools governance}
\pgfkeyssetvalue{/hp/q36/code/approved-tools/group}{Governance}
\pgfkeyssetvalue{/hp/q36/code/accountability/n}{5}
\pgfkeyssetvalue{/hp/q36/code/accountability/pct}{33.3}
\pgfkeyssetvalue{/hp/q36/code/accountability/label}{Developer accountability}
\pgfkeyssetvalue{/hp/q36/code/accountability/group}{Governance}
\pgfkeyssetvalue{/hp/q36/code/privacy-security/n}{5}
\pgfkeyssetvalue{/hp/q36/code/privacy-security/pct}{33.3}
\pgfkeyssetvalue{/hp/q36/code/privacy-security/label}{Privacy/security/IP controls}
\pgfkeyssetvalue{/hp/q36/code/privacy-security/group}{Governance}
\pgfkeyssetvalue{/hp/q36/code/usage-tracking/n}{1}
\pgfkeyssetvalue{/hp/q36/code/usage-tracking/pct}{6.7}
\pgfkeyssetvalue{/hp/q36/code/usage-tracking/label}{Usage tracking/measurement}
\pgfkeyssetvalue{/hp/q36/code/usage-tracking/group}{Governance}
\pgfkeyssetvalue{/hp/q36/code/tool-inventory/n}{2}
\pgfkeyssetvalue{/hp/q36/code/tool-inventory/pct}{13.3}
\pgfkeyssetvalue{/hp/q36/code/tool-inventory/label}{Tool inventory mentioned}
\pgfkeyssetvalue{/hp/q36/code/tool-inventory/group}{Other}
\pgfkeyssetvalue{/hp/q36/code/not-finalized/n}{1}
\pgfkeyssetvalue{/hp/q36/code/not-finalized/pct}{6.7}
\pgfkeyssetvalue{/hp/q36/code/not-finalized/label}{Policy not finalized}
\pgfkeyssetvalue{/hp/q36/code/not-finalized/group}{Other}
\pgfkeyssetvalue{/hp/q36/group/mandate/label}{Mandate}
\pgfkeyssetvalue{/hp/q36/group/mandate/n}{10}
\pgfkeyssetvalue{/hp/q36/group/mandate/pct}{66.7}
\pgfkeyssetvalue{/hp/q36/group/governance/label}{Governance}
\pgfkeyssetvalue{/hp/q36/group/governance/n}{14}
\pgfkeyssetvalue{/hp/q36/group/governance/pct}{93.3}
\pgfkeyssetvalue{/hp/q36/group/other/label}{Other}
\pgfkeyssetvalue{/hp/q36/group/other/n}{3}
\pgfkeyssetvalue{/hp/q36/group/other/pct}{20.0}
\pgfkeyssetvalue{/hp/q36/top1/label}{Encouraged use}
\pgfkeyssetvalue{/hp/q36/top1/slug}{encouraged-use}
\pgfkeyssetvalue{/hp/q36/top1/n}{5}
\pgfkeyssetvalue{/hp/q36/top1/pct}{33.3}
\pgfkeyssetvalue{/hp/q36/top2/label}{Developer accountability}
\pgfkeyssetvalue{/hp/q36/top2/slug}{accountability}
\pgfkeyssetvalue{/hp/q36/top2/n}{5}
\pgfkeyssetvalue{/hp/q36/top2/pct}{33.3}
\pgfkeyssetvalue{/hp/q36/top3/label}{Privacy/security/IP controls}
\pgfkeyssetvalue{/hp/q36/top3/slug}{privacy-security}
\pgfkeyssetvalue{/hp/q36/top3/n}{5}
\pgfkeyssetvalue{/hp/q36/top3/pct}{33.3}

%% file: values/hiring-q4-12-values.tex
\pgfkeyssetvalue{/hm/q412/n}{13}
\pgfkeyssetvalue{/hm/q412/n-full-sample}{22}
\pgfkeyssetvalue{/hm/q412/avg-codes-per-respondent}{1.15}
\pgfkeyssetvalue{/hm/q412/total-code-applications}{15}
\pgfkeyssetvalue{/hm/q412/code/ch-over-reliance/n}{1}
\pgfkeyssetvalue{/hm/q412/code/ch-over-reliance/pct}{7.7}
\pgfkeyssetvalue{/hm/q412/code/ch-over-reliance/label}{Challenge: Candidate over-reliance / shallow understanding}
\pgfkeyssetvalue{/hm/q412/code/ch-over-reliance/group}{Challenges}
\pgfkeyssetvalue{/hm/q412/code/ch-competence-not-verifiable/n}{4}
\pgfkeyssetvalue{/hm/q412/code/ch-competence-not-verifiable/pct}{30.8}
\pgfkeyssetvalue{/hm/q412/code/ch-competence-not-verifiable/label}{Challenge: Competence not verifiable}
\pgfkeyssetvalue{/hm/q412/code/ch-competence-not-verifiable/group}{Challenges}
\pgfkeyssetvalue{/hm/q412/code/ch-debugging-hidden/n}{2}
\pgfkeyssetvalue{/hm/q412/code/ch-debugging-hidden/pct}{15.4}
\pgfkeyssetvalue{/hm/q412/code/ch-debugging-hidden/label}{Challenge: Debugging skills hidden}
\pgfkeyssetvalue{/hm/q412/code/ch-debugging-hidden/group}{Challenges}
\pgfkeyssetvalue{/hm/q412/code/ch-rule-enforcement/n}{1}
\pgfkeyssetvalue{/hm/q412/code/ch-rule-enforcement/pct}{7.7}
\pgfkeyssetvalue{/hm/q412/code/ch-rule-enforcement/label}{Challenge: Rule enforcement difficulty}
\pgfkeyssetvalue{/hm/q412/code/ch-rule-enforcement/group}{Challenges}
\pgfkeyssetvalue{/hm/q412/code/ch-inexperience/n}{1}
\pgfkeyssetvalue{/hm/q412/code/ch-inexperience/pct}{7.7}
\pgfkeyssetvalue{/hm/q412/code/ch-inexperience/label}{Challenge: Inexperience with GenAI-era assessment}
\pgfkeyssetvalue{/hm/q412/code/ch-inexperience/group}{Challenges}
\pgfkeyssetvalue{/hm/q412/code/nc-live-interactive/n}{2}
\pgfkeyssetvalue{/hm/q412/code/nc-live-interactive/pct}{15.4}
\pgfkeyssetvalue{/hm/q412/code/nc-live-interactive/label}{Non-challenge: Live/interactive assessment formats}
\pgfkeyssetvalue{/hm/q412/code/nc-live-interactive/group}{Non-challenges (mitigating strategies)}
\pgfkeyssetvalue{/hm/q412/code/nc-process-tracking/n}{1}
\pgfkeyssetvalue{/hm/q412/code/nc-process-tracking/pct}{7.7}
\pgfkeyssetvalue{/hm/q412/code/nc-process-tracking/label}{Non-challenge: Process tracking}
\pgfkeyssetvalue{/hm/q412/code/nc-process-tracking/group}{Non-challenges (mitigating strategies)}
\pgfkeyssetvalue{/hm/q412/code/nc-higher-order/n}{1}
\pgfkeyssetvalue{/hm/q412/code/nc-higher-order/pct}{7.7}
\pgfkeyssetvalue{/hm/q412/code/nc-higher-order/label}{Non-challenge: Higher-order question focus}
\pgfkeyssetvalue{/hm/q412/code/nc-higher-order/group}{Non-challenges (mitigating strategies)}
\pgfkeyssetvalue{/hm/q412/code/nc-additive-ai-skill/n}{1}
\pgfkeyssetvalue{/hm/q412/code/nc-additive-ai-skill/pct}{7.7}
\pgfkeyssetvalue{/hm/q412/code/nc-additive-ai-skill/label}{Non-challenge: Additive AI-skill assessment}
\pgfkeyssetvalue{/hm/q412/code/nc-additive-ai-skill/group}{Non-challenges (mitigating strategies)}
\pgfkeyssetvalue{/hm/q412/code/ai-banned-na/n}{1}
\pgfkeyssetvalue{/hm/q412/code/ai-banned-na/pct}{7.7}
\pgfkeyssetvalue{/hm/q412/code/ai-banned-na/label}{AI banned — question not applicable}
\pgfkeyssetvalue{/hm/q412/code/ai-banned-na/group}{Other}
\pgfkeyssetvalue{/hm/q412/group/challenges/label}{Challenges}
\pgfkeyssetvalue{/hm/q412/group/challenges/n}{9}
\pgfkeyssetvalue{/hm/q412/group/challenges/pct}{69.2}
\pgfkeyssetvalue{/hm/q412/group/non-challenges-mitigating-strategies/label}{Non-challenges (mitigating strategies)}
\pgfkeyssetvalue{/hm/q412/group/non-challenges-mitigating-strategies/n}{5}
\pgfkeyssetvalue{/hm/q412/group/non-challenges-mitigating-strategies/pct}{38.5}
\pgfkeyssetvalue{/hm/q412/group/other/label}{Other}
\pgfkeyssetvalue{/hm/q412/group/other/n}{1}
\pgfkeyssetvalue{/hm/q412/group/other/pct}{7.7}
\pgfkeyssetvalue{/hm/q412/top1/label}{Challenge: Competence not verifiable}
\pgfkeyssetvalue{/hm/q412/top1/slug}{ch-competence-not-verifiable}
\pgfkeyssetvalue{/hm/q412/top1/n}{4}
\pgfkeyssetvalue{/hm/q412/top1/pct}{30.8}
\pgfkeyssetvalue{/hm/q412/top2/label}{Challenge: Debugging skills hidden}
\pgfkeyssetvalue{/hm/q412/top2/slug}{ch-debugging-hidden}
\pgfkeyssetvalue{/hm/q412/top2/n}{2}
\pgfkeyssetvalue{/hm/q412/top2/pct}{15.4}
\pgfkeyssetvalue{/hm/q412/top3/label}{Non-challenge: Live/interactive assessment formats}
\pgfkeyssetvalue{/hm/q412/top3/slug}{nc-live-interactive}
\pgfkeyssetvalue{/hm/q412/top3/n}{2}
\pgfkeyssetvalue{/hm/q412/top3/pct}{15.4}

%% file: figures/q47-stacked.tex
\newcommand{\edqfourseventfigure}{%
\begin{figure}[t!]
  \centering
  \includegraphics[width=0.49\textwidth]{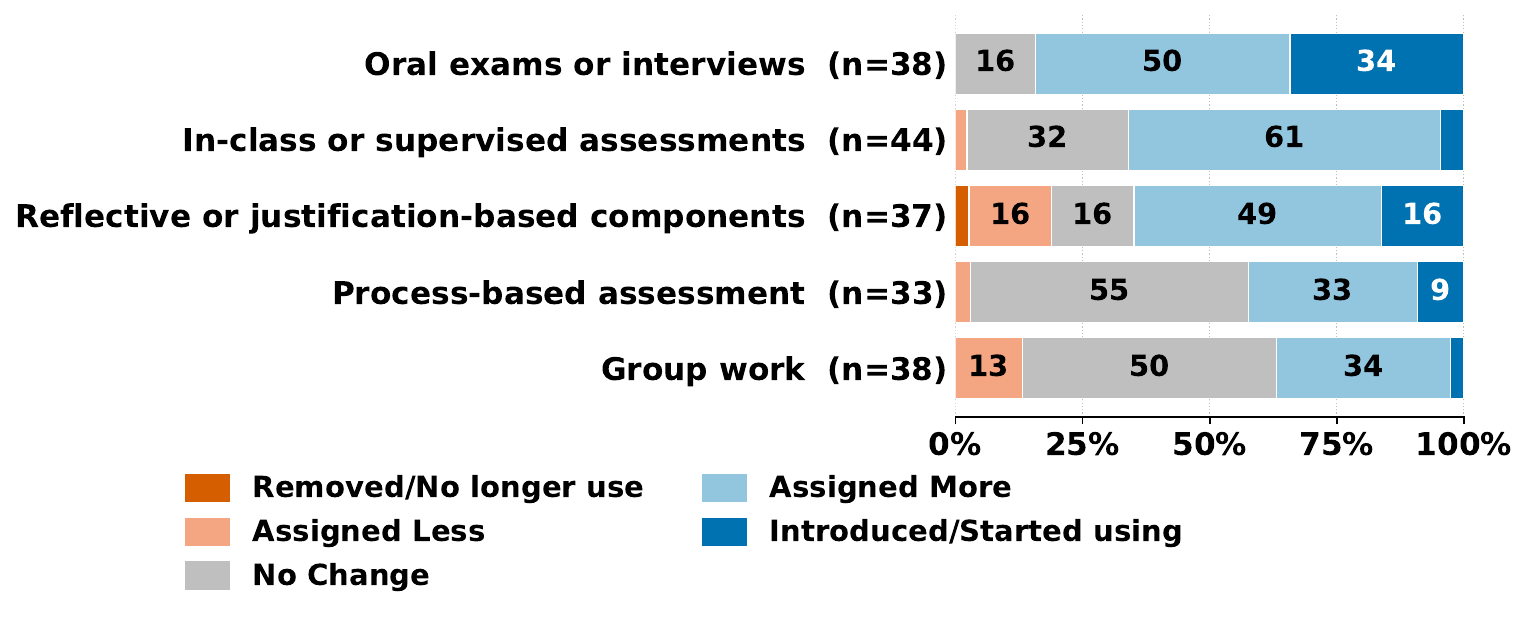}
 \caption{Educator changes to assessment formats (ED-Q4.7), ordered by share reporting an increase (\textit{Assigned More}/\textit{Introduced/Started using}).}
  \label{fig:ed-q47-stacked}
\vspace{-0.6cm}
\end{figure}
}

%% file: figures/skills-importance.tex
\newcommand{\skillimpfigure}{%
\begin{figure}[t!]
  \centering
  \includegraphics[width=0.49\textwidth]{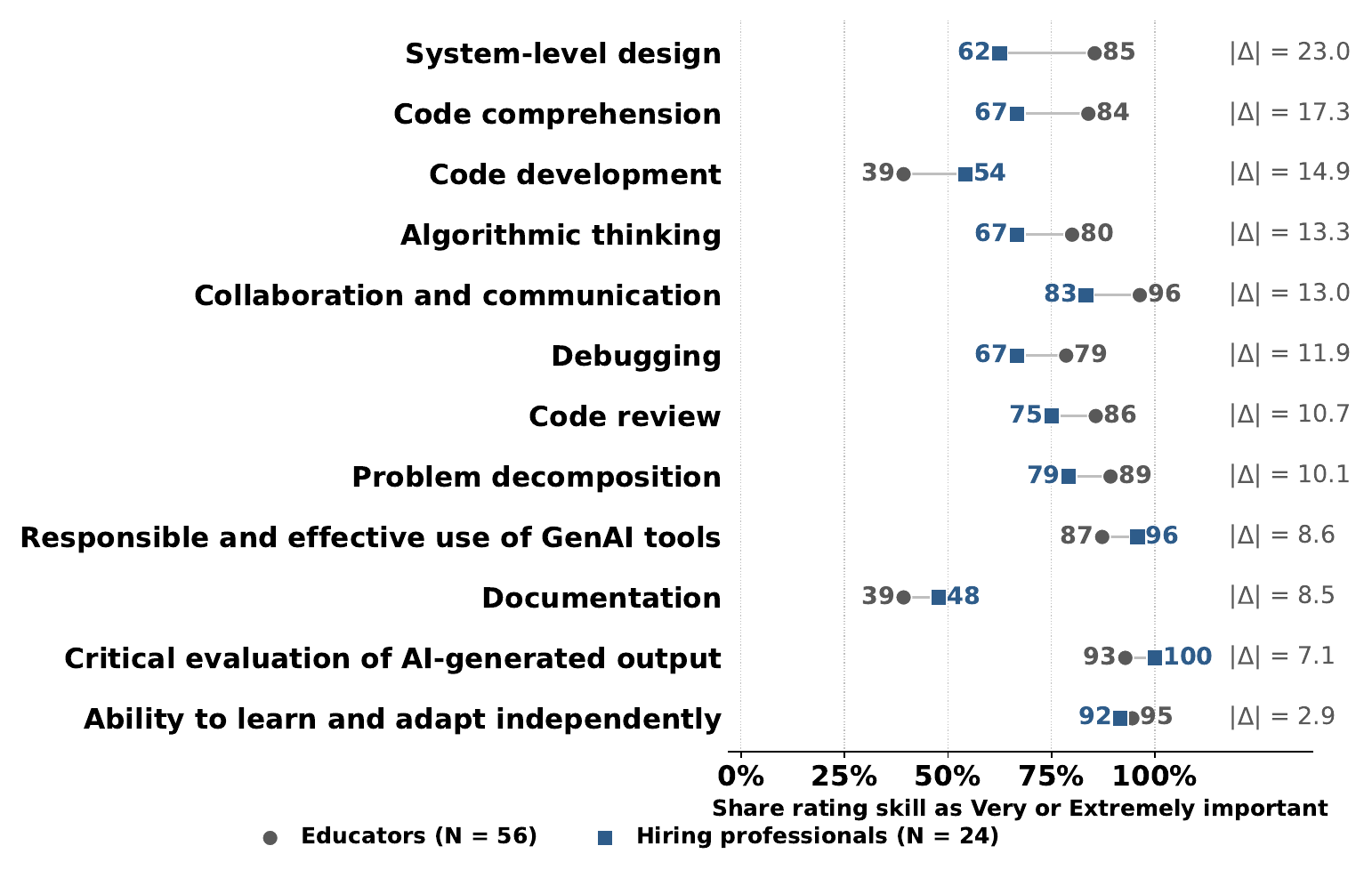}
 \caption{Educator vs.\ hiring-professional skill-importance ratings. Share of substantive respondents per population rating the skill \textit{Very} or \textit{Extremely important}, ordered by descending gap (shown at right).\vspace{-0.5cm}}
\label{fig:skills-imp}
\end{figure}
}

%% file: figures/skills-readiness.tex
\newcommand{\skillreadyfigure}{%
\begin{figure}[t!]
  \centering
  \includegraphics[width=0.49\textwidth]{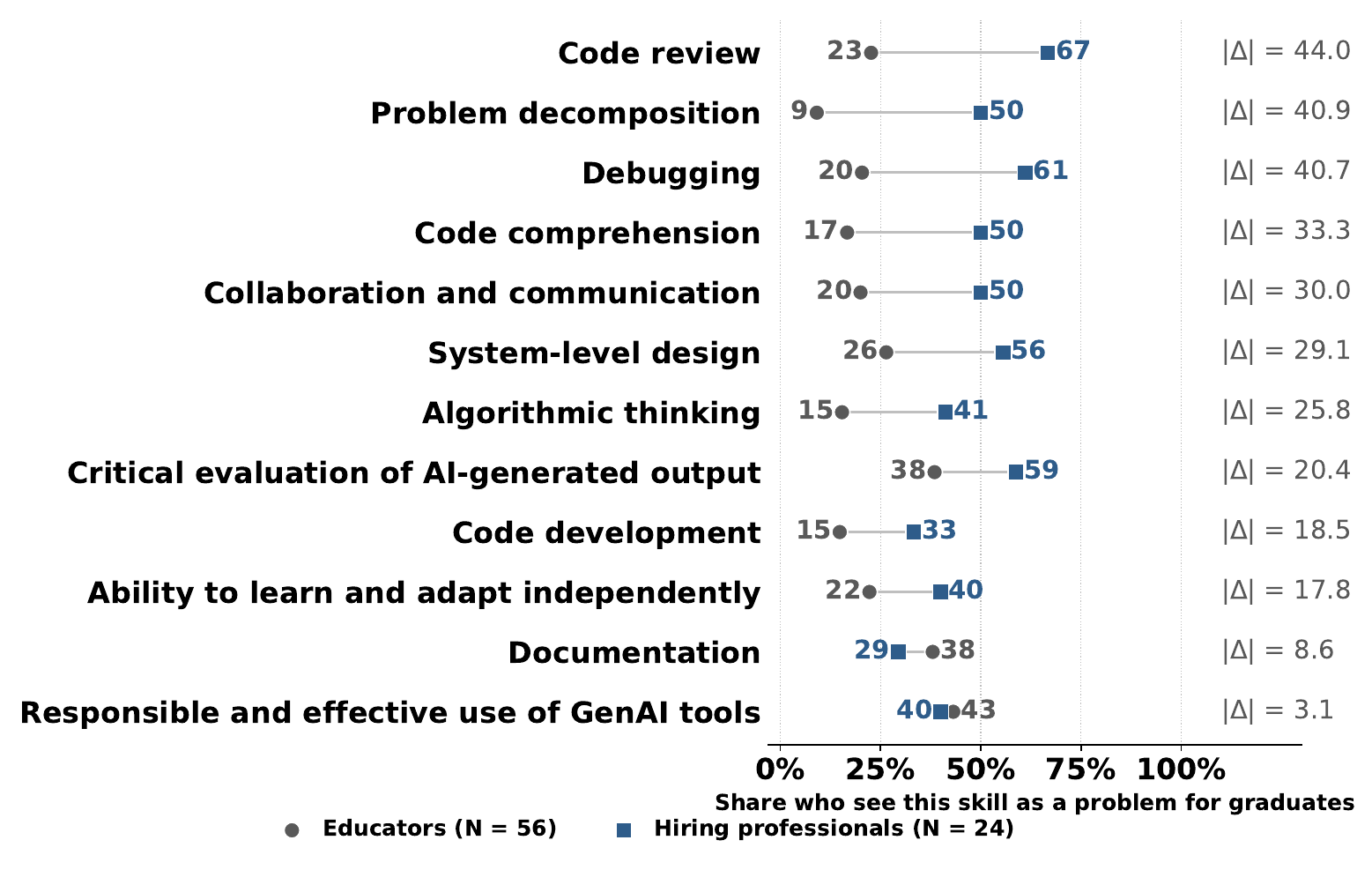}
\caption{Educator vs.\ hiring-professional readiness ratings for 12 skills. Share of substantive responses per population viewing the skill as a problem for graduates, ordered by descending gap (shown at right).}
  \label{fig:skills-ready}
  \vspace{-0.7cm}
\end{figure}
}

%% file: findings/findingsrq1.tex
\newcommand{\findingrqone}{%
\begin{figure}[h!]
\centering
\begin{findingsbox}[RQ1 -- Teaching and Assessment Practices]
\small
GenAI adoption in teaching prep is common (\valpct{q3/q31/users/pct}\%), focused on content creation over student interaction. All educators guide students on GenAI use, but few (\valpct{q43/code/citation-format/pct}\%) specify how to disclose it. Assessments are shifting to GenAI-resistant formats: \valpct{q4/q47/oral/substantive/increased-pct}\% increased oral exams and \valpct{q4/q47/in-class/substantive/increased-pct}\% in-class assessments. Most would accept GenAI-written code if students fully understand it (\valpct{q4/q410/positive/pct}\%), yet only \valpct{q4/q412/positive/pct}\% can confidently detect it.
\end{findingsbox}
\vspace{-0.3cm}
\end{figure}
}

%% file: findings/findingsrq2.tex
\newcommand{\findingrqtwo}{%
\begin{figure}[h!]
\centering
\begin{findingsbox}[RQ2 -- Hiring Practices Changes]
\small
Individual GenAI use is near-universal (\hmvalpct{q3/q32/top-box/pct}\%). Only \hmvalpct{q3/q35/clear-formal/pct}\% report a clear, formal policy, and these govern \textit{how} GenAI is used (\hmvalpct{q36/group/governance/pct}\%) more than \textit{whether} it must be used (\hmvalpct{q36/group/mandate/pct}\%). Candidate-evaluation practices are still emerging. Hiring professionals often ask about GenAI experience (\hmvalpct{q4/q48/positive/pct}\%), yet only \hmvalpct{q4/q49/positive/pct}\% are confident their practices distinguish genuine skill. Interviews are shifting toward observable, higher-order tasks where AI offers little help, paralleling educators' move to oral and in-class formats.
\end{findingsbox}
\vspace{-0.8cm}
\end{figure}
}

%% file: findings/findingsrq3.tex
\newcommand{\findingrqthree}{%
\begin{figure}[h!]
\centering
\begin{findingsbox}[RQ3 -- Graduate Preparedness Alignment]
\small
Both populations agree on which skills matter most, rating \emph{Critical evaluation of AI-generated output}, \emph{Responsible and effective use of GenAI tools}, and \emph{Ability to learn and adapt independently} highest. They disagree on graduate readiness for those skills: hiring professionals report larger gaps on 10 of 12, concentrated on foundational, non-GenAI-specific skills such as \emph{code review} and \emph{problem decomposition}.
\end{findingsbox}
\vspace{-0.7cm}
\end{figure}
}